\documentclass[twocolumn]{aastex631}
\usepackage{xspace}
\usepackage[version=4]{mhchem}
\usepackage{graphicx}
\usepackage{multirow}
\usepackage{txfonts}
\usepackage{xcolor}
\usepackage{float}

\newcommand{\mystar}{GJ\,3090\xspace}
\newcommand{\myplanet}{GJ\,3090\,b\xspace}

\newcommand\microns{\textmu m\xspace}

\begin{document}

\title{{Escaping Helium and a Highly Muted Spectrum Suggest a Metal-Enriched Atmosphere} on Sub-Neptune \myplanet from JWST Transit Spectroscopy}

\correspondingauthor{Eva-Maria Ahrer}
\email{ahrer@mpia.de}

\newcommand{\umontreal}{Department of Physics and Trottier Institute for Research on Exoplanets, Universit\'{e} de Montr\'{e}al, Montréal, QC, Canada}
\newcommand{\mpia}{Max-Planck-Institut f\"{u}r Astronomie, K\"{o}nigstuhl 17, 69117 Heidelberg, Germany}

\author[0000-0003-0973-8426]{Eva-Maria Ahrer}
\affil{\mpia}

\author[0000-0002-3328-1203]{Michael Radica} 
\altaffiliation{NSERC Postdoctoral Fellow}
\affiliation{Department of Astronomy \& Astrophysics, University of Chicago, 5640 South Ellis Avenue, Chicago, IL 60637, USA}
\affil{\umontreal}

\author[0000-0002-2875-917X]{Caroline Piaulet-Ghorayeb}
\altaffiliation{E. Margaret Burbridge Postdoctoral Fellow}
\affiliation{Department of Astronomy \& Astrophysics, University of Chicago, 5640 South Ellis Avenue, Chicago, IL 60637, USA}
\affil{\umontreal}

\author[0009-0002-2380-6683]{Eshan Raul}
\affil{Department of Astronomy, University of Wisconsin--Madison, Madison, WI 53706, USA}

\author[0000-0002-3295-1279]{Lindsey Wiser} 
\affil{School of Earth \& Space Exploration, Arizona State University, Tempe, AZ 85257, USA}

\author[0000-0003-0156-4564]{Luis Welbanks}
\affil{School of Earth \& Space Exploration, Arizona State University, Tempe, AZ 85257, USA}

\author[0000-0002-9147-7925]{Lorena Acu\~{n}a}
\affil{Max Planck Institute for Astronomy, K\"{o}nigstuhl 17, D-69117 Heidelberg, Germany}

\author[0000-0002-1199-9759]{Romain Allart}
\affil{\umontreal}

\author[0000-0002-2195-735X]{Louis-Philippe Coulombe} 
\affil{\umontreal}

\author[0000-0002-3191-2200]{Amy Louca}
\affil{SRON Netherlands Institute for Space Research, Niels Bohrweg 4, 2333 CA Leiden, The Netherlands}
\affil{Leiden Observatory, Leiden University, P.O.\ Box 9513, 2300 RA Leiden, The Netherlands}

\author[0000-0003-4816-3469]{Ryan MacDonald}
\affil{Department of Astronomy, University of Michigan, Ann Arbor, MI, USA}

\author[0000-0001-9518-9691]{Morgan Saidel}
\affil{Division of Geological and Planetary Sciences, California Institute of Technology, Pasadena, CA 91125, USA}

\author[0000-0001-5442-1300]{Thomas M. Evans-Soma}
\affiliation{School of Information and Physical Sciences, University of Newcastle, Callaghan, NSW, Australia}
\affil{\mpia}

\author[0000-0001-5578-1498]{Bj\"{o}rn Benneke} 
\affil{\umontreal}

\author[0000-0002-4997-0847]{Duncan Christie}
\affil{\mpia}

\author[0000-0002-9539-4203]{Thomas G. Beatty}
\affiliation{Department of Astronomy, University of Wisconsin--Madison, Madison, WI 53706, USA}

\author[0000-0001-9291-5555]{Charles Cadieux}
\affil{\umontreal}

\author[0000-0001-5383-9393]{Ryan Cloutier}
\affil{Department of Physics \& Astronomy, McMaster University, 1280 Main St W, Hamilton, ON, L8S 4L8, Canada}

\author[0000-0001-5485-4675]{René Doyon}
\affil{\umontreal}

\author[0000-0002-9843-4354]{Jonathan J. Fortney}
\affil{Department of Astronomy \& Astrophysics, University of California, Santa Cruz, CA 95064, USA}

\author[0009-0003-2576-9422]{Anna Gagnebin}
\affil{Department of Astronomy \& Astrophysics, University of California, Santa Cruz, CA 95064, USA}

\author[0009-0007-9356-8576]{Cyril Gapp}
\affil{\mpia}
\affiliation{Department of Physics and Astronomy, Heidelberg University, Im Neuenheimer Feld 226, D-69120 Heidelberg, Germany}

\author[0000-0001-5271-0635]{Hamish Innes}
\affil{Department of Earth Sciences, Freie Universit\"{a}t Berlin, Malteserstr. 74-100, 12249 Berlin, Germany}
\affil{Institute of Planetary Research, German Aerospace Center (DLR), Rutherfordstraße 2, 12489 Berlin, Germany}

\author[0000-0002-5375-4725]{Heather A.\ Knutson}
\affil{Division of Geological and Planetary Sciences, California Institute of Technology, Pasadena, CA 91125, USA}

\author[0000-0002-9258-5311]{Thaddeus Komacek}
\affil{Department of Physics, University of Oxford, Oxford, OX1 3PW, UK}

\author[0000-0001-6878-4866]{Joshua Krissansen-Totton}
\affil{Department of Earth and Space Sciences/Astrobiology Program, University of Washington, Seattle, WA, USA}


\author[0000-0002-0747-8862]{Yamila Miguel}
\affil{Leiden Observatory, Leiden University, P.O. Box 9513, 2300 RA Leiden, The Netherlands}
\affil{SRON Netherlands Institute for Space Research, Niels Bohrweg 4, 2333 CA Leiden, The Netherlands}


\author[0000-0002-5887-1197]{Raymond Pierrehumbert}
\affil{Department of Physics, University of Oxford, Oxford, OX1 3PW, UK}

\author[0000-0001-6809-3520]{Pierre-Alexis Roy} 
\affil{\umontreal}

\author[0000-0002-0298-8089]{Hilke E. Schlichting}
\affil{Department of Earth, Planetary, and Space Sciences, University of California, Los Angeles, Los Angeles, CA 90095, USA}

\begin{abstract}
Sub-Neptunes, the most common planet type, remain poorly understood. Their atmospheres are expected to be diverse, but their compositions are challenging to determine, even with JWST. Here, we present the first JWST spectroscopic study of the warm sub-Neptune \myplanet (2.13\,R$_\oplus$, T$_\mathrm{eq, A=0.3}$$\sim$700\,K) which orbits an M2V star, making it a favourable target for atmosphere characterization. We observed four transits of \myplanet; two each using JWST NIRISS/SOSS and NIRSpec/G395H, yielding wavelength coverage from 0.6 -- 5.2 \microns.
{  We detect the signature of the 10833\,\AA\ metastable Helium triplet at a statistical significance of 5.5$\sigma$ with an amplitude of 434$\pm79$\,ppm, marking the first such detection in a sub-Neptune with JWST. This amplitude is significantly smaller than predicted by solar-metallicity forward models, suggesting a metal-enriched atmosphere which decreases the mass-loss rate and attenuates the Helium feature amplitude. 
Moreover, we find that stellar contamination, in the form of the transit light source effect, dominates the NIRISS transmission spectra, with unocculted spot and faculae properties varying across the two visits separated in time by approximately six months.
Free retrieval analyses on the NIRSpec/G395H spectrum find tentative evidence for highly muted features and a lack of CH$_4$.
These findings are best explained by a high metallicity atmosphere ($>$100$\times$ solar at 3$\sigma$ confidence, for clouds at $\sim$\textmu bar pressures) using chemically-consistent retrievals and self-consistent model grids.
Further observations of \myplanet are needed for tighter constraints on the atmospheric abundances, and to gain a deeper understanding of the processes that led to its potential metal enrichment.}
\end{abstract}



\section{Introduction} 
\label{sec:intro}

In the decades since the first exoplanets were found around Sun-like stars \citep[e.g.,][]{Mayor1995AStar,Charbonneau2000DetectionStar,Henry2000APlanet}, thousands of new planets have been detected. One of the most impactful outcomes of these discoveries is that planets intermediate in mass and radius to Earth and Neptune, so-called ``sub-Neptunes'', are in fact, the most common type of planet in the galaxy \citep{borucki_kepler_2010, fulton_california-kepler_2017, fulton_california-kepler_2018}. Moreover, population-level studies have found that the radius distribution of these small planets shows a gap in occurrence rate between 1.8 -- 2.0 Earth radii around FGK stars known as the ``radius valley'' \citep{fulton_california-kepler_2017, petigura_california-kepler_2022}. This is commonly thought to divide these exoplanets into a smaller, likely rocky population and a larger, gaseous population \citep[e.g.,][]{owen_evaporation_2017, ginzburg_core-powered_2018, lee_breeding_2016}. However, the picture is blurrier for sub-Neptunes around M dwarf stars, which show a less well-defined radius valley \citep{cloutier_evolution_2020, ho_shallower_2024, venturini_fading_2024}, and a significant population that have densities suggestive of a substantial volatile content \citep{luque_density_2022, Rogers2023}. 

 {Previous modelling efforts have illustrated the impacts of atmospheric evolution processes (e.g., core-powered mass loss; \citealp{ginzburg_super-earth_2016, ginzburg_core-powered_2018, gupta_sculpting_2019}, and photoevaporation; \citealp{lopez_how_2012,owen_evaporation_2017,rogers_photoevaporation_2021}) on the present day composition and structure of sub-Neptunes. Such mass loss processes in particular, are well traced \citep[if not necessarily well constrained; e.g.,][]{Zhang2024} by observations of the metastable 10833\,\AA\ He triplet \citep{oklopcic_2018_new, oklopcic_helium_2019}. To date, numerous detections of escaping atmospheres have been made \citep[e.g.,][]{spake_helium_2018, Nortmann2018Helium, Mansfield2018HeliumHATP11, allart_high-resolution_2019,Kirk2020ConfirmationII/NIRSPEC,spake_posttransit_2021,ZhangTOI560,Guilley2024Heliumsurvey,Orell-Miquel2024Helium,vissapragada_helium_2024,Zhang2024}, with He observations of young mini-Neptunes in particular highlighting the impacts of mass loss in sculpting the radius valley around Sun-like stars \citep{Zhang2022, ZhangTOI560, zhang_detection_2023}. Recently, JWST observations, particularly with NIRISS/SOSS \citep[e.g.,][]{fu_water_2023, fournier-tondreau_near-infrared_2024}, have proved efficient at detecting escaping He from exoplanet atmospheres, potentially providing a new avenue to examine the effects of atmosphere loss on the population of sub-Neptunes orbiting late-type stars.} 

 {In addition to atmosphere loss, a range of possible formation and migration histories \citep[e.g.,][]{burn_radius_2024}, as well as, interactions with the surface \citep[e.g., with a magma ocean;][]{Chachan2018magma, kite_superabundance_2019, lichtenberg_vertically_2021, schlichting_chemical_2022} can also result in a wide range of bulk compositions for sub-Neptunes. Critically though, it has long been known that the nature of sub-Neptune exoplanets cannot be uniquely revealed by bulk density measurements alone \citep{valencia_composition_2010, rogers_framework_2010}, with atmosphere observations being the necessary piece of information to unveil the nature of this most common class of planets in the galaxy. }


Prior to the JWST era, observational studies of sub-Neptune atmospheres { (aside from Helium escape)} were routinely hindered by insufficient wavelength coverage and sensitivity, resulting in degeneracies in atmosphere composition due to clouds \citep[e.g.,][]{bean_ground-based_2013, berta_flat_2012, kreidberg_clouds_2014, benneke_sub-neptune_2019, roy_water_2023}, overlap in molecular absorption \citep[e.g.,][]{benneke_water_2019, tsiaras_water_2019, Bezard2022},  {or stellar contamination \citep[e.g.,][]{edwards_2021_hubble, barclay_stellar_2021, mikal-evans_hubble_2023}}.

However, with the launch of JWST we have now begun to uniquely constrain the chemical composition, and thereby uncover the surprising diversity of sub-Neptune atmospheres. \citet{madhusudhan_carbon-bearing_2023} reported carbon-bearing molecules in the atmosphere of the cool, 255\,K sub-Neptune K2-18 b from a JWST NIRISS and NIRSpec transmission spectrum, and suggested that the inferred composition was consistent with predictions for temperate ocean-covered planets with H2-rich atmospheres \citep[i.e., Hycean planets, e.g.,][]{madhusudhan_habitability_2021}. Planetary conditions that are consistent with predictions for Hycean environments were also reported by \citet{holmberg_possible_2024} for the 354\,K sub-Neptune TOI-270\,d (which was also previously studied by HST \citealp{mikal-evans_hubble_2023})  using  NIRSpec observations. When combining the NIRSpec observations analyzed by \citet{holmberg_possible_2024} with additional NIRISS data, \citet{benneke_jwst_2024} find a similar chemical composition (i.e., H$_2$O and CO$_2$ abundances) for the atmosphere of TOI-270d than that reported by \citet{holmberg_possible_2024},  but interpret the chemical conditions as signatures of a volatile-rich, miscible-envelope scenario, where roughly half the mass of TOI-270\,d’s envelope is composed of high-mean-molecular-weight volatiles well mixed with H$_2$/He. The interpretation of the inferred chemical composition for sub-Neptunes, and their implications for the diversity in their climate and habitable conditions is a highly active and growing area of research \citep[e.g.,][]{innes_runaway_2023, glein2024, Cooke2024,  Rigby2024, Shorttle2024K2-18b, wogan_jwst_2024, Leconte_moist_convection_2024, Schmidt2025}.

More recently, \citet{Piaulet_Ghorayeb_2024} revealed an H$_2$O-rich, ``steam atmosphere'' on the warm, 616\,K sub-Neptune GJ 9827\,d, and evidence for S-bearing compounds have been found in the atmospheres of GJ 3470\,b by \citet{beatty_sulfur_2024} and L~98-59\,d by \citet{Gressier2024L98-59d} and \citet{Banerjee2024L98-59d}. While the previous five planets have suggested metallicities in the range of $\sim$100--500$\times$ solar, a much larger, $1000\times$ solar metallicity was found for the canonical sub-Neptune GJ 1214\,b \citep{Schlawin2024GJ1214b} along with significant amounts of haze \citep[]{kempton_reflective_2023, Gao2023GJ1214, Ohno2024GJ1214b, Schlawin2024GJ1214b}. 

These early results demonstrate that the diversity of sub-Neptunes extends well beyond typical ideas of rocky `super-Earths' and H$_2$/He-rich `mini-Neptunes'. {  They have also shown that robustly constraining atmospheric metallicity is a challenging feat due to the increasing variety of unknowns when it comes to sub-Neptune atmospheres.} To further develop our understanding of sub-Neptunes as a population, we require a larger sample size of planets with well-characterized atmospheres. 

\subsection{The GJ 3090 System}
\label{sec:gj3090b_system}

\mystar is an M2V star hosting one confirmed planet --- the sub-Neptune GJ 3090\,b \citep{Almenara2022GJCharacterisation}. The star was determined to have a mass of $0.519 \pm 0.013$\,M$_\odot$ and a radius of $0.516 \pm 0.016$\,R$_{\odot}$ (see Table\,\ref{tab:stellar_parameters}). It is a relatively bright and nearby star with J$_{\rm{mag}}$ = 8.168$\pm$0.021 \citep[2MASS;][]{Skrutskie2006The2MASS}, and a distance of $22.444\pm0.013$\,pc \citep{Almenara2022GJCharacterisation}. 

\begin{figure}
    \centering
    \includegraphics[width=.48\textwidth]{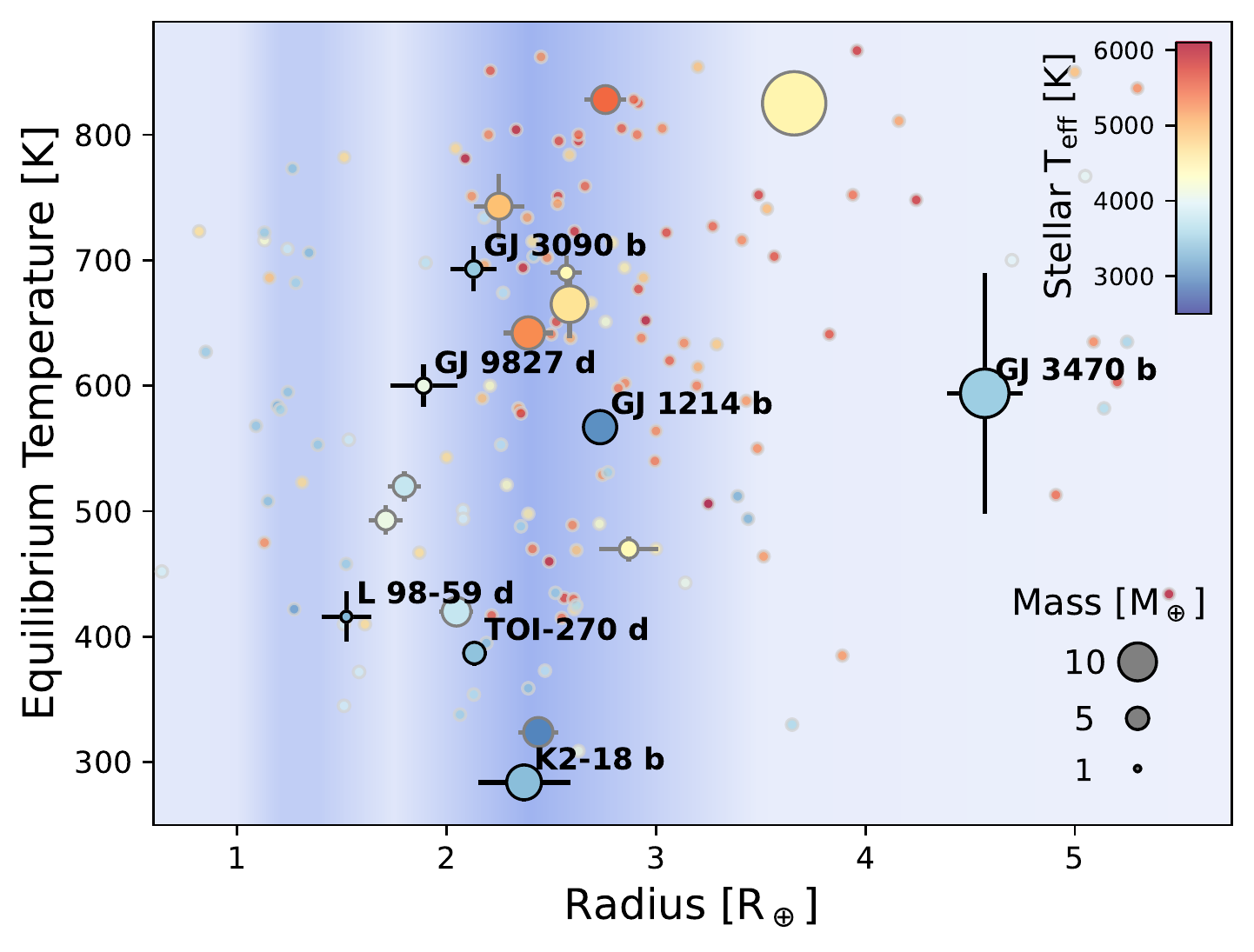}
    \caption{GJ 3090\,b in context with the population of sub-Neptune planets in equilibrium temperature vs.\ planet radius space. Markers are coloured by the effective temperature of the stellar host and marker sizes reflect the planet mass. The background shading represents the distribution of sub-Neptune planets from \citet{fulton_california-kepler_2017}, highlighting the radius valley near 1.7\,R$_\oplus$. Planets with published JWST near-infrared atmosphere spectra are outlined in black and labelled. Planets that will have JWST atmosphere observations by the end of Cycle 3 are outlined in grey.}
    \label{fig: Population}
\end{figure}

Using observations with the Transiting Exoplanet Survey Satellite \citep[TESS; ][]{Ricker2015TransitingTESS}, \citet{Almenara2022GJCharacterisation} identified \myplanet with a radius of $2.13 \pm 0.11$\,R$_\oplus$ and a period of $2.8531054\pm0.0000023$\,days. In addition, the High Accuracy Radial Velocity Planet Searcher (HARPS) spectrograph \citep{Mayor2003SettingHARPS} was used to determine the planet's mass as $3.34 \pm 0.72$\,M$_\oplus$, resulting in a mean density of $1.89_{-0.45}^{+0.52}$\,g\,cm$^{-3}$ \citep{Almenara2022GJCharacterisation}. The equilibrium temperature was determined to be T$_\mathrm{eq, A=0.3} = 693 \pm 18$\,K. The authors also found some evidence for an eccentric orbit, with a 95\% confidence upper limit of $ e < 0.32$. In addition to planet b, their radial velocity (RV) data suggests an outer planet candidate at a period of 13 days and a minimum mass of $17.1_{-3.2}^{+8.9}$\,M$_\oplus$, whose transits were not detected with TESS. We show GJ 3090\,b in context with the population of sub-Neptunes in Figure~\ref{fig: Population}. In particular, GJ 3090\,b is the warmest of all sub-Neptunes with published atmosphere spectra; $\sim$100\,K warmer than GJ 9827\,d and nearly double the temperature of K2-18\,b. Recently, \citet{Parker2025GJ3090} found no atmospheric signatures in \myplanet's atmosphere using four CRIRES+ transits in the K-Band (sensitive to \ce{CH4}, \ce{H2O}, \ce{NH3}, and \ce{H2S}). Their study suggests a metal-rich atmosphere with $>150 \times$ solar metallicity at a mean molecular weight $> 7.1$\,g/mol and an unconstrained cloud layer.

\begin{table}
    \centering
        \caption{Stellar parameters of \mystar as determined by \citet{Almenara2022GJCharacterisation} and adopted throughout this work.}
    \label{tab:stellar_parameters}
    \begin{tabular}{lc}
    \hline
    \hline 
        Parameter & Value \\ \hline
        Effective temperature, T$_\mathrm{eff}$ (K) & $3556 \pm 70$\\
        Spectral Type & M2V \\
        Metallicity, $Z$ ([Fe/H]) & $-0.060 \pm 0.120$ \\
        Surface gravity, $\rm \log (g)$ (log$_{10}$(cm/s$^2$)) & $4.727 \pm 0.029$\\
        Radius, $\rm R_s$ (R$_\odot$)  &  $0.516 \pm 0.016$\\
        Mass, $\rm M_s$ (M$_\odot$) & $0.519 \pm 0.013$\\
        Rotation period, P$_\mathrm{rot}$ (days) & $17.65 \pm 0.48$ \\
        \hline 
        \hline
    \end{tabular}
\end{table}

\citet{Almenara2022GJCharacterisation} also showed that \mystar exhibits photometric variations in the TESS light curves at a $\sim$1.5\% level with a period of $17.65 \pm 0.48$\,d, which is consistent with the stellar rotation period inferred from the RV ($\sim$17.73\,d) and archival WASP data ($18.20 \pm 0.40$\,d). Using the star's rotation period and mass, \citet{Almenara2022GJCharacterisation} estimated the system to be relatively young ($1.02_{-0.23}^{+0.15}$\,Gyrs). Their analysis of common activity indicators in the HARPS spectra confirms that \mystar is moderately active.

\subsection{Structure of this Work}

In this work we add to the growing sample of sub-Neptunes with precise atmosphere observations by presenting the first JWST transmission spectrum of \myplanet, a sub-Neptune at the outer edge of the radius valley. We outline our reduction and analysis of two transits with NIRISS/SOSS and two transits with NIRSpec/G395H in Sections~\ref{sec: observations} and \ref{sec: light curve fits}, with special emphasis on metastable helium in Section~\ref{sec: Helium}. We present our atmosphere modelling in Section~\ref{sec:atmosphere} and discuss our results in Section~\ref{sec:discussion} before concluding in Section~\ref{sec:conclusions}.

\section{Observations and Data Analysis}
\label{sec: observations}

The observations presented here were taken as part of JWST program GO 4098 (PIs: Benneke \& Evans-Soma), with the purpose of ``Exploring the existence and diversity of volatile-rich water worlds''. Four transits of \myplanet were observed with JWST, two with the NIRISS instrument \citep{Doyon2023} in SOSS mode \citep{albert_near_2023} on 6 December 2023 and 4 July 2024 covering the 0.6 -- 2.8 \textmu m waveband, and two with the NIRSpec instrument \citep{birkmann_near-infrared_2022} using the G395H grating on 2 Aug 2023 and 23 Sep 2023, adding the 2.7 -- 5.2 \textmu m wavelength range. 

\subsection{NIRISS/SOSS}
For the NIRISS/SOSS observations we used two groups per integration and a total of 779 integrations, resulting in an overall observing time of 3.6\,hours which includes an out-of-transit time of $\sim$2.3\,hours and 1.281\,hours of transit time (including ingress and egress). The readout mode was NISRAPID and the subarray was SUBSTRIP256, providing the full 0.6--2.85\,µm wavelength coverage \citep{albert_near_2023}.

\subsubsection{\texttt{exoTEDRF} Reduction}
\label{sec:exoTEDRF}

We reduce the NIRISS/SOSS observations using \texttt{exoTEDRF} \citep{Radica2024b, Radica2023, feinstein_early_2023}, closely following the procedure laid out in \citet{radica_muted_2024} and \citet{benneke_jwst_2024}. We follow the standard \texttt{Stage1} and \texttt{Stage2} steps, as has been done in many other works \citep[e.g.,][]{Radica2023, lim_atmospheric_2023, fournier-tondreau_near-infrared_2024, Cadieux2024, radica_promise_2025, Piaulet_Ghorayeb_2024}, and note any specific alterations here. In particular, due to the low number of groups observed up-the-ramp, we use the time-domain cosmic ray detection algorithm outlined in \citet{radica_muted_2024}, with an outlier threshold of 5$\sigma$. Furthermore, we perform a piece-wise background subtraction, finding optimal pre- and post-step scaling values of 0.62686 and 0.64511 compared to the reference background model, and we perform the 1/$f$ noise correction at the integration-level using the \texttt{scale-achromatic-window} method identically to \citet{benneke_jwst_2024}.

We extract the stellar spectra using a simple box aperture with a width of 30 pixels, which we find minimizes the out-of-transit baseline scatter in the white light curve. Any dilution resulting from the order 1 and order 2 self-contamination is expected to be negligible \citep{Radica2022, Darveau-Bernier2022}, and indeed, we find no difference in the resulting transmission spectra when applying the \texttt{ATOCA} extraction algorithm  {which explicitly models the order overlap}. Due to the judicious choice of telescope aperture position angle for these observations, there are no background contaminants, either dispersed or undispersed, affecting these spectra.

\subsubsection{\texttt{NAMELESS Reduction}}

The \texttt{NAMELESS} reduction pipeline \citep{coulombe_broadband_2023,Coulombe2025highlyreflectivewhiteclouds} starts with the uncalibrated files and allows us to run \texttt{jwst}'s Stages\,1 and 2 \citep{bushouse_2023_8157276} with modifications and perform other calibrations in between the default steps. We apply the following default steps: super-bias subtraction, reference pixel correction, non-linearity correction, ramp-fitting, and flat-fielding. We continue with a manual bad pixel correction, masking the ones that are consistently different to the surrounding values (e.g., by being negative or abnormally high). We correct the value of these bad pixels at all integrations using the \texttt{scipy.interpolate.griddata} bicubic interpolation function.  

This is followed by a background subtraction where the background model provided by STScI\footnote{\url{https://jwst-docs.stsci.edu/}} is scaled to match the observed background. For that we also use two independent scaling values, accounting for the characteristic abrupt transition in the intensity of the background flux.  

Because the integrations consist of only two groups for this dataset, the jump detection step is automatically skipped by the \texttt{jwst} pipeline. We thus correct for cosmic rays using the same method as described in \citet{Coulombe2025highlyreflectivewhiteclouds}, in which we compute the running median of all pixels, considering a window of 11 integrations, and clip all counts that deviate by more than 4$\sigma$ from its median.

The 1/$f$ noise is corrected by using a column-by-column scaling as described in \citet{coulombe_broadband_2023,Coulombe2025highlyreflectivewhiteclouds}. Given that the 1/$f$ noise can vary over the length of a single column, we compute this scaling factor separately for order 1 and order 2 considering only pixels that are within a 30-pixel window from the centre of the traces. When computing the scaling values for order 2, we do not consider pixels that overlap with the window of order 1. The 1/$f$ values computed from order 1 are subtracted from the full columns whereas those computed from order 2 are subtracted only from its window.

We extract the NIRISS/SOSS spectroscopic light curves from the first and second order using a simple box aperture with a width of 36\,pixels.  

\subsection{NIRSpec/G395H}

For each observation of \myplanet with NIRSpec/G395H four groups and 2845 integrations were used, summing to an overall exposure time of 3.6\,hours which consists of $\sim$2.3\,hours of out-of-transit time and 1.28\,hours of transit time (including ingress and egress). NRSRAPID was the readout mode used for taking these NIRSpec observations in combination with the SUB2048 subarray. 

\subsubsection{\texttt{exoTEDRF} Reduction}

We also reduce both NIRSpec/G395H visits with \texttt{exoTEDRF} which has been recently upgraded to also handle NIRSpec data \citep{Radica2024b} applying many of the same routines originally developed for SOSS observations. For both visits, we apply the standard \texttt{Stage1} steps as for the SOSS observations, skipping the reference pixel correction which we find imparts row-correlated noise, particularly to the NRS2 detector. We perform a 1/$f$ noise correction (which also serves to subtract background emission) at the group level by subtracting the median value of each column, masking all bad pixels as well as pixels within eight pixels above or below the target spectral trace. As with the SOSS observations, we apply a time-domain cosmic ray detection due to the low number of groups.

We then apply the standard \texttt{exoTEDRF} \texttt{Stage2}, including a new step, heavily based on the \texttt{Extract2DStep} within \texttt{jwst} \citep{bushouse_2023_8157276} to extract the wavelength solution of the observations given the specific position of the target star within the NIRSpec slit. We then repeat the 1/$f$ and background subtraction at the integration level, using the same parameters as the group-level subtraction, though we find that the addition of this step results in a negligible improvement to the noise properties of the final data frames. We also apply \texttt{exoTEDRF}'s principal component analysis-based \texttt{TracingStep} to NIRSpec data for the first time. This step reveals a sub-pixel drift in the y-position of the target trace over the course of the time series observations for both detectors and both visits, as well as a slight rotation about pixel (600, 15) for NRS1 and (500, 8) for NRS2. Finally, we locate the positions of the NRS1 and NRS2 spectral traces using the \texttt{edgetrigger} algorithm \citep{Radica2022}, and extract the stellar spectra using a box aperture extraction with a width of eight pixels around the target trace. 

\subsubsection{\texttt{Eureka!} Reduction}
\texttt{Eureka!}~is an open-source pipeline for reducing and analysing JWST and Hubble Space Telescope (HST) exoplanet transit, eclipse and phase curve observations \citep{Bell2022Eureka:Observations}\footnote{\url{https://github.com/kevin218/Eureka}}. It has been extensively applied to JWST observations \citep[see e.g.,][]{Ahrer2023EarlyNIRCam,  Moran2023HighObservations, Zieba2023Trappist1c, Beatty2024GJ3470, Xue2024HD209}. 

Starting with the uncalibrated fits files, we first run \texttt{Eureka!}'s Stages\,1 and 2, which are wrappers around the regular \texttt{jwst} pipeline and allow for changes as well as additions before running individual steps. We opted to use this capability to remove 1/$f$ noise before the groups are combined into integrations by conducting a mean column-by-column background subtraction. Conducting a background subtraction at the group-level has shown significant improvement in the noise properties for NIRSpec/G395H observations \citep[e.g.,][]{Alderson2023EarlyG395H}. 

\texttt{Eureka}'s Stage\,3 extracts the individual stellar spectra from each frame. First, we performed an outlier rejection with a $3\sigma$ rejection threshold along the time axis on the full frame. We then corrected for the curvature of the spectral trace and subtracted the background using the weighted mean value of each column, masking 10 times the median as outliers, and using a background region that is six pixels from the centre of the trace. Finally, we extracted the spectra within an aperture half width of four pixels. In Stage\,4 we generated light curves for each NRS1 and NRS2 and calculated stellar limb-darkening for each bin using \texttt{ExoTiC-LD} \citep{Grant2024} and the 1D \texttt{MPS-ATLAS} library \citep{kostogryz2023mps}, using the stellar parameters in Table\,\ref{tab:stellar_parameters}.

\section{Light Curve Analysis}
\label{sec: light curve fits}

\subsection{Joint White Light Curve Fitting}
\label{sec:joint-fit}

\begin{figure*}
    \centering
    \includegraphics[width=\linewidth]{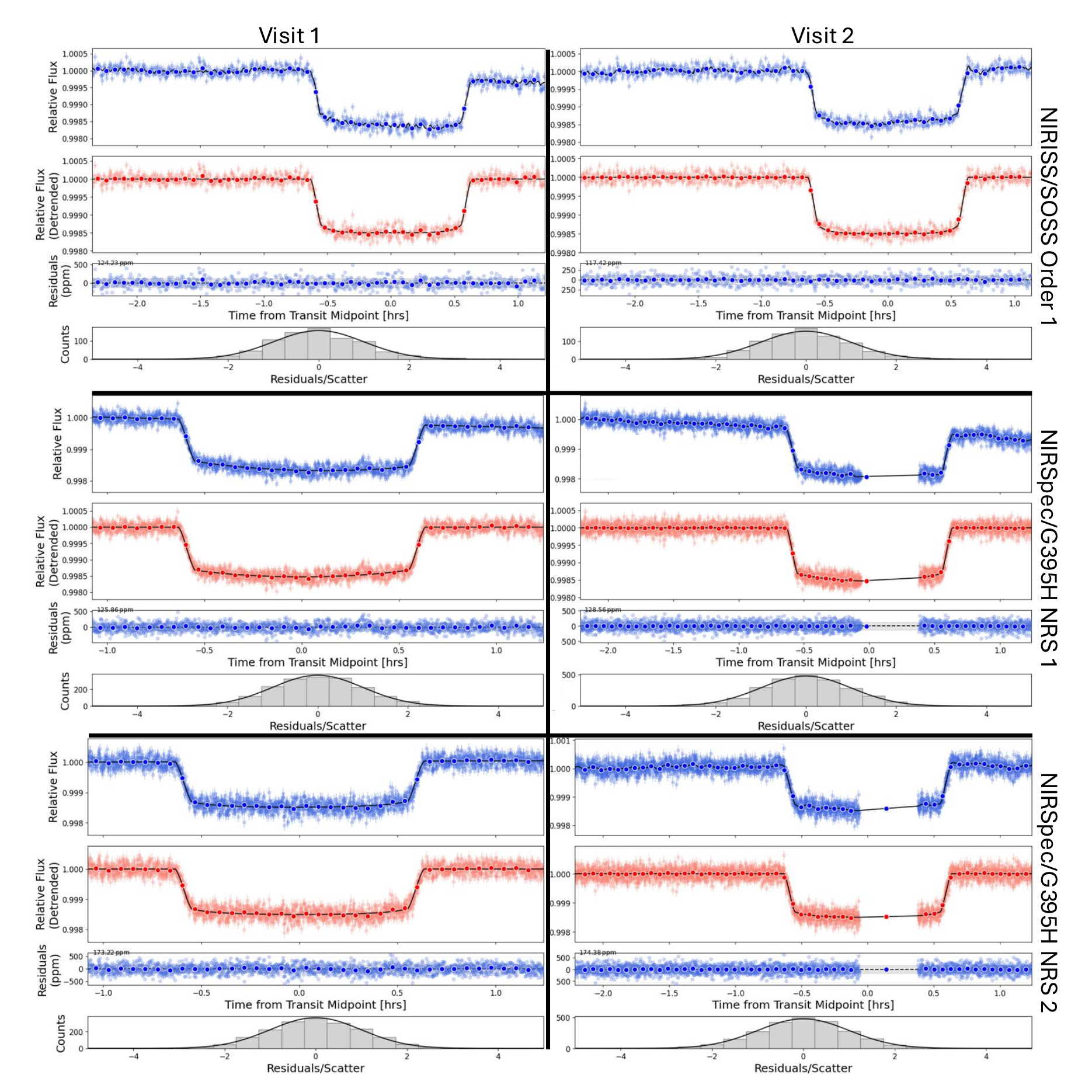} 
    \caption{Results of the joint white light curve fitting. In each of the six panels, the first row shows the raw white light curve in blue, with the best-fitting model overplotted in black. The second row shows the systematics-corrected white light curve, with the best-fitting astrophysics model in black. The third row shows the fit residuals, and the bottom row the histogram of the residuals. 
    In the outer grid, the first column of panels corresponds to the first visit with each instrument and the second column to the second visit. The first row of panels shows the order 1 white light curves from NIRISS/SOSS. Note that NIRISS/SOSS order 2 is left out in this figure as it is much noisier compared to order 1 but shows similar systematics. The second and third rows are the NIRSpec/G395H white light curves for the NRS1 and NRS2 detectors respectively. Note, integrations around the flare have been masked in both detectors of the second NIRSpec visit, and the first 1000 integrations of NIRSpec visit 1 are cut to remove large amounts of correlated noise. The raw, uncut NIRSpec light curves are shown in Figure~\ref{fig:raw nirspec}.}
    \label{fig:WLC}
\end{figure*}

In order to obtain the most accurate possible orbital solution for GJ 3090\,b, we jointly fit the light curves produced by \texttt{exoTEDRF} from all four visits; that is, we jointly fit a total of eight light curves: the first and second order white light curves from each of the two NIRISS/SOSS visits, and the NRS1 and NRS2 light curves from each NIRSpec visit. For SOSS, we construct the order 1 white light curve by summing all the extracted flux, whereas we only consider wavelengths from 0.6 -- 0.85\,µm for order 2 \citep[e.g.,][]{feinstein_early_2023, Radica2023, albert_near_2023, radica_muted_2024, fournier-tondreau_near-infrared_2024}. For NIRSpec, we use wavelengths $\lambda \in [2.8, 3.7]$\,\microns and $\lambda \in [3.8, 5.1]$\,\microns for NRS1 and NRS2 respectively \citep[e.g.,][]{Alderson2024}.    

We use the same general methodology as \citet{radica_muted_2024} for the joint fit, and employ the flexible light curve fitting library \texttt{exoUPRF}\footnote{\url{https://github.com/radicamc/exoUPRF}} \citep{Radica2024c}. The full light curve model consists of two components: an astrophysical model and a systematics model. For the astrophysical model, we use a classic \texttt{batman} \citep{Kreidberg2015BatmanPython} transit model. The achromatic orbital parameters (the mid-transit time, $T_0$, orbital period, $P$, orbital eccentricity, $e$, scaled semi-major axis, $a/R_*$, and orbital inclination, $i$) are shared between all eight light curves. Chromatic parameters (scaled planet radius, $R_p/R_*$, two parameters of the quadratic limb-darkening law) are fit to both light curves of a given order or detector (i.e., one single value of $R_p/R_*$ is fit to both SOSS order 1 white light curves). We fix the orbital period to 2.8531054\,d \citep{Almenara2022GJCharacterisation}, and use wide, uninformative priors for all other parameters, except for the eccentricity for which we put an upper bound at 0.32 (based on the 3$\sigma$ upper limit in \citet{Almenara2022GJCharacterisation} --- though c.f., Appendix~\ref{sec: eccentricity fits} for the impacts of different assumed eccentricities). We also test freely fitting the orbital period, but find a value exactly in agreement with, but marginally less precise than, that determined by \citet{Almenara2022GJCharacterisation}.   

The systematics models handle everything else in the light curves that is not the planet's transit.  {For each light curve, the optimal systematics model is determined by comparing the Bayesian Information Criterion}. For SOSS, in addition to a linear slope fit independently to each order and each visit, we also linearly detrend against the beating pattern picked up by the principal component analysis (PCA) and commonly seen in SOSS observations \citep[e.g.,][]{albert_near_2023, coulombe_broadband_2023, Cadieux2024, radica_muted_2024}. In order to handle the $\sim$100\,ppm-amplitude correlated noise that is visible by eye in the transit baseline, we also include a Gaussian process (GP) for each visit using a Matérn 3/2 kernel as implemented by \texttt{celerite} \citep{foreman-mackey_fast_2017}. We share the GP timescale between both orders, but fit the GP amplitude separately \citep[e.g.,][]{radica_muted_2024}. Moreover, we train the GP on the y-position of the spectral trace, as determined through the PCA performed by \texttt{exoTEDRF}, as we find that this performs better than a GP with time in removing residual correlated noise.

For NIRSpec, the systematics model consists of a linear slope \citep[e.g.,][]{Alderson2023EarlyG395H, Moran2023HighObservations, wallack_jwst_2024} with time that we fit independently to each detector and each visit. For visit 2, we also include a GP with time, again using the Matérn 3/2 kernel. Since the correlated noise structures are consistent between the two detectors, we share the characteristic frequency between NRS1 and NRS2, but fit the GP amplitude separately to each. In visit 1, the pre-transit baseline was plagued by large systematics which we were not able to adequately correct. We therefore cut the first 1000 integrations for the NRS1 and NRS2 light curves for NIRSpec visit 1. Moreover, there was a clear flare just after mid-transit during the second NIRSpec visit. Flares have proven to be non-trivial to model in transit light curves \citep[e.g.,][]{lim_atmospheric_2023, Howard2023, radica_promise_2025}, and after several attempts to fit the flare (see Appendix~\ref{sec:appendix_stellar_flare_modelling}) we opt to simply mask the affected integrations (integrations 1810 -- 2160). Figure~\ref{fig:raw nirspec} shows the NRS2 light curves for both visits without any integrations cut. 

Finally, for each light curve (both NIRISS and NIRSpec), we include an error inflation term added in quadrature to the flux error. As with the astrophysical model, we use wide, uninformative priors for each parameter. Our total model, therefore, has 54 free parameters. We explore the posterior space with \texttt{emcee} \citep{foreman-mackey_emcee_2013}, using 110 chains and 100,000 steps per chain, the first 80\% of which we discard as burn-in. The light curves and best-fitting models for each instrument and visit are shown in Figure~\ref{fig:WLC}, and the constraints on the most relevant parameters are presented in Table~\ref{tab:fitted_lc_params}.

\begin{table}
    \caption{Retrieved planetary parameters from the joint light curve fitting of the four transits of \myplanet.}
    \centering
    \begin{tabular}{lcc}
    \hline
    \hline
    Fitted parameters & \\
    \hline

    Mid-transit time, $T_0$ (MJD$_\mathrm{TDB}$) & $60158.812806 \pm 0.000074$ \\
    Planet-to-star radius ratio, $R_p/R_*$: & \\
    ~~~~NIRISS/SOSS order 1 & $0.03789^{+0.00044}_{-0.00038}$ \\
    ~~~~NIRISS/SOSS order 2 & $0.03911^{+0.00046}_{-0.00047}$ \\
    ~~~~NIRSpec/G395H NRS1 & $0.03878^{+0.00025}_{-0.00021}$ \\
    ~~~~NIRSpec/G395H NRS2 & $0.03835^{+0.00023}_{-0.00020}$ \\
    Semi-major axis, $a/R_*$ & $12.96 \pm 0.63$ \\
    Inclination, $i$ & $86.86 \pm 0.35$ \\
    Eccentricity, $e$ & $0.057^{+0.072}_{-0.041}$ \\
    \hline
    \hline
    \label{tab:fitted_lc_params}
    \end{tabular}
\end{table}

\subsection{Spectroscopic Light Curve Fitting}

\subsubsection{NIRISS/SOSS}

\begin{figure}
    \centering
    \includegraphics[width=.48\textwidth]{ 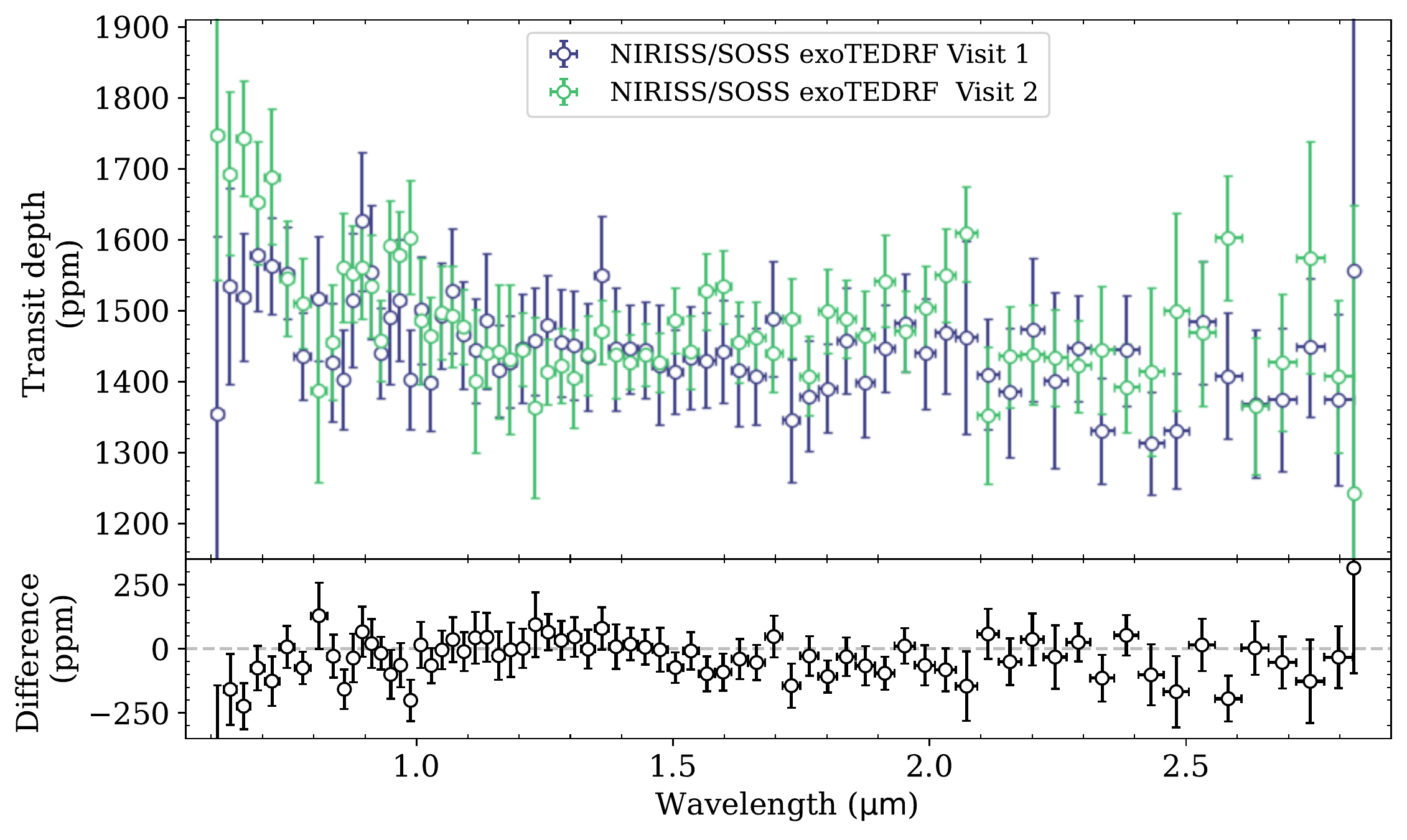}
    \caption{\texttt{exoTEDRF} transmission spectra of \myplanet from both NIRISS/SOSS visits. Visit 1 is displayed in blue, Visit 2 in light green and the differences between them are shown in the bottom panel in black.}
    \label{fig:niriss-transmission-spectrum-nameless-differences}
\end{figure}

For the \texttt{exoTEDRF} light curves, we continued to use \texttt{exoUPRF}, and binned the light curves to $R$=100 before fitting in order to increase the S/N and better handle the substantial systematics. We fixed the orbital parameters to the values in Table~\ref{tab:fitted_lc_params}, and left the scaled planet radius free. We freely fit the two parameters of the quadratic limb darkening law in the range $u_1,u_2 \in [-1,1]$. For the systematics model, we detrended against the linear slope with time and beating pattern. We also included the additive error inflation term and the GP model, where we fixed the GP timescale to the best-fitting value from the white light curve fit and allowed the amplitude to vary freely for each wavelength bin. This has proved an effective method to remove correlated noise when the noise structures do not vary strongly with wavelength \citep[e.g.,][]{radica_muted_2024}. We followed this same prescription for both visits. The posterior exploration was carried out using the dynamic nested sampling implemented through \texttt{dynesty} \citep{Speagle2020Dynesty:Evidences} using 1000 live points. The final \texttt{exoTEDRF} NIRISS/SOSS transmission spectra for both visits (which we treat as our fiducial spectra for the remainder of the analysis) are shown in Figure~\ref{fig:niriss-transmission-spectrum-nameless-differences}. {  Note that the two spectra show differences (e.g.\ offset and slope) which are discussed in detail in Section\,\ref{sec:atmosphere}.}

For fitting the \texttt{NAMELESS} spectroscopic light curves we used the \texttt{Tiberius} pipeline fitting stage \citep{Kirk2017RayleighHAT-P-18b,Kirk2021ACCESSWASP-103b}\footnote{\url{https://github.com/JamesKirk11/Tiberius}} with the nested sampling algorithm \texttt{PolyChord} \citep{Handley2015PolyChord:Sampling,Handley2015PolyChord:Cosmology} adaption \citep{Ahrer2022LRG-BEASTS:NTT/EFOSC2}. In both visits we used a spectral resolution of $R$=100 for the binning of the light curves. The system parameters were fixed to the retrieved values from the joint white light analysis in Table\,\ref{tab:fitted_lc_params}, but we left the transit depth free as well as the limb-darkening parameters $u_1, u_2$ using the quadratic limb-darkening law and a large uninformative prior range \citep[e.g., see ][]{Coulombe2024LDbiases}. To detrend our light curves we also fitted for a linear trend in time, a linear dependency on the beating pattern (PCA component) and a noise multiplying factor to account for any additional white noise. Contrary to the \texttt{exoTEDRF} light curve fitting, we did not include a GP model to fit our \texttt{NAMELESS} light curves. The \texttt{NAMELESS} transmission spectra are in excellent agreement with the fiducial \texttt{exoTEDRF} spectra as can be seen in the Appendix, Figure~\ref{fig:other-reductions-differences}. 

Despite the large differences in transit depth uncertainties, we ultimately select the \texttt{exoTEDRF} reduction as our fiducial spectra for two reasons: firstly, to have consistently reduced and fit spectra across our entire NIRISS and NIRSpec wavelength range, and secondly as we believe the larger error bars on the \texttt{exoTEDRF} SOSS spectra (most likely introduced through the use of a GP) compared to the \texttt{NAMELESS} spectra better encapsulate the uncertainties in the measured transit depths due to the significant amounts of correlated noise visible in the light curves. 

\subsubsection{NIRSpec/G395H}

\begin{figure}
    \centering
        \includegraphics[width=0.48\textwidth]{ 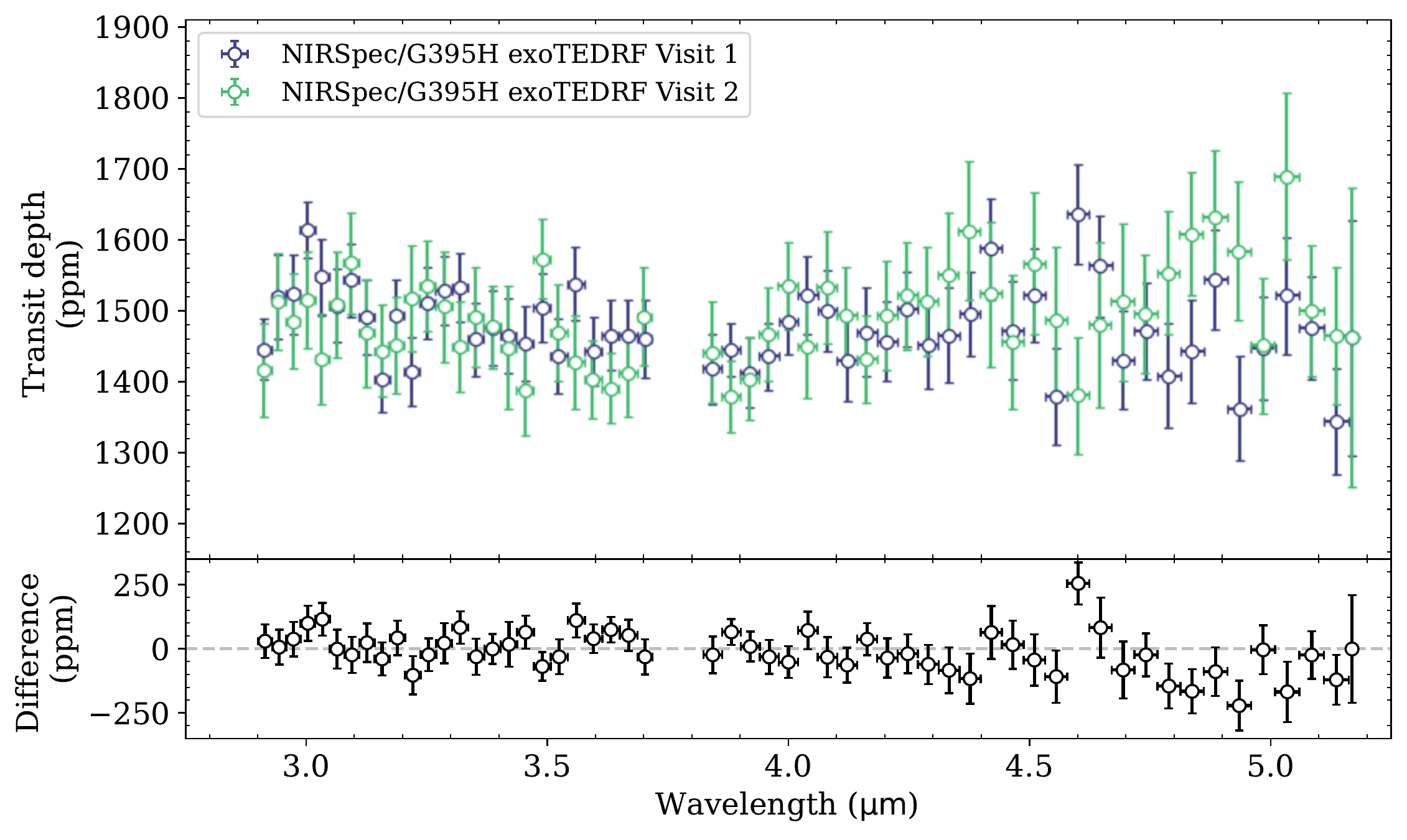}
        \caption{\texttt{exoTEDRF} transmission spectra of \myplanet from both NIRSpec/G395H visits. Visit 1 is displayed in the darker colour, visit 2 in the lighter colour and the differences between them are shown in the bottom panel in black.}
    \label{fig:nirspec-transmission-spectrum-visit-differences}
\end{figure}

For the \texttt{exoTEDRF} spectra, we followed the same procedure as outlined above for SOSS, except we altered the systematics models as befitted the NIRSpec observations. For visit 1, the systematics model consisted of a linear slope with time and the additive error inflation term. For visit 2, we also include the GP model, but unlike with SOSS, we freely fit both the amplitude and timescale to each bin as the correlated noise structures displayed some variations with wavelength. We again show the final transmission spectra for each visit in Figure~\ref{fig:nirspec-transmission-spectrum-visit-differences}. {  The two spectra from the two visits show some differences, especially in the redder wavelength ranges where the signal-to-noise is lower. However, we have not found evidence that the two visits are statistically different (at 1-sigma, $>68\%$ agreement) which would prohibit a combined analysis.} We focus on results from the \texttt{exoTEDRF} spectra in the remainder of the analysis in order to have a consistently-reduced and fit spectrum. 

For our \texttt{Eureka!}~reduction of the NIRSpec data, we also fit the light curves using \texttt{Eureka!}'s Stage\,5. Again we fixed all system parameters according to our joint white light analysis (Table\,\ref{tab:fitted_lc_params}). For the limb-darkening we used the quadratic limb-darkening law and left $u_2$ free, while we fixed $u_1$ to the generated value from Stage\,4 \citep[\texttt{ExoTiC-LD;MPS-ATLAS}][]{Grant2024,kostogryz2023mps} to avoid degeneracy, but we also investigated leaving both $u_1$ and $u_2$ free, which did not change our retrieved transmission spectrum. Therefore, the fiducial model for spectroscopic light curve fitting using the NIRSpec data consists of the transit model using the parameters R$_\mathrm{p}$/R$_\mathrm{s}$, $u_2$ and a linear trend in time, as well as a white noise parameter in the form of a multiplier. In the case of visit 2, we also added a GP model to account for the additional variability seen in the light curves. We freely fit the GP covariance amplitude for each spectroscopic light curve, but fix the length scale to the value obtained when fitting the white-light curve. Again, we compare the \texttt{Eureka!} NIRSpec/G395H transmission spectra with the fiducial \texttt{exoTEDRF} ones in the Appendix, Fig.\,\ref{fig:other-reductions-differences}.

\section{Analysis of the Metastable He I Triplet}
\label{sec: Helium}

\begin{figure}
    \centering
    \includegraphics[width=.5\textwidth]{ 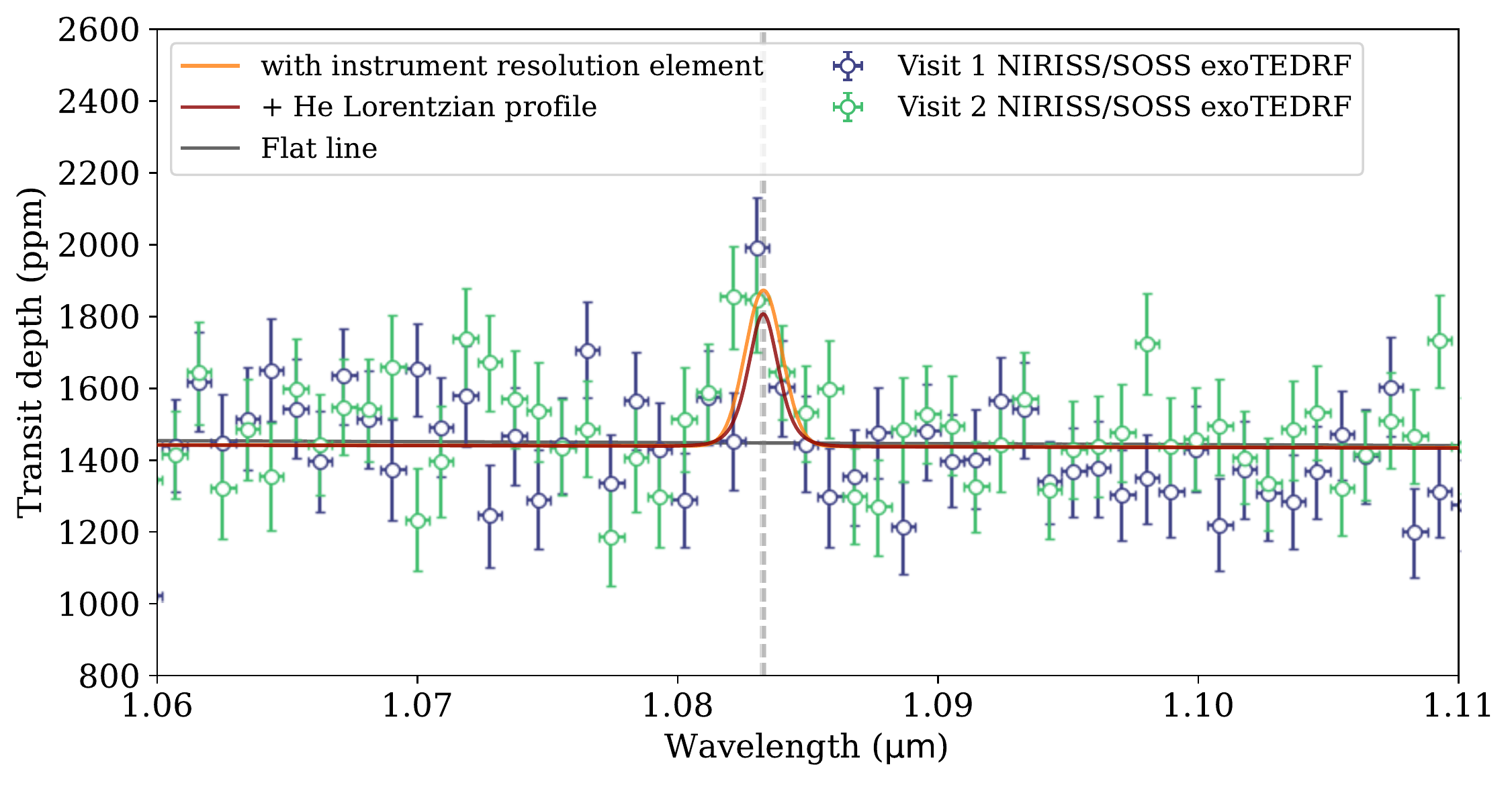}
    \caption{Pixel-level NIRISS/SOSS transmission spectrum of \myplanet using the \texttt{exoTEDRF} pipeline in the wavelength range surrounding the He I triplet (gray dashed line). Our best fits to the data via three model considerations are: (1) a Gaussian model of the He lines with a width dictated by the instrument resolution (orange), (2) a Gaussian for the instrument response convolved with the helium triplet line profiles treated as Lorentzians (dark red), and (3) a flat line (black).}
    \label{fig:soss-helium}
\end{figure}

NIRISS/SOSS has wavelength coverage which includes the metastable helium I triplet at 1.0833\,µm; a tracer of escaping H$_2$/He atmospheres. For this reason we also compute a transmission spectrum of \myplanet at pixel-level resolution for our two visits with SOSS using both the \texttt{exoTEDRF} pipeline and \texttt{NAMELESS} --- to better observe this narrow feature. Figure~\ref{fig:soss-helium} shows the spectra for both visits using our fiducial \texttt{exoTEDRF} reduction. 

 {As we do not expect to resolve the He line shapes, we use simple Gaussian and Lorentzian models to} investigate three fitting scenarios using both pixel-level spectra simultaneously and the nested sampling algorithm \texttt{dynesty} \citep{Speagle2020Dynesty:Evidences}  {to quantify the He detection}: (1) a Gaussian function with a full-width-half-maximum (FWHM) equal to the resolution of NIRISS/SOSS (R$\sim$650 at 1.083\,\textmu m) centred on the helium triplet with three free parameters (offset, slope and amplitude), under the assumption that the individual lines of the He triplet do not contribute to the shape of the observed absorption; (2) Lorentzian helium line profiles convolved with the instrumental resolution with four free parameters (offset, slope, amplitude and He line width); (3) a flat line, with two free parameters (offset and a slope) --- i.e., a null result for He detection. Note that model (2) is just a simple test of whether the He absorption is significantly broadened beyond the instrumental resolution. 

Using the \texttt{exoTEDRF} reduction, we find that model (1) is favoured over (2), with $\Delta \log(\mathcal{Z)} \sim 2.5$, i.e., there is no indication that the width of the feature needs to be modelled with more than the instrumental resolution element. This is supported by running the identical analysis on the \texttt{NAMELESS} reduction, where we find a Bayesian evidence difference of $\Delta \log(\mathcal{Z)} \sim 1.2$ in favour of model (1). 

Moreover, the flat line (model 3) is rejected compared to the Gaussian (model 1) by a Bayesian evidence difference of $\Delta \log(\mathcal{Z)} = 12.8 \pm 0.2$ for \texttt{exoTEDRF} and $\Delta \log(\mathcal{Z)} = 15.8 \pm 0.2$ for \texttt{NAMELESS}. This shows that the Gaussian Helium model is preferred to a flat line at $\sim$5.5$\sigma$ significance (\texttt{exoTEDRF}) and $\sim$5.9$\sigma$ significance (\texttt{NAMELESS}). The best-fit amplitudes from both reductions are $435\pm79$\,ppm (\texttt{exoTEDRF}) and  $460\pm77$\,ppm (\texttt{NAMELESS}), entirely consistent with each other and $>$5$\sigma$ inconsistent with zero. This result marks the first detection of escaping helium from a sub-Neptune with JWST.

We further use a Gaussian model with a fixed width (0.75\,\AA) convolved at the resolution of NIRISS/SOSS to provide an estimate of the resolved helium signature as seen at high spectral resolution \citep[e.g.,][]{fournier-tondreau_near-infrared_2024, radica_muted_2024, Piaulet_Ghorayeb_2024}. The best-fit value for the helium absorption amplitude expected at high resolution (convolved at R$\sim$100,000) is 1.0$\pm$0.2\,\%. This is in reach of current high-resolution spectrographs such as NIRPS \citep{Bouchy_2017NIRPS} in 3--4 transits for robust detection and interpretation of the helium line shape.

\begin{figure}
    \centering
    \includegraphics[width=.48\textwidth]{ 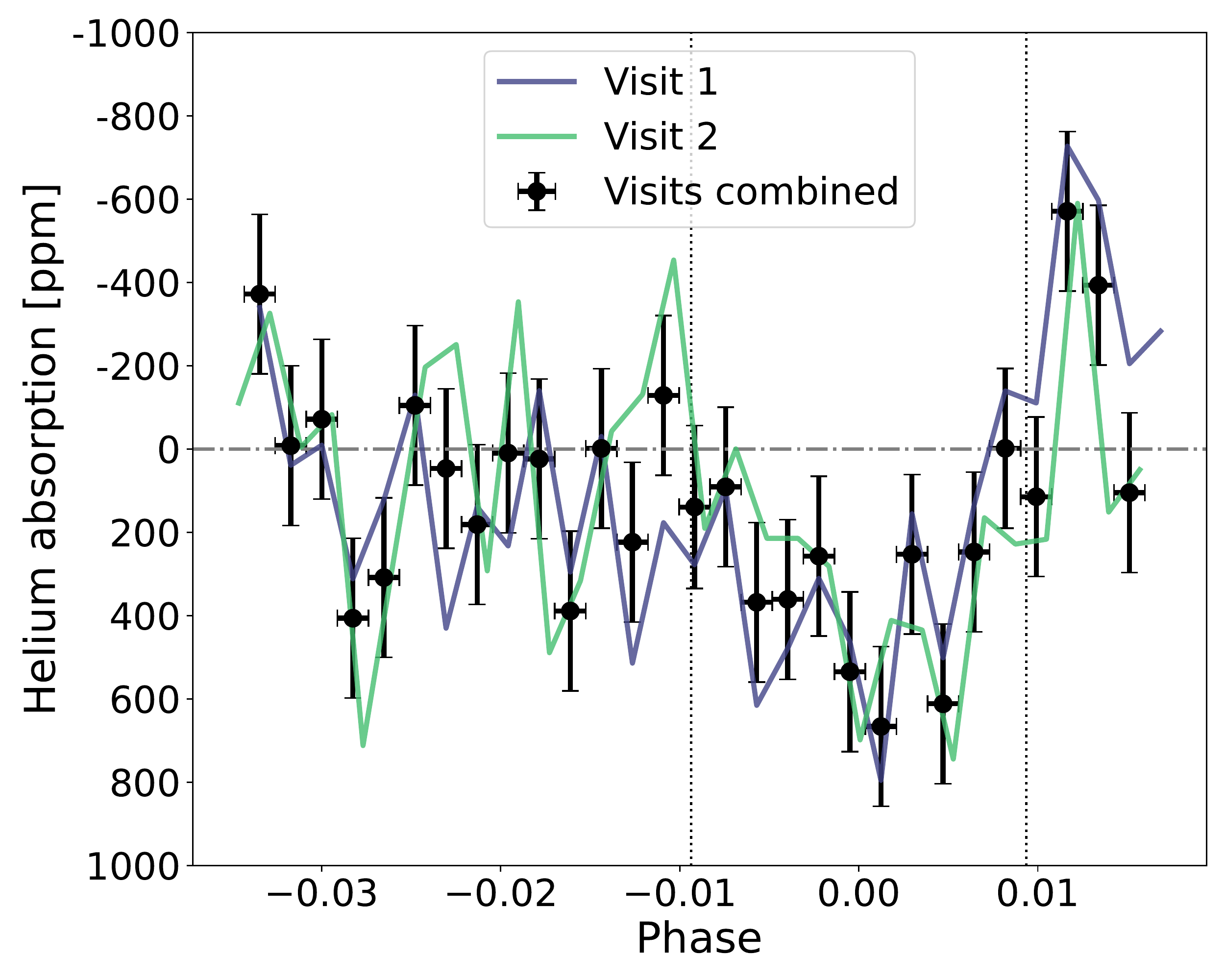}
    \caption{NIRISS/SOSS relative helium light curve showing clear absorption during the transit for visit 1 (dark blue) and visit 2 (light green), and the two visits combined (black). The transit ingress and egress are marked with the veritcal dotted lines. The relative light curve was computed by dividing the light curve which includes the He triplet by the average of the surrounding light curves (from 1.062 to 1.078\,µm and from 1.088 to 1.105\,µm).}
    \label{fig:soss-helium-lc}
\end{figure}

We then proceed to confirm that the He signal is stable throughout the whole transit and is not only produced during a fraction of it. We use the light curves at the pixel resolution of NIRISS/SOSS for the 47 pixels around the helium line (centered at 1.083\,µm), covering the 1.062--1.105\,µm wavelength range. Each light curve is first normalized by its average out-of-transit flux. A baseline white light curve surrounding the helium triplet is then generated by averaging all the light curves (a total of 38) from 1.062 to 1.078\,µm and from 1.088 to 1.105\,µm. The helium light curve is then computed by dividing the light curve covering the 1.083\,µm He triplet by the previously produced white light curve. We show the results in Figure~\ref{fig:soss-helium-lc} for both visits as well as the two combined. We find clear absorption of $\sim$500\,ppm during the whole transit. With the relatively short baseline at hand, no significant absorption is measured pre- and/or post-transit that could have traced an atmospheric escape tail. 

\section{Atmosphere and Stellar Contamination Modelling}
\label{sec:atmosphere}

\subsection{Methodology}
\label{sec:modelling-methodology}

The NIRISS spectra display 150--200\,ppm slopes towards short wavelengths, especially shortwards of $\sim$1.1\,\textmu m (Figure \ref{fig:niriss-transmission-spectrum-nameless-differences}) and are also systematically shifted up compared to the NIRSpec spectra (from both visits) by $\sim$95\,ppm. We were unable to trace this back to inconsistent system parameters used during light curve fitting, limb-darkening choices or any specific choices made during the data reduction (e.g., bias scale, outlier rejections, aperture size, background subtraction method). Instead, we interpret these features as telltale signs of unocculted stellar heterogeneities, with time-varying properties that result in offsets between the NIRISS and NIRSpec spectra, and even between the two NIRISS spectra. 

The distributions of heterogeneities on stellar surfaces, and therefore the impacts of stellar contamination on transmission spectra in the form of the transit light source (TLS) effect \citep{Rackham2018ThePlanets, Rackham2019TheStars}, are not constant in time. There is a significant time gap between the NIRISS and NIRSpec observations compared to the stellar rotation period, as well as between individual visits with NIRISS, meaning that we cannot reasonably assume that the distribution of spots and faculae on GJ 3090, and thus the impacts of the TLS effect, will be the same in one visit compared to another. The differing behaviour of the NIRISS spectra at the shortest wavelengths, as well as the offsets between the NIRISS and NIRSpec spectra also support this point. 

In light of the impacts of the TLS effect, the most conservative approach would be to jointly fit the spectra from all four visits, assuming the same underlying planetary atmosphere, and different TLS realizations for each. However, this level of complexity is currently out of reach of most modelling codes. Therefore, we take a step-by-step approach, and fit the two NIRISS visits separately, as well as NIRSpec separately from NIRISS. 

For NIRISS/SOSS we find the spectra to be entirely dominated by the effects of stellar contamination (see Section \ref{sec:tls_effect} and Appendix \ref{sec:appendix_TLS_modeling}) and thus explore the properties of the stellar heterogeneities and that give rise to the observed features. Particularly, given that the two NIRISS visits are separated by $>$200 days, we explore the differences between the properties of the stellar surface at these two epochs. With NIRSpec/G395H, however, we do not find strong evidence that the spectra are affected by stellar contamination. Therefore, combine the spectra from the two visits via a weighted average, in order to probe the properties of the planetary atmosphere.

\subsection{Overview of Models}

We use \texttt{SCARLET}, \texttt{POSEIDON} and \texttt{AURORA} to perform 1D atmosphere retrievals on both the NIRISS/SOSS and NIRSpec/G395H spectra, accounting for the potential impact of unocculted stellar surface heterogeneities as well as degeneracies between stellar and planetary properties. Specifically, we perform stellar heterogeneity-only (described in more detail in Appendix~\ref{sec:appendix_TLS_modeling}) and joint planet atmosphere-stellar heterogeneity retrievals on the NIRISS/SOSS data, whereas for the NIRSpec/G395H data we perform planet atmosphere-only retrievals and combined atmosphere-stellar heterogeneity retrievals using either free chemistry or chemically-consistent compositions. Finally, we put our atmosphere results in context using self-consistent chemical equilibrium and disequilibrium forward model grids. For each retrieval code, inferences on the stellar photosphere from NIRISS are presented in Table~\ref{tab:niriss_results} and constraints on the planet atmosphere from NIRSpec are shown in Figure~\ref{fig:corner_nirspec_molecules}.

\subsubsection{Atmosphere-Only and Joint Retrievals}
\label{sec:atm_joint_retrievals}

\paragraph{\texttt{Aurora}}

We perform flexible ``free" retrievals using \texttt{Aurora} \citep{Welbanks2021Aurora:Spectra}. With \texttt{Aurora}, a transmission spectrum is generated based on a model parameterized by constant-with-altitude molecular gas volume mixing ratios (H$_2$O, CH$_4$, CO$_2$, SO$_2$, H$_2$S), an isothermal vertical temperature profile, cloud properties following the two sector prescription from \citet{Welbanks2021Aurora:Spectra} using the linear combination approach from \citet{Line2016THESPECTRA}, and the reference radius at a reference pressure \citep{welbanks_degeneracies_2019}. Gases included are selected based on those expected in warm atmospheres and with cross-sections described in previous works \citep{Tsai2022, Welbanks2024}. We additionally retrieve for the impact of stellar heterogeneities through the TLS effect following the description of \citet{Pinhas2018RetrievalAURA} as implemented in \citet{Welbanks2021Aurora:Spectra}. We consider both hydrogen-rich and secondary atmospheres by using the center-log-ratio transformation as described in \citet{Welbanks2021Aurora:Spectra} and scenarios where H and He in a solar mixture \citep{asplund_2009} are the background gas of the planet atmosphere. Additionally, we allow for the possibility of instrumental offsets between both NIRSpec detectors and between NIRISS orders. 

\texttt{Aurora} estimates planet parameters using \texttt{PyMultiNest} nested sampling \citep{Buchner2014} and 2000 live points. Models assume a one-dimensional hydrostatic equilibrium atmosphere spanning pressures 10$^{-9}$ to 10$^{2}$ bar in 100 uniformly space layers in logarithmic pressure space. The model performs line-by-line opacity sampling at a spectral resolution of 20,000 then bins the spectrum down to the resolution of the observations. Our fiducial case for the NIRSpec observations has 13 free parameters: five molecules, one for an isothermal pressure-temperature profile, four for inhomogeneous clouds and hazes, one for the radius of the planet at the one reference pressure parameter, and one for the instrumental offset between NRS1 and NRS2. 

\paragraph{\texttt{POSEIDON}}

We perform an additional independent atmospheric retrieval analysis using the open-source retrieval code \texttt{POSEIDON} \citep{macdonald_hd_209458b_2017, macdonald_poseidon_2023}. \texttt{POSEIDON} makes use of 2000 \texttt{PyMultiNest} live points in all of our retrievals, a nested sampling algorithm used for our parameter estimation and model comparison \citep{Feroz2009MULTINEST:Physics, Buchner2014X-rayCatalogue}. A more comprehensive description of the radiative transfer technique and forward model (\texttt{TRIDENT}) used by \texttt{POSEIDON} and its corresponding opacity database can be found in \citet{macdonald_trident_2022}.
Using the \texttt{exoTEDRF} reductions of the NIRISS/SOSS and NIRSpec/G395H datasets specified in Section \ref{sec:exoTEDRF}, we generate model spectra from both 0.6 to 2.9\,µm and 2.9 to 5.3\,µm, respectively, at R=20,000. This is convolved with a Gaussian kernel to the native resolution of each instrument and multiplied by the corresponding instrument sensitivity function so that it can then be binned to the desired wavelength spacing of each spectrum.

We include a wide array of atmospheric and stellar parameters in our retrievals of GJ 3090\,b. In each of our retrievals, we fit for a reference radius R$_\mathrm{p,ref}$ at a designated reference pressure of 10 bar. To account for the difference in observations between instruments and in order to compute detection significances, we conduct a combination of nested retrievals, including flat lines models and multi-gas models with or without stellar contamination (see Section \ref{sec:tls_effect}) for both our NIRISS/SOSS and NIRSpec/G395H reductions. Our base multi-gas atmospheric models include the gases H$_2$, He, H$_2$O, CH$_4$, CO$_2$, SO$_2$, CO, NH$_3$, and H$_2$S,  {where H$_2$ and He are calculated as the background gas at a fixed He/H$_2$ ratio of 0.17}. Each of these models also include clouds and hazes, where we utilize a two-parameter (log a$_\mathrm{haze}$ and $\gamma$) power-law prescription for hazes \citep{macdonald_hd_209458b_2017}. We assume an optically-thick gray opacity, in which all cloud layers deeper than P$_\mathrm{surf}$ are set to infinite opacity in our retrievals. We also assume both a fixed surface gravity of the stellar regions of log g $=$ 4.727 $\pm$ 0.029 (cgs) \citep{Almenara2022GJCharacterisation} and an isothermal atmospheric temperature for our models.
The priors for each of our models of GJ 3090\,b are specified as the following: R$_\mathrm{p,ref}$ (R$_\oplus$) $=$ $\mathcal{U}$ (1.28, 2.45), T (K) $=$ $\mathcal{U}$ (100, 1000), log a$_\mathrm{haze}$ $=$ $\mathcal{U}$ (-4, 8), $\gamma_\mathrm{haze}$ $=$ $\mathcal{U}$ (-20, 2), log P$_\mathrm{cloud}$ $=$ $\mathcal{U}$ (-7, 2), log X $=$ $\mathcal{U}$ (-12, 0).

We additionally utilize \texttt{POSEIDON} to determine the extent of the TLS effect on the NIRISS spectra of GJ 3090\,b.
\texttt{POSEIDON} accounts for stellar contamination by multiplying a bare-rock transmission term, (R$_p$/R$_*$)$^2$, by the wavelength-dependent stellar contamination factor from two discrete stellar heterogeneities (spots and faculae).
\texttt{POSEIDON} creates the model spectra of the active stellar regions by utilizing the \texttt{PyMSG} package, through which \texttt{POSEIDON} interpolates across the PHOENIX grid of stellar atmosphere models \citep{husser_new_2013}. 
We include five extra free parameters in addition to the aforementioned parameters in these models: f$_\mathrm{fac}$ $=$ $\mathcal{U}$ (0.0, 0.5), f$_\mathrm{spot}$ $=$ $\mathcal{U}$ (0.0, 0.5), T$_\mathrm{fac}$ (K) $=$ $\mathcal{U}$ (T$_*$-36, 1.2T$_*$), T$_\mathrm{spot}$ (K) $=$ $\mathcal{U}$ (2300, T$_*$+36), T$_\mathrm{phot}$ (K) $=$ $\mathcal{U}$ (T$_*$, 12), where T$_*$ $=$ 4236 $\pm$ 12. This results in a total of 16 free parameters tested in both our NIRSpec and NIRISS retrievals of GJ 3090\,b.

 {We lastly use \texttt{POSEIDON} to conduct several Bayesian comparisons between the different models delineated above for both NIRISS and NIRSpec. To provide more robust Bayesian comparisons between our atmosphere models and each other nested model, we construct additional "minimal atmosphere" models. These models limit the number of atmospheric free parameters by getting rid of any completely unconstrained molecules. For our NIRISS minimal atmosphere models, we include CO$_2$, SO$_2$, CO, and H$_2$S, while for our corresponding NIRSpec models, we include CO$_2$, SO$_2$, and CO. Allowing the cloud top pressure (log P$_\mathrm{cloud}$) to vary, this results in 9 free parameters in our NIRISS minimal atmosphere model (12 when star spots are jointly included), and 8 free parameters in our NIRSpec minimal atmosphere model (11 with star spots).}

\paragraph{\texttt{SCARLET}}

We also perform retrievals with \texttt{SCARLET} to test other assumptions not supported by other codes (e.g., fitting shared atmosphere properties and visit-specific TLS parameters to the NIRISS/SOSS spectra) for free retrievals, and to put the results from the free retrievals in context with a further exploration involving chemically-consistent retrievals. We use a version of \texttt{SCARLET} \citep{benneke_atmospheric_2012,benneke_how_2013,benneke_strict_2015,benneke_sub-neptune_2019, benneke_water_2019, pelletier_where_2021} adapted for observations of sub-Neptunes with JWST \citep{piaulet_evidence_2023, Piaulet_Ghorayeb_2024} to perform retrievals on the transmission spectra of GJ 3090\,b. 

In our forward modelling framework for free retrievals, the molecular abundances are assumed to be constant with altitude in the atmosphere. For chemically-consistent retrievals, the chemistry is dictated by chemical equilibrium in each atmospheric layer given the local temperature and pressure conditions, and parameterized by the C/O ratio and metallicity of the atmosphere. We explore the presence of higher SO$_2$ abundances than predicted in equilibrium \citep[e.g.,][]{beatty_sulfur_2024} by performing retrievals where the chemical-equilibrium SO$_2$ abundance is overwritten by a vertically-constant abundance profile with the abundance as a free parameter (while the other molecular abundances are scaled to keep the total equal to unity). 

Since transmission spectra are largely insensitive to the temperature profile, we either assume that it is an isotherm (where the temperature fitted in the retrieval is representative of the planet's photosphere at the terminator), or parameterize it following \citet{MadhusudhanSeager2009apjRetrieval} in order to assess the sensitivity of our retrieved abundances to the choice of T-P profile parameterization. For the set T-P profile and abundances, we compute the atmospheric structure in hydrostatic equilibrium and the radiative transfer calculation for the slant transmission geometry. 

For each model evaluation within the Bayesian retrieval framework, we use \texttt{scipy.minimize} to fit the radius at a pressure of 10 mbar in our atmosphere model that provides the best match to the observed spectrum (iterating over the hydrostatic equilibrium and radiative transfer steps for each test radius for consistency). Our baseline free retrievals include H$_2$, He, H$_2$O, CO, CO$_2$, CH$_4$, HCN, H$_2$S, SO$_2$, and NH$_3$, but we perform a few retrievals with only a subset of these molecules to assess Bayesian detection significances \citep{trotta_bayes_2008,benneke_how_2013}. We assume a mix of H$_2$ and He as the filler gas, with a Jupiter-like He/H$_2$ ratio of 0.157 \citep{vonZahn1996}. Our opacities are computed from the \texttt{HELIOS-K} \citep{grimm_helios-k_2015} cross-sections for H$_2$O \citep{ExoMol_H2O}, CO \citep{Hargreaves2019}, CO$_2$ \citep{ExoMol_CO2},  CH$_4$ \citep{HargreavesEtal2020apjsHitempCH4}, HCN \citep{Harris_HCN_2006}, H$_2$S \citep{ExoMol_H2S}, SO$_2$ \citep{ExoMol_SO2}, and NH$_3$ \citep{ExoMol_NH3}. For the baseline model, we parameterize clouds as a gray opacity source with the cloud-top pressure $p_\mathrm{cloud}$ fitted in the retrieval. We also fit the cloud covering fraction $f_\mathrm{cloud}$ as implemented in \citet{Piaulet_Ghorayeb_2024} to account for potential patchy cloud coverage. To account for the potential impact of small-particle hazes on our spectrum, we parameterize them using the slope enhancement parameter $c_\mathrm{haze}$ that multiplies the Rayleigh scattering slope. Beyond planetary parameters, we account for potential instrument offsets between orders 1 and 2 of NIRISS/SOSS, between the spectra obtained from the first and second NIRISS/SOSS visit, and between the NRS1 and NRS2 detectors of NIRSpec/G395H.

We use the \texttt{SCARLET} implementation of \texttt{nestle}\footnote{\url{http://kylebarbary.com/nestle}} \citep{skilling_nested_2004, skilling_nested_2006} to sample the parameter space. We compute the models at a resolving power of 15,625 (or 31,250 for a higher resolution test) and convolve them in each observed bandpass assuming uniform throughput for the likelihood evaluation.

Finally, for retrievals where we fit for the impact of both the planetary atmosphere and stellar heterogeneities on the observed transmission spectrum, we use the \texttt{SCARLET} implementation of TLS effect modelling which leverages the \texttt{MSG} module to obtain stellar models (see \citealp{Piaulet_Ghorayeb_2024} for a description). For these joint retrievals, we simultaneously sample the parameters of the planetary atmosphere and of the stellar surface, as parameterized in the \texttt{stctm} implementation. The stellar contamination factor $\epsilon_{\lambda, \, \rm{het}}$ multiplies the predicted transmission spectrum given the sampled atmosphere parameters prior to the likelihood evaluation. We test cases where we consider no stellar contamination, spots only (with a single population having a shared $T_{\rm{spot}}$), and both spots and faculae with an associated temperature contrast and covering fraction for each heterogeneity population. In this case, we use an updated version of \texttt{SCARLET} (Piaulet-Ghorayeb et al., in prep) which supports the modelling of visit-specific stellar contamination signatures with a shared visit-independent planetary atmosphere composition.

\begin{table}
    \caption{\label{tab:niriss_results} Constraints on the stellar contamination parameters affecting the NIRISS/SOSS transmission spectrum for free retrievals including both stellar heterogeneities and the planetary atmosphere contribution. Retrievals were applied to both visits separately using \texttt{SCARLET}, \texttt{POSEIDON}, and \texttt{Aurora}. Lower limits reported are 2$\sigma$.}
    \centering
    \begin{tabular}{lcccc}
    \hline
    \hline
    Parameter & \texttt{POSEIDON} & \texttt{SCARLET} & \texttt{Aurora} \\
    \hline
    \multicolumn{3}{l}{\textbf{NIRISS Visit 1}} &  \\
    T$_\mathrm{phot, star}$ [K] & $3564.14^{+54.62}_{-56.57}$ &$3556.17^{+67.33}_{-69.22}$ &$3564.48^{+80.37}_{-84.87}$\\ 

    T$_\mathrm{spot}$ [K]\footnote{For the \texttt{SCARLET} retrieval, derived from the samples on $T_\mathrm{phot}$ and $\Delta T_\mathrm{spot}$.} & $2960.64^{+345.03}_{-301.51}$ &$2964.61^{+320.89}_{-259.97} $&$2839.66^{+413.39}_{-373.06}$\\ 

    f$_\mathrm{spot}$ [K] & $0.12^{+0.05}_{-0.06}$ &$0.14^{+0.07}_{-0.06}$ &$0.10^{+0.05}_{-0.05}$\\ 
    Order 2 offset & $-$ &$-17.83^{+31.82}_{-30.97}$ &$-23.73_{-32.36}^{+30.6}$\\ 
    \multicolumn{3}{l}{\textbf{NIRISS Visit 2}} &  \\
    T$_\mathrm{phot, star}$ [K] & $3575.31^{+52.32}_{-49.54}$ &$3591.1^{+62.25}_{-62.61}$ &$3612.62^{+73.9}_{-71.95}$\\ 

    T$_\mathrm{spot}$ [K]$^a$ & $3223.19^{+94.3}_{-129.71}$ &$3194.19^{+84.84}_{-103.26}$ &$3195.61^{+101.87}_{-110.89}$\\ 

    f$_\mathrm{spot}$ [K] & $>0.23$ &$>0.31$ &$>0.31$\\ 
    Order 2 offset & $-$ &$69.87^{+31.19}_{-32.48}$ &$105.29_{-35.15}^{+36.33}$\\ 
    \hline
    \hline
    \end{tabular}
\end{table}

\subsubsection{Self-Consistent Model Grids}

\paragraph{\texttt{ScCHIMERA}}

Finally, we estimate the atmospheric parameters of GJ 3090\,b with ``grid-based retrievals” using the atmosphere modeling code, \texttt{ScCHIMERA}. From a grid of pre-computed models, we estimate planet parameters using Bayesian nested sampling following a process similar to \citet{Welbanks2024}. Models are one-dimensional, varying along a vertical pressure-temperature profile, and assume the atmosphere thermal structure and composition are in radiative-convective-thermochemical equilibrium (1D-RCTE). The model grid includes a range of values for planet irradiation temperature (T$_\mathrm{irr}$=550--700\,K in 25\,K increments, a proxy for energy redistribution allowing for cooler temperatures near the day-night terminator), atmospheric carbon-to-oxygen ratio (C/O=0.2--0.6 in 0.1 increments, although we did explore allowing C/O to extend up to 0.8 with a coarse grid prior to limiting C/O$<$0.6), and atmospheric metallicity ([M/H] = 2.5--4.5 in 0.25 increments, in which M accounts for all non-H/He elements and [] denotes log$_{10}$ relative to solar ratios). 

To produce each model, \texttt{ScCHIMERA} iteratively computes the pressure-temperature structure (10$^{-8}$--10$^{2}$ bar in 10$^{0.2}$ bar layers) from an intrinsic temperature (T$_{int}$~=~100\,K, this choice does not impact the RCTE model at low pressures probed by transmission observations) and the top-of-atmosphere incident stellar flux computed from a PHOENIX stellar model \citep[][T$_\mathrm{star}$=3556\,K, $\rm \log(g)$(c.g.s)=4.727]{husser_new_2013}. Then, the \texttt{NASA CEA2} routine for Gibbs free energy minimization \citep{GordonMcbride1994} solves the equilibrium gas volume mixing ratios of thousands of molecular/atomic species along the pressure-temperature profile. We include opacity sources for major radiative species: \ce{H2}-He collision-induced absorption, H/\ce{e-}/\ce{H-} bound/free-free continuum, and the line opacities for H$_2$O, CO, CO$_2$, CH$_4$, NH$_3$, H$_2$S, PH$_3$, HCN, C$_2$H$_2$, OH, TiO, VO, SiO, FeH, CaH, MgH, CrH, ALH, Na, K, Fe, Mg, Ca, C, Si, Ti, O, \ce{Fe+}, \ce{Mg+}, \ce{Ti+}, \ce{Ca+}, \ce{C+} \citep[for details of line sources, see][]{Mansfield2021, Iyer2023}. 

To estimate the properties of GJ 3090\,b, we use Bayesian nested sampling with \texttt{PyMultiNest} \citep{Buchner2014}. We allow for 500 live points within the grid parameter space (T$_{irr}$, [M/H], and C/O) and tri-linearly interpolate the temperature structure and gas mixing ratio profiles between grid models. During this parameter estimation stage, we compute transmission spectra with the addition of for a vertically uniform grey cloud opacity ($\kappa _\mathrm{cloud}$) post-processed onto the spectrum, a cloud covering fraction, an instrument offset allowing the NRS1 transit depth to move up/down, and a multiplier on planet radius ($\times$R$_\mathrm{p}$). Overall, we include free parameters for T$_{irr}$, [M/H], C/O, $\kappa _\mathrm{cloud}$, cloud covering fraction, instrument offset, and $\times$R$_\mathrm{p}$. We report constraints on [M/H], C/O, instrument offset, and $\times$R$_\mathrm{p}$ in Table \ref{tab:mol_constr}. The remaining parameters are unconstrained.

To test the impact of disequilibrium chemistry due to vertical mixing and photochemistry, we additionally processed models using the \texttt{VULCAN} kinetics code \citep{Tsai2017}, following the process in \citet{Welbanks2024}. This disequilibrium model grid spans a smaller parameter space than the 1D-RCTE grid, including only models with C/O=0.4, 0.5, and 0.6. We assume a vertically constant $K_{zz}$ profile, $K_{zz}=10^{9}$ cm$^2$s$^{-1}$, and use the UV-stellar spectrum of GJ 832 \citep{Youngblood2016ApJ...824..101Y} as a proxy for GJ 3090. We adopt a zenith angle of 83 degrees for the terminator region following \citet{Tsai2023PhotochemicallyWASP-39b}. With the introduction of vertical mixing, T$\rm _{int}$ does influence the model. Age-luminosity relations predict that the T$_\mathrm{int}$ of GJ~3090\,b may be lower than 100\,K (see Section\,\ref{sec: interior models}), however, a higher temperature combined with vertical mixing can serve to quench CH$_4$, perhaps contributing to the observed muted features in GJ~3090\,b's spectrum. We allow a conservatively high T$_\mathrm{int}$ for this reason. The full posterior distributions for the equilibrium and disequilibrium grid-based retrievals are shared on Zenodo. 

\begin{table}
    \caption{\label{tab:mol_constr} Chemically-consistent retrieval (SCARLET) and grid constraints on the atmospheric composition of GJ 3090\,b from the R=100 version of the \texttt{exoTEDRF} NIRSpec/G395H visit 1+2 spectrum. We only report constraints on the parameters that were meaningful to explain the data from a Bayesian model comparison standpoint.  The `grid-trieval' results were obtained with \texttt{ScCHIMERA} using chemical and disequilibrium grids. Free parameters in the grid not reported here (T, $\kappa_\mathrm{cloud}$, and cloud covering fraction) were unconstrained. Upper and lower limits are 2$\sigma$, except for C/O reported from the \texttt{ScCHIMERA} grid-based retrievals. C/O is unconstrained by \texttt{ScCHIMERA}, but we report the 1$\sigma$ preference.}
    \centering
    \begin{tabular}{lcccc}
    \hline
    \hline
    \hline
    Parameter & \texttt{ScCHIMERA}&  \texttt{ScCHIMERA}& \texttt{SCARLET}\\
     & eq. grid & diseq. grid  &  eq. retrieval\\
    \hline
        Met. [$\times$solar] & $>1,348$ & $>1,230$& $>776$ \\
        C/O  & $<0.43$ & $<0.54$ & $<0.42$\\      
        $\times$R$_{p}$ & 0.99$\pm$0.01 &0.99$\pm$0.01 & 0.99$\pm$0.01 \\
    \hline
    \hline
    \end{tabular}
\end{table}

\section{Results and Discussion}
\label{sec:discussion}

\subsection{Impact of Stellar Activity on the NIRISS Observations}
\label{sec:tls_effect}

\mystar is a relatively active M dwarf which affects both the light curve analysis as well as the retrieved transmission spectra. Particularly for the NIRSpec transits, flares and spots produced correlated noise structures in our light curves, resulting in our choice to cut integrations for which we did not manage to find an adequate fit with any of our models (see Appendices \ref{sec:appendix_stellar_flare_modelling} and \ref{sec:appendix_cut_integrations}). Though we note that our explorations into fitting models to the observed stellar flux variations did not show any effect on the resulting transmission spectrum (see Appendix\,\ref{sec:appendix_stellar_flare_modelling}).

\begin{figure*}
    \centering
    \includegraphics[width=0.9\textwidth]{ 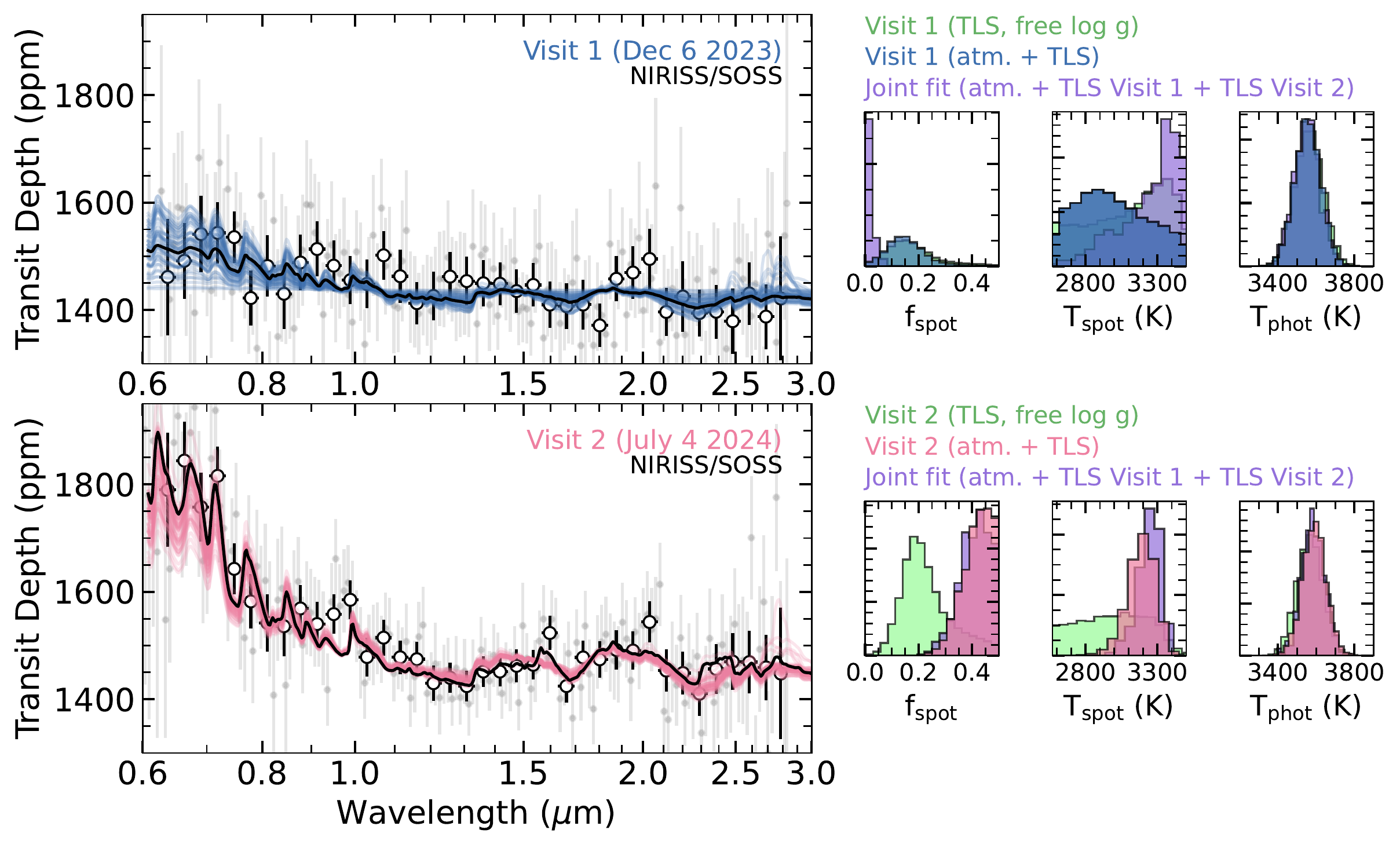}
    \caption{Results from the \texttt{stctm} stellar contamination retrievals, and \texttt{SCARLET} joint retrieval of the atmosphere and stellar contamination components for retrievals performed on individual or on both NIRISS/SOSS spectra of GJ 3090\,b.
    \textit{Left panels:} Measured transmission spectrum (black data points) for the first (top panel) and the second (bottom panel) NIRISS/SOSS transit, along with the best-fitting model (black line) and sample spectra from the posterior distribution (semi-transparent colored lines) from the individual atmosphere + TLS retrievals. We performed the retrievals on the R=100 \texttt{exoTEDRF} spectra (gray points), also shown binned by a factor of four (black points) on the figure for visualization purposes. For the Visit 1 (2) spectrum, the best-fit order 2 offset of -14.4 (+94.4) ppm was applied on the data points displayed.
    \textit{Right panels:} Marginalized posterior distributions on the stellar heterogeneity properties for the first (top) and second NIRISS/SOSS visit (bottom). We show the results from \texttt{SCARLET} retrievals performed on individual visits (visit 1 in blue, visit 2 in pink) and on both visits assuming a shared atmosphere component but different stellar heterogeneity components (in purple). We also display the posterior distributions from the \texttt{stctm} retrievals to individual visits (green) which only accounts for the impact of the TLS effect on the observations (see Appendix~\ref{sec:appendix_TLS_modeling}) The inferred heterogeneity properties are different for each visit, as evidenced by the difference in the blueward slope observed in the transmission spectrum. For visit 1, where stellar contamination is not confidently detected (see text) the joint fit to both visits only provides an upper limit on the spot covering fraction and a lower limit on the spot temperature.}
    \label{fig:spots_niriss}
\end{figure*}

On the other hand, the observations in the bluer wavelength range with NIRISS/SOSS showed features that are well explained by stellar heterogeneities, burying any features and inferences about \myplanet's atmosphere (see Appendix \ref{sec:appendix_TLS_modeling}). Both visits demonstrated different inferred heterogeneity parameters irrespective of the retrieval code used (Figure \ref{fig:spots_niriss}; Table \ref{tab:niriss_results}), highlighting the stochasticity of appearing/disappearing spots on the surface within the time between the two SOSS visits (6 months / 10 stellar rotation periods).  

We also cross-check the stellar contamination retrieval results by performing fits directly to the out-of-transit NIRISS stellar spectra for both visits \citep[e.g.,][]{wakeford_exoplanet_2019, Moran2023HighObservations, radica_promise_2025}, and the heterogeneity parameters we infer from the out-of-transit analysis largely agree with those retrieved from the in-transit fits. More information on these fits can be found in Appendix \ref{sec:stellar_spectrum_modelling} and the results are summarized in Figure~\ref{fig: stellar spectrum fit}.

{ In what follows, we quote results from the \texttt{SCARLET} retrievals, since \texttt{SCARLET} supports the joint fitting of a planetary atmosphere to two NIRISS visits while accounting for visit-specific stellar contamination components. Though, we note that our conclusions from individual visit fits, in terms of the visit-to-visit changes in the stellar contamination component, are consistent across all three retrieval frameworks (Table \ref{tab:niriss_results}).}
The results from the \texttt{SCARLET} retrievals including stellar contamination retrievals are shown in Figure~\ref{fig:spots_niriss}. We find strong ($5.07\sigma$) evidence for stellar contamination shaping the spectrum for the second visit. The evidence for the presence of spots is even stronger (6.4$\sigma$) if hazes are not included in the atmosphere model. The inference of spots is also supported by the large inferred spot covering fractions, even when trying to explain both visits with the same atmosphere model, as well as the strong temperature contrast between the spots and photosphere (see Figure \ref{fig:spots_niriss}). The model with spots favors an offset of $69.87^{+31.19}_{-32.48}$ ppm between order 1 and order 2 (Table\,\ref{tab:niriss_results}), but spots are detected regardless of whether an offset is included in the retrieval. We find no evidence for faculae from a Bayesian model comparison standpoint. 

The conclusions for the first visit are more nuanced, which is expected given the less-pronounced short-wavelength slope (Fig. \ref{fig:spots_niriss}). When a haze slope is included in our atmosphere model, we do not significantly detect spots in the visit 1 spectrum. In the absence of a Rayleigh slope enhancement from small-particle hazes, spots are detected at 2.1$\sigma$.  
The inferred spot properties are also less constrained than for the second visit, with large uncertainties on the spot temperature (Fig.\,\ref{fig:spots_niriss}), and we only obtain an upper limit on the spot covering fraction (and a lower limit on the spot temperature) when the retrieval is performed under the assumption of a shared atmosphere component across both visits. Contrary to visit 2, even when fitting for an offset between order 1 and order 2 in the visit 1 spectrum, we do not retrieve a value different from zero at the 1$\sigma$ level. 

When performing retrievals with a shared atmosphere and a TLS component that is either shared or visit-specific, the \texttt{SCARLET} retrievals have a slight (2.0$\sigma$) preference for visit-specific rather than shared stellar heterogeneity properties, which further highlights the time-variable nature of stellar heterogeneities. Finally, we find that consistent heterogeneity parameters can explain the spectrum in stellar-contamination-only retrievals (that is, if we assume that there is no contribution from GJ 3090\,b's atmosphere to the transmission spectra). Although, lower spot covering fractions are allowed in visit 2 when the gravity of the stellar photosphere and heterogeneity components are allowed to adopt values larger than the literature value --- something preferred both in the \texttt{stctm} fit and in the fit to the out-of-transit stellar spectra. { We additionally perform tests using \texttt{POSEIDON} where we use two nested models to explain each NIRISS visit: one including only colder stellar heterogeneities (spots), and another where both spots and a minimal planetary atmosphere are considered (as defined in the \texttt{POSEIDON} section of Section~\ref{sec:atm_joint_retrievals}). We find that heterogeneity parameters consistent to those of the \texttt{SCARLET} joint retrieval are inferred in each scenario, and that the addition of the atmospheric component is not required from a Bayes factor standpoint (favored by $<2\sigma$ for both of the visits). This highlights the fact that no meaningful atmospheric constraints can be placed from our joint TLS-atmosphere retrievals on the NIRISS spectra (Table \ref{tab:niriss_allpara}).}

Finally, while stellar contamination can also impact transmission spectra at longer wavelengths (although to a lesser extent), we find that it does not affect the main conclusions we draw on the atmospheric inferences (Figure \ref{fig:corner_nirspec_molecules}). Our constraints on SO$_2$ and CH$_4$ remain largely unchanged when considering an additional spot component in our model, while our constraints on the water abundance become less meaningful (since the star can produce ``fake'' water features), with a broad posterior compared to the lower limits from atmosphere-only fits. Our inference of high metallicity is not impacted, however, as the retrieval compensates overall lower water abundances with higher CO$_2$ abundances to retain a high atmospheric mean molecular weight (MMW), similar to that inferred from atmosphere-only modeling (Figure \ref{fig:corner_nirspec_molecules}).

We recommend for future observation planning that NIRISS and NIRSpec observations be scheduled close in time such that a consistent TLS model can reasonably be assumed to hold for both instruments, which can thereby help break the atmosphere-TLS degeneracy present in spectra from both instruments. A similar experimental design was tested for LHS 1140\,b by \citet{Cadieux2024}, who observed two consecutive transits separated by 24.7\,days with NIRISS. The two spectra showed consistent TLS configurations over this time period, which represents $\sim$20\% of stellar rotation period \citep{Cadieux2024}. In our case though, stellar contamination broadly prevents us from placing meaningful constraints on the planetary atmosphere from the NIRISS observations alone (with broad unconstrained molecular abundance posteriors). 

\subsection{The Atmospheric Composition of GJ 3090 b}

{ We use three types of retrievals and models to interpret the NIRSpec/G395H spectrum of GJ 3090 b in terms of its atmospheric composition: (1) free retrievals with \texttt{SCARLET}, \texttt{POSEIDON}, and \texttt{Aurora}, (2) chemical equilibrium retrievals with \texttt{SCARLET} and (3) forward models using the \texttt{ScCHIMERA} ``grid-based retrieval'' approach, and chemical disequilibrium forward model calculations using \texttt{ScCHIMERA} paired with \texttt{VULCAN}. The free retrieval approach is the most agnostic, as it makes no assumption about the atmospheric chemistry and instead infers atmospheric abundances independently from each other, from the observed spectrum. Still, for sub-Neptunes with muted spectral features, chemical equilibrium models have proved useful to assess the range of atmospheric scenarios compatible with the data (see e.g., \citealp{wallack_toi836_2024,teske_toi776c_2025}). We recognize, however, the atmosphere of GJ 3090 b may not be in chemical equilibrium, as seen for TOI-270 d \citep{benneke_jwst_2024}, and use disequilibrium models to assess the impact of the chemical equilibrium assumption on our findings. }

{ Our free retrievals find no evidence for CH$_4$ in the atmosphere of GJ 3090 b (Figure  \ref{fig:corner_nirspec_molecules}), despite our sensitivity to the prominent CH$_4$ band near 3.3$\mu$m. We also find that an abundance of heavy molecular species (H$_2$O, CO$_2$ or SO$_2$) is preferred as an explanation for the muted spectral features, even when stellar contamination is jointly fitted with atmospheric properties (Table~\ref{tab:mol_constr}, Figure \ref{fig:corner_nirspec_molecules}). }The \texttt{SCARLET}, \texttt{POSEIDON}, and \texttt{Aurora} retrievals consistently yield lower limits on the abundances of H$_2$O and CO$_2$, two-sided (broad) constraints on the SO$_2$ abundance ($VMR-4.83^{+1.16}_{-2.54}$ from SCARLET, $-4.34^{+1.65}_{-3.28}$ from POSEIDON and $-5.37^{+1.61}_{-2.94}$ from Aurora)  
, and a 2$\sigma$ upper limit of $\sim$100\,ppm on the CH$_4$ abundance (Figure \ref{fig:corner_nirspec_molecules}). Although H$_2$O, CO$_2$ or SO$_2$ are not individually significantly detected from a model comparison standpoint, the model requires the presence of at least one of these heavy molecules at 3.4$\sigma$ (significance from the \texttt{SCARLET} retrieval). { The low signal-to-noise of the atmospheric signatures does not enable us to infer the atmospheric metallicity from free retrievals, and obtaining a directly data-driven metallicity constraint from this chemistry-agnostic approach would require repeat observations}. 

{ As noted in the previous subsection, our conclusion of a high metallicity from the NIRSpec transmission spectrum is robust to the consideration of stellar contamination as it cannot falsely cause a detection of SO$_2$ or CO$_2$. Indeed, we verify with an additional \texttt{POSEIDON} retrieval, identical to the free chemistry retrievals described above, but with the addition of the effects of TLS, that when we include TLS in our NIRSpec retrievals, the abundance of H$_2$O is unconstrained, however the high-abundances of CO$_2$ and SO$_2$, as well as the non-detection of CH$_4$, remain.  }

\begin{figure*}
    \centering    
    \includegraphics[width=0.9\textwidth]{ 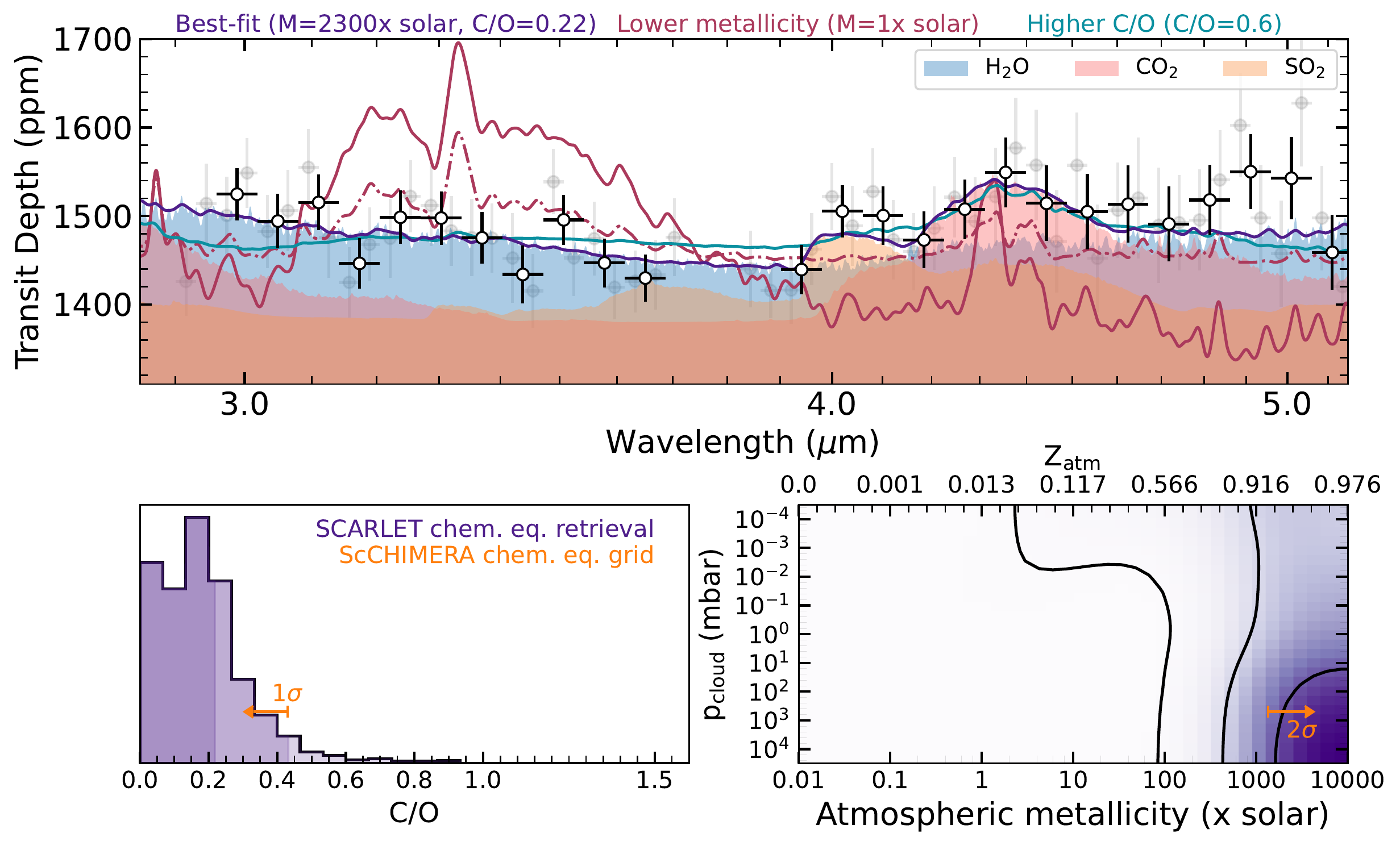}
    \caption{The confident non-detection of CH$_4$ at $\sim$3.4\,µm in the NIRSpec/G395H spectrum drives our inference of a high-metallicity atmosphere, while the tentative SO$_2$ feature suggests a low C/O atmosphere on GJ 3090\,b. \textit{Top panel:} NIRSpec/G395H Visit 1+2 transmission spectrum from \texttt{exoTEDRF} (gray points, used for retrieval), and the binned spectrum (black points), with representative atmosphere models. The purple line is the best-fitting chemically-consistent model, while the pink model highlights the impact of lowering the metallicity to 1$\times$ solar (solid line: cloud-free model, dash-dotted model: clouds at 1\,mbar), and the blue model illustrates the expectation for a higher C/O (C/O=0.6) atmosphere. For the low-metallicity case, CH$_4$, rather than H$_2$O absorption, dominates over the wavelength range covered by the NRS1 detector. The colour shadings illustrate the contributions of H$_2$O, CO$_2$, and SO$_2$ to the model spectrum. \textit{Bottom panels:} Posterior distributions on the atmospheric C/O ratio and metallicity from the \texttt{SCARLET} chemical equilibrium retrieval (purple) as well as the \texttt{ScCHIMERA} equilibrium grid fits (1$\sigma$ upper limit on the C/O ratio, and 2$\sigma$ lower limit on the metallicity quoted in orange). The left panel is the C/O ratio, with the 1, 2, and 3$\sigma$ upper limits shown in different colour shadings. The right panel is the joint posterior distribution on the atmospheric metallicity and cloud-top pressure (1, 2, and 3$\sigma$ contours for the \texttt{SCARLET} retrievals outlined in black), which highlights that the small-amplitude features in our spectrum favour a high metallicity atmosphere on \myplanet.}
    \label{fig:atmosphere_plot}
\end{figure*}

The results from the chemically-consistent retrievals are shown in Figure~\ref{fig:atmosphere_plot}. Despite having fewer parameters and making the assumption that all molecular abundances follow chemical equilibrium expectations (except for SO$_2$ when it is fitted independently), the \texttt{SCARLET} retrievals enable us to start drawing more meaningful conclusions about the atmospheric chemistry than the free retrievals, since we obtain limited information on individual molecular abundances due to the highly-muted features in the spectrum of \myplanet. { However, even if the spectral features have low amplitudes, they are significant enough to enable us to rule out clouds as the origin of their low amplitudes, at least in the most metal-poor atmospheres (at the 3$\sigma$ level, our observations could still be explained by clouds at pressures of less than 10$\mu$bar in a 10$\times$ solar metallicity atmosphere; Figure \ref{fig:atmosphere_plot}). The preference for high-metallicity, and generally sub-solar C/O ratios as the explanation for the transmission spectrum is independently supported by the \texttt{ScCHIMERA} chemical equilibrium grid-based retrieval results, that incorporate an additional level of consistency with pressure-temperature profiles calculated self-consistently in chemical equilibrium rather than prescribed by the observations (as in the \texttt{SCARLET} retrievals). The C/O is similarly unconstrained in the grid-based chemical equilibrium retrievals, although sub-solar values are preferred (Figure \ref{fig:gridretrieval}). In agreement with the chemically consistent retrieval findings, the lack of constraints on cloud parameters ($\kappa_\mathrm{cloud}$ and a cloud covering fraction) demonstrates that clouds cannot sufficiently mute spectral features to explain the observed NIRSpec/G395H spectrum without invoking high metallicity.}

\begin{figure*}
    \centering
    \includegraphics[width=0.9\linewidth]{ 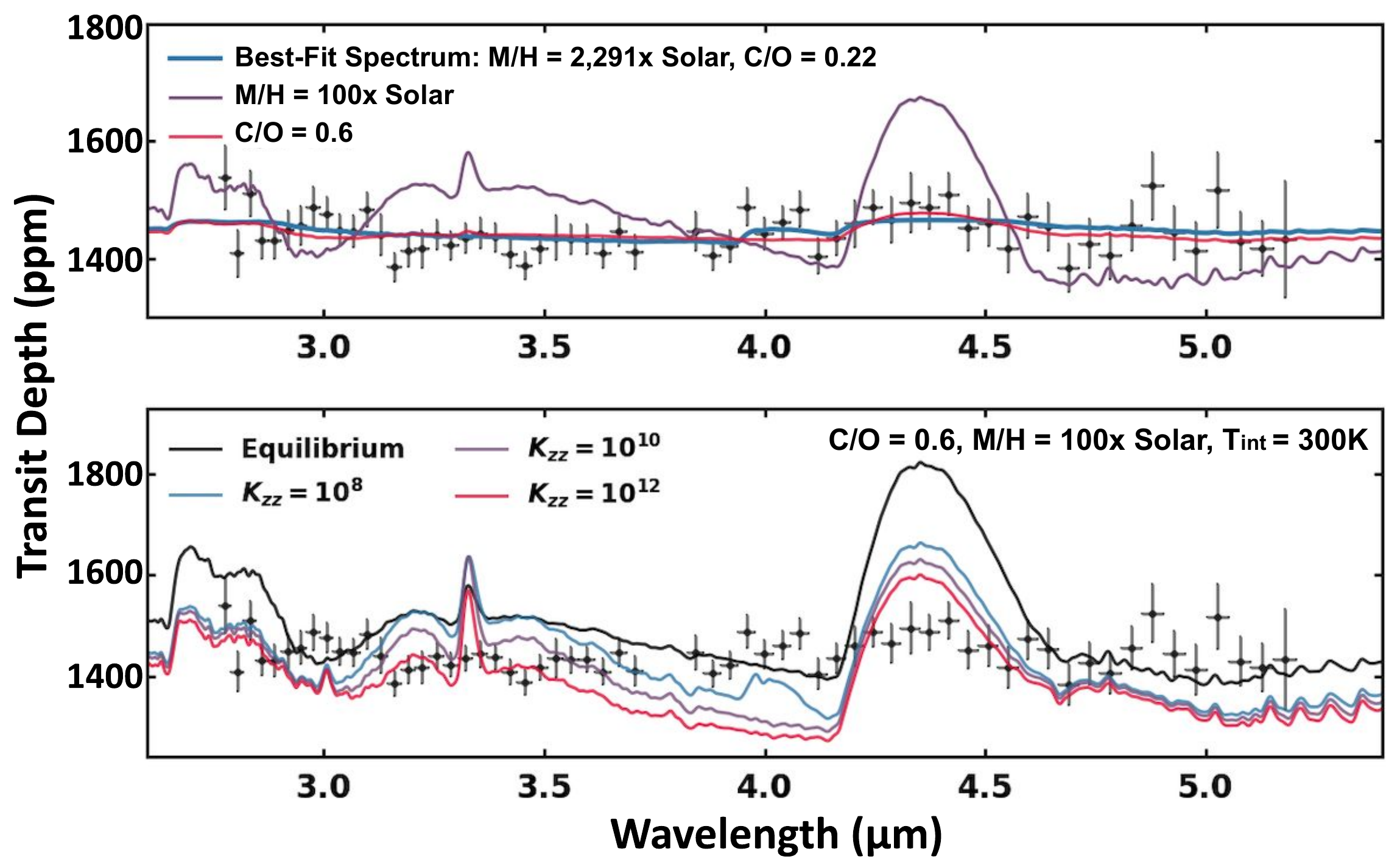} 
    \caption{The effects of varying the atmosphere chemistry on the NIRSpec/G395H transmission spectrum of \myplanet.
    \emph{Top:} Plotted in blue is the best-fit spectrum from the grid-based 1D RCTE retrieval with \texttt{ScCHIMERA}. In purple and blue, are model spectra with the same parameters as the best-fit model but with M/H set to 100$\times$ solar or C/O set to 0.6, respectively. 
    \emph{Bottom:} Models with adjusted K$_{zz}$, demonstrating that even a large degree of vertical mixing combined with a high internal temperature (in this case, plotted models assume T$_\mathrm{int}$=300\,K, while our grid-based retrievals assume a lower T$_\mathrm{int}$=100 K) cannot sufficiently deplete CH$_4$ even in a lower-metallicity atmosphere (C/O=0.6, M/H=100$\times$ solar) to match the lack of observed spectra features.}
    \label{fig:gridmodels}
\end{figure*}

{ Overall, the inference of high metal enrichment from chemical equilibrium modeling is driven by the muted spectral features in the infrared (Figure \ref{fig:atmosphere_plot}), with a potential spectroscopic signature of CO2 at $\sim$4.2
\textmu m and the lack of \ce{CH4}. Note that the absence of the methane feature is not necessarily an indicator of high metallicity in exoplanet atmospheres, as it can be achieved even in low-metallicity atmospheres provided that methane is depleted by disequilibrium chemistry, such as photochemistry, or quenching due to interactions with a hot interior \citep[e.g.,][]{fortney_beyond_2020}. We examine the impact of disequilibrium chemistry using the \texttt{ScCHIMERA} self-consistent forward modeling approach. Even when considering a high interior temperature of 300\,K (which may be allowable due to tidal heating from a non-zero orbital eccentricity; see Section~\ref{sec: interior models}) and strong mixing with a C/O of 0.6 (relatively high compared to what was found by the retrievals, see Figure \ref{fig:atmosphere_plot}), we find that quenching in a low-metallicity atmosphere cannot reproduce the level of methane depletion we observe (Figure \ref{fig:gridmodels}). This modeling demonstrates the robustness of the high-metallicity conclusion in the self-consistent modeling approach, even in the presence of quenching driven by a hot interior. This high metallicity would not be representative of an H$_2$ rich envelope. Meanwhile, the low-C/O conclusion requires further investigation as it hinges more strongly on the presence of SO$_2$, for which we do not achieve a significant detection.}

{ Finally, we once again use \texttt{POSEIDON} to compare the data-model fit statistics from our NIRSpec/G395H retrievals to a flat-line fit using the minimal atmosphere models outlined in the \texttt{POSEIDON} section of Section~\ref{sec:atm_joint_retrievals}. Our nested Bayesian model comparison results in a $\sim$2.1$\sigma$ preference for our minimal atmosphere-only model to a flat line, demonstrating moderate evidence for the presence of atmospheric features. Similarly, we find a $\sim$2.1$\sigma$ preference for our joint minimal atmosphere-TLS model over our spots-only model, suggesting that our atmospheric signal cannot merely be attributed to stellar contamination. This indicates that the presence of both an atmosphere and stellar spots best explains the observed NIRSpec/G395H spectrum.}

\subsection{Helium Absorption Supports a Metal-Enriched Atmosphere}
\label{sec:helium-discussion}

We find strong evidence for helium absorption in the transmission spectrum of \myplanet, though the resolution of NIRISS/SOSS is not high enough to constrain the width and shape of the helium triplet absorption and as such is not sufficient to provide constraints on the hydrodynamic escape of \myplanet's atmosphere. This follows the prediction by \citet{DosSantos2023JWSTAtmosphericEscape} and has also been borne out in the analysis of the NIRISS/SOSS observations of the warm-Jupiter HAT-P-18\,b \citep{fu_water_2023, fournier-tondreau_near-infrared_2024}.

We model the predicted excess metastable helium absorption for \myplanet using the \texttt{p-winds} \citep{DosSantosVidotto2022} python wrapper for the one-dimensional photo-ionization hydrodynamic code \texttt{ATES} \citep{CaldiroliHaardt2021, BiassoniCaldiroli2024}. \texttt{ATES} assumes a non-isothermal atmosphere, allowing us to self-consistently calculate the temperature profile as a function of radial distance by obtaining solutions that do not violate energy conservation \citep{CaldiroliHaardt2021}. Using \texttt{ATES}, the planet parameters, and a proxy stellar spectrum of GJ\,832 from the MUSCLES survey \citep{Youngblood2016ApJ...824..101Y} which matches \mystar closely in terms of stellar parameters ($T_\mathrm{eff}$, $\log g$, $\log R'_\mathrm{HK}$), we calculate the predicted metastable helium absorption and mass-loss rate for \myplanet assuming a 90/10 H$_2$/He ratio. We find a predicted excess absorption of $\sim$3.5\% at the metastable helium wavelengths, consistent with the strong absorption signals predicted for planets orbiting M-type hosts \citep{BiassoniCaldiroli2024}. However, convolving our signal with the resolution of NIRISS/SOSS at the metastable helium wavelengths, we find a predicted excess absorption for \myplanet with NIRISS/SOSS of 0.51$\%$, corresponding to a mass-loss rate of log $\dot{M}$ of 10.1 (cgs). If assumed to be constant in time, this would yield an upper limit of $\sim$50\,Gyr for photoevaporation to completely strip the planet of its atmosphere.

However, it is critical to note that this modelled absorption is a factor of 10 higher than we have detected in our observations ($434 \pm 79$\,ppm or $0.0434 \pm 0.0079 \%$). The models used to predict the excess absorption and mass-loss rate are optimized for solar metallicity atmospheres \citep{CaldiroliHaardt2021, BiassoniCaldiroli2024}. While most mass-loss models do not have metallicity as a tunable parameter, previous works \citep[e.g.,][]{owen_planetary_2012, owen_metallicity_2018, ZhangTOI560, Vissapragada2024, Zhang2024} have shown that as metallicity increases, particularly beyond $\sim$100$\times$ solar, the excess absorption signal and mass-loss rate are significantly reduced. This, therefore, provides an independent line of evidence that the metallicity of GJ 3090\,b's atmosphere is indeed highly elevated, and the envelope lifetime derived above should be taken as a lower limit as a decreased mass-loss rate due to elevated atmosphere metallicity will increase the atmospheric lifetime.

 {A potential caveat to this conclusion is that the comparative weakness of the He signal that we detect could be due to He depletion, as has been inferred for Jupiter \citep[e.g.,][]{vonzahn_helium_1998} as well as WASP-80\,b \citep[e.g.,][]{fossati_possible_2023}. However, sub-Neptunes at the upper edge of the radius valley are instead predicted to potentially be He-enriched \citep[e.g.,][]{malsky_helium_2022, Cherubim2024, cherubim_oxidation_2025}, making the He-depletion scenario less plausible \textit{a-priori}.}

Further observations with high-resolution instruments will allow us to resolve the metastable helium line, providing better constraints on the mass loss rate, outflow temperature and velocity structure of the outflow. This marks another sub-Neptune helium detection which is lower than expected, consistent with ground-based observations \citep{ZhangTOI560,Zhang2022}, { though the first using JWST.}

\subsection{Potential Scenarios Leading to Metal-Enrichment}

\label{sec:metal-enrichment}

There are multiple pathways through which sub-Neptunes can become metal-enriched. Formation theories propose that these planets may acquire their metal enrichment by accreting metal-rich planetesimals \citep[e.g.,][]{Fortney2013}. A large amount of solids in the form of water ice could have been accreted onto \myplanet's atmosphere if it formed beyond the ice line \citep{leger_new_2004,luger_habitable_2015,alibert_formation_2017,kite_habitability_2018,venturini_nature_2020,bitsch_dry_2021}. With planet migration mechanisms the planet could have later moved closer to its host star and become warm enough that a fraction of the accreted material sublimates and ends up as water vapour in the atmosphere, increasing the metallicity. Alternatively, volatiles could even sublimate directly upon their initial accretion into the growing atmosphere. 

For warm and hot sub-Neptunes, however, it is also expected that their envelopes experience significant mass loss due to hydrodynamic atmospheric escape processes \citep{owen_evaporation_2017,ginzburg_core-powered_2018}, which can create a high-metallicity atmosphere. Hydrodynamic escape can be powered by the high XUV irradiation from the host star \citep[photoevaporation, e.g.,][]{Watson1981, VidalMadjar2003, Ehrenreich2015, Lammer2003, Erkaev2007, Salz2016} and/or a planet's internal energy from formation \citep[core-powered mass-loss, e.g.,][]{ginzburg_core-powered_2018,gupta_sculpting_2019,Gupta2020corepowered_signatures}. Given the $\sim$1 Gyr age of \mystar \citep{Almenara2022GJCharacterisation} and the observed helium escape (see Section~\ref{sec: Helium}), one plausible explanation for its observed metal-rich atmosphere could be the extensive history of hydrodynamic atmospheric escape that has progressively removed a H$_2$/He envelope. 

Recent works by \citet{malsky_helium_2022}, \citet{Cherubim2024}, and Louca et al.~(submitted) underscore the transformative role of hydrodynamic escape on the atmospheric evolution for sub-Neptune and Neptune-sized exoplanets. For example, the findings from Louca et al.\ indicate that extreme hydrodynamic escape can significantly enrich atmosphere metal content for planets with equilibrium temperatures between 700 and 1000\,K, resulting in a water-enriched atmosphere after $\sim$300\,Myr. They also showed a notable decrease in the atmospheric C/O ratio due to the drag of heavier elements.

Another possible scenario is to enrich the atmosphere by interactions between a magma ocean and the atmosphere, producing significant amounts of molecules such as H$_2$O in the H$_2$/He-rich envelopes directly above the magma ocean \citep[e.g.][]{kite_water_2021,schlichting_chemical_2022,Rogers2024,lichtenberg_super-earths_2024}. {  While we would expect H$_2$O features in \myplanet's atmosphere,} we are not able to robustly constrain the water abundance in the atmosphere of \myplanet due to degeneracies with stellar contamination{.   So} we cannot make any statements about the hydrogen budget in the atmosphere {  at this point in time}. Nevertheless, this mechanism suggests that oxidised planets enrich the atmosphere with O-rich species, also including \ce{CO2} or SO$_X$-compounds, in line with our {  potential findings of these species} in the spectrum of \myplanet.

\subsection{Potential Implications for the Interior}
\label{sec: interior models}

The interior structure of sub-Neptunes is highly degenerate between the composition and mass of the core and the amount of H/He in the envelope \citep{fortney_framework_2013,otegi_impact_2020} and to resolve this degeneracy, atmospheric characterization data is crucial. Here we estimate conclusions on the mass of the core of \myplanet we can draw when we assume the metal mass fraction of the envelope ($Z_\mathrm{env}$) to be constrained by the one derived from our atmospheric modelling analysis (see Figure~\ref{fig:atmosphere_plot}, bottom right panel). 

We generate a suite of interior structure models with the GAS gianT modeL for Interiors \citep[\texttt{GASTLI},][]{acuna21,acuna24}. \texttt{GASTLI} models the planetary interior as two layers: a metal-rich core and an envelope. The core, which in this work we refer to as the mass of the region where metals are found outside the envelope, comprises rocks and water in a 1:1 mass ratio, while the envelope consists of H/He and water. A 1:1 water-to-rock ratio for the core has previously been assumed in interior models of gas giants and sub-Neptunes \citep{fortney_planetary_2007}. The exact core compositions of volatile-rich planets remain uncertain, even for the Solar System gas and ice giants, despite the availability of Love numbers and gravitational field data \citep{SS_interiors_Helled}. Furthermore, sub-Neptunes can exhibit a wide range of ice-to-rock ratios depending on their formation locations \citep{Mousis19_icelines,Mah21}. Thus, the 1:1 water-to-rock core composition provides a simple and reasonable starting assumption. The metal mass fraction of the envelope, $Z_\mathrm{env}$, and the core mass fraction (CMF) are user-defined parameters. In this work, we use the publicly available atmospheric grid from \cite{kempton_reflective_2023} (K23) to calculate the interior-atmosphere boundary temperature at 1000 bar. This grid is suitable for \myplanet given its equilibrium temperature, low surface gravity, and M dwarf host star. K23 adopt water as the only metal absorber in their opacity calculations, rather than scaling absorbers such as CH$_{4}$, CO and CO$_{2}$ according to solar abundances. To assess opacity-driven differences, we compared PT profiles from K23 and \texttt{GASTLI}’s default atmospheric grid, which includes a solar-scaled mix of absorbers \citep{molliere_model_2015,acuna24}. Using the same set of equilibrium and internal temperatures, and surface gravity consistent with \myplanet, K23's grid predicts a temperature 250 K warmer at 1000 bars than the \texttt{petitCODE} grid. \cite{acuna_interior-atmosphere_2023} report that such temperature differences (200 K) result in radius variations of 1\% for sub-Neptunes, well below the 5\% observational uncertainty in \myplanet’s radius.

\begin{figure}[h]
    \centering
    \includegraphics[width=0.48\textwidth]{ 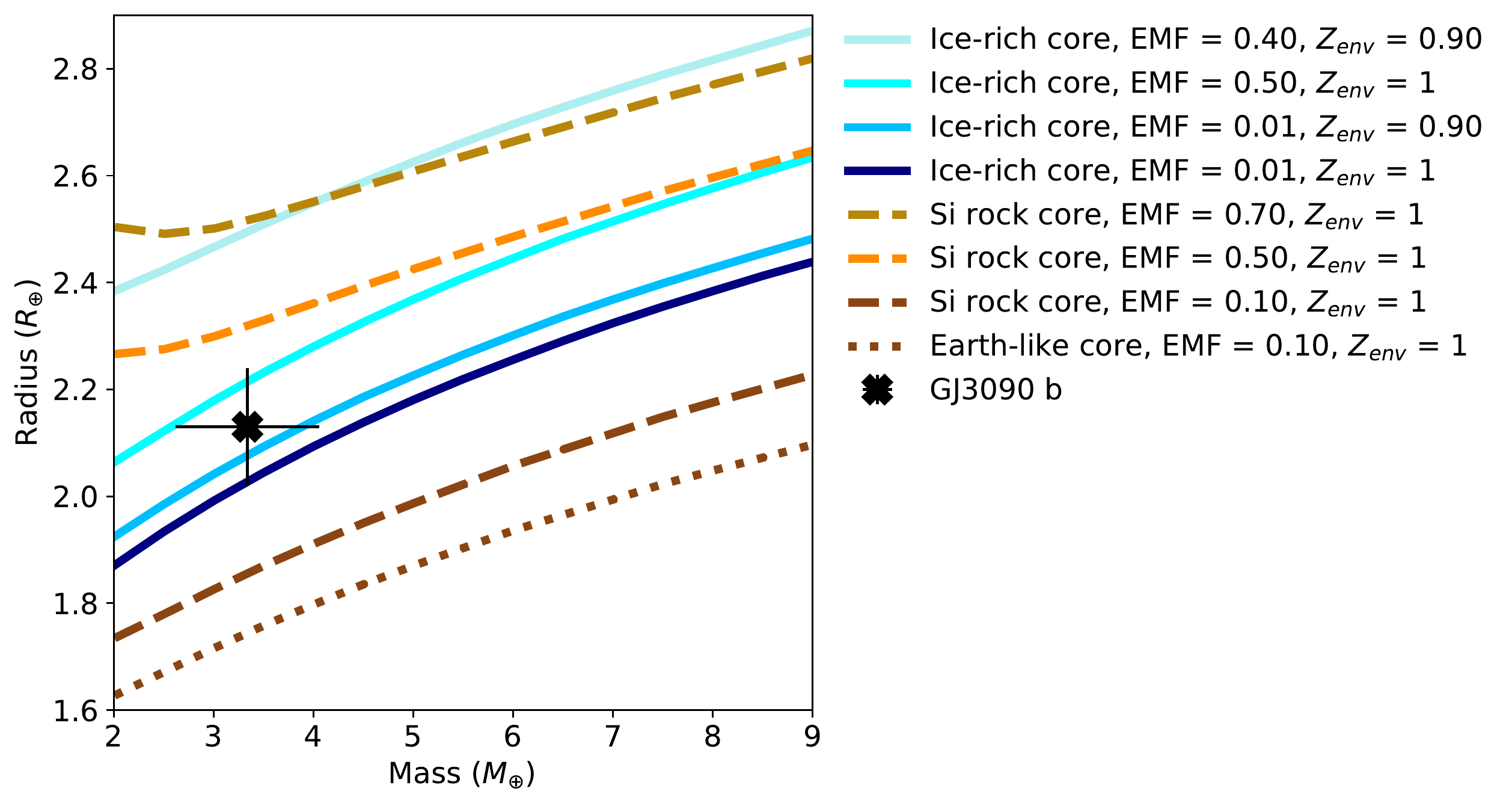}
    \caption{Mass-radius curves for \myplanet ($T_\mathrm{eq}$ = 693 K) at varying core compositions (ice-rich, pure silicate rock and Earth-like iron-to-rock ratio), envelope mass fractions (EMFs), and mass fraction of metals in the envelope ($Z_\mathrm{env}$). The \texttt{GASTLI} models (solid lines) assume a miscible 1:1 ice and rock core, whereas the A21 models \citep[dashed and dotted lines]{aguichine_mass-radius_2021} assume dry, pure rock (no Fe) and Earth-like (32\% Fe) core compositions.}
    \label{fig:interior_MRrel}
\end{figure}

In addition to the CMF and envelope metal mass fraction, the intrinsic (or internal) temperature can strongly influence the planet's radius, particularly for younger planets \citep{lopez_understanding_2014}. The age of \mystar is  1.02$^{+0.23}_{-0.15}$\,Gyr \citep{Almenara2022GJCharacterisation}. We estimate the internal temperature of \myplanet by comparing its radius and age (including uncertainties) with thermal evolution curves calculated at different CMFs and $Z_\mathrm{env}$ at a constant mass of M$_\mathrm{p}$ = 3.34 $M_{\oplus}$ with \texttt{GASTLI}. We set a lower bound of  $Z_\mathrm{env}$ = 0.90, as the 3$\sigma$ limit of the envelope metal mass fraction from atmospheric retrievals is $\sim$0.916. This analysis constrains \myplanet's internal temperature to $T_\mathrm{int} = $ 20--40\,K and 20--70\,K within 1$\sigma$ and 3$\sigma$, respectively. Within the 20--70\,K range, the planet radius increases by 0.05 $R_{\oplus}$, which is below the observational radius uncertainty, $\sigma(R) = 0.11 \ R_{\oplus}$. Therefore, we adopt a nominal value of $T_\mathrm{int} = 40$\,K for the computation of mass-radius curves. 

Figure~\ref{fig:interior_MRrel} presents the mass-radius curves for \myplanet, assuming $T_\mathrm{eq}$ = 693\,K, $T_\mathrm{int}$ = 40\,K and envelope metal mass fractions ranging from $Z_\mathrm{env}$ = 1 (pure water) to 0.90. The mass and radius of \myplanet are compatible with a CMF $\sim$ 0.99 to 0.50 for a pure water envelope ($Z_\mathrm{env}=$ 1) at the 1$\sigma$ level. For envelopes containing H/He, a CMF of 0.60 remains consistent with the mass and the radius at the 3$\sigma$ level. Thus, we can conclude that if \myplanet has a well-mixed core of rock and ice \citep[i.e., water in supercritical and/or superionic phases, see ][]{haldemann_aqua_2020,mousis_irradiated_2020}, its mass fraction lies between 0.50 and $\sim$1. 


Sub-Neptunes that form close to the water ice line may accrete ice-rich material, resulting in cores where ice and rock are miscible \citep{vazan22, luo_majority_2024}. In contrast, formation inside the ice line or close to the Fe and Si lines may produce dry cores dominated by silicates and Fe \citep{lee_breeding_2016, aguichine_craddles_20}. To explore this scenario, we include mass-radius curves for pure rock and Earth-like cores \citep{aguichine_mass-radius_2021}, which correspond to no Fe and 32\% Fe by mass, respectively. In these models, the envelope is entirely constituted by water. Figure~\ref{fig:interior_MRrel} shows that \myplanet's mass and radius are consistent with CMFs = 0.50--0.70 (1$\sigma$) for pure rock cores, while CMF $< $0.30 are ruled out at the 3$\sigma$ level. For Earth-like cores, CMF $>$ 0.90 are inconsistent with the mass and radius data beyond 3$\sigma$.

\section{Summary and Conclusions}
\label{sec:conclusions}

In this paper, we presented four transit observations of \myplanet with JWST, two each with NIRISS/SOSS and NIRSpec/G395H as part of a JWST sub-Neptune survey (GO\,4098, PIs: B. Benneke, T. Evans-Soma). 

We detect absorption from the He I triplet at 1.0833\,µm in the NIRISS/SOSS pixel-level transmission spectra in both independent reductions and both visits at a statistical significance of $>$5$\sigma$. The observed amplitude is an order of magnitude smaller than predicted by forward models with solar metallicity, independently supporting the inferred high metallicity for \myplanet's atmosphere. We also predict that this absorption feature is observable with ground-based high-resolution spectrographs which can further constrain the mass-loss and outflow temperature. 

We detect strong ($\sim$5$\sigma$) evidence for the TLS effect in the NIRISS/SOSS wavelength range, preventing any robust constraints on the planetary atmosphere. We find that the two SOSS visits, taken 6 months ($\sim$10 stellar rotation periods) apart, showed a difference in stellar heterogeneity parameters, highlighting the variable nature of stellar contamination signals. {  This prohibited any atmospheric inferences from the NIRISS/SOSS data as TLS-only retrieval models were preferred over TLS+atmosphere models.} 

Moreover, offsets between the NIRISS and NIRSpec spectra and between the two NIRISS visits, which could not be traced back to data reduction or light curve fitting differences, also suggest different stellar contamination realizations between spectra from the two instruments. These variable impacts of the TLS effect all but precluded joint atmosphere analyses of the NIRISS and NIRSpec spectra. Therefore, we recommend that future exoplanet observations which need both NIRISS and NIRSpec for their science are scheduled close enough in time that the star's photosphere will not have significantly changed and a single TLS realization can be assumed for spectra from both instruments. 

Using the two NIRSpec/G395H visits we presented atmospheric retrievals in addition to chemically-consistent retrievals and a grid analysis to explore the impact of disequilibrium chemistry. We find that \myplanet's spectrum is best explained by a high metallicity atmosphere. The best-fitting grid models prefer a $>$1000$\times$ solar metallicity atmosphere (best fit at mean molecular weight $\sim$ 27 amu), and chemically consistent retrievals indicate a metallicity $>$100$\times$ at 3-$\sigma$ confidence for clouds at $<$µbar pressures, or $>$700$\times$ solar at 2-$\sigma$ irrespective of the presence of clouds. {  The free retrievals were not able to constrain any one molecule, though the presence of at least one high-MMW molecule is favoured by $3.4\sigma$}. 
Both the high metallicity and the sub-solar C/O ratio could be explained by ice-rich formation or atmospheric evolution under chemical exchange with the interior and/or atmospheric mass loss. 

However, our findings are subject to certain caveats. Firstly, while we conducted a comprehensive analysis regarding stellar contamination and the systematics in our data, we note that any underestimation (or overestimation) of the TLS effect, or inaccuracies in the stellar models can affect our retrieved atmospheric constraints. In addition, due to the almost featureless nature of the transmission spectrum of \myplanet in the NIRSpec wavelength range and the stellar contamination in NIRISS, our retrieved atmospheric constraints {  may be biased based on model assumptions. The chemically consistent modeling may be limited, while the free chemistry may be hindered by its flexibility. The chemically consistent models are largely informed by the muted features  and potential hints of heavier molecules, like CO$_2$ and SO$_2$, in the atmosphere.} However, the lack of CH$_4$ could also be explained by chemistry we do not consider in our modelling, e.g., associated with 3D effects as have been explored for hot Jupiters \citep{zamyatina_2023, Zamyatina2024WASP-96bCH4}. {  On the other hand, free chemistry retrievals favour at least one of the heavy molecules to be present in \myplanet's atmosphere at 3.4$\sigma$.  Therefore, our both chemically consistent and free chemistry modelling are suggest a high-metallicity atmosphere which is in line with what was found in recent CRIRES+ observations \citep{Parker2025GJ3090}}, though further observations are needed to definitively detect absorption from carbon- and sulfur-bearing species in the atmosphere of \myplanet. \\

\noindent
\textbf{Acknowledgements}\\
This work is based on observations with the NASA/ESA/CSA James Webb Space Telescope, obtained at the Space Telescope Science Institute (STScI) operated by AURA, Inc. All of the data presented in this paper were obtained from the Mikulski Archive for Space Telescopes (MAST) at the Space Telescope Science Institute. The data used in this paper can be found in MAST at the following DOIs: \dataset[10.17909/r2bz-wb41]{http://dx.doi.org/10.17909/r2bz-wb41}.
EA is grateful for the support from the Max Planck Gesellschaft (MPG) and the Deutsches Zentrum für Luft- und Raumfahrt (DLR). 
MR acknowledges support from the Natural Sciences and Engineering Research Council of Canada (NSERC) and the Fonds de recherche de Québec -- nature et technologie (FRQNT).
CPG acknowledges support from the NSERC Vanier scholarship, and the Trottier Family Foundation. CPG also acknowledges support from the E. Margaret Burbidge Prize Postdoctoral Fellowship from the Brinson Foundation.
HES gratefully acknowledges support from JWST grant \#JWST-GO-04098.005-A.
R.A. acknowledges the Swiss National Science Foundation (SNSF) support under the Post-Doc Mobility grant P500PT\_222212 and the support of the Institut Trottier de Recherche sur les Exoplan\`{e}tes (iREx).\\

\noindent
\textbf{Data Availability}\\
Data products (light curves, transmission spectra) as well as modelling outputs (posterior distributions and best-fit models) are availabe on Zenodo, under \url{https://doi.org/10.5281/zenodo.15273592}.

\vspace{5mm}
\facilities{JWST (NIRSpec and NIRISS)}

\software{\texttt{astropy} \citep{astropy:2013, astropy:2018}, 
\texttt{batman} \citep{Kreidberg2015BatmanPython}, 
\texttt{celerite} \citep{foreman-mackey_fast_2017},
\texttt{emcee} \citep{Foremak-Mackey2013}, 
\texttt{Eureka!} \citep{Bell2022Eureka:Observations},
\texttt{exoTEDRF} \citep{Radica2024b},
\texttt{ExoTiC-LD} \citep{Grant2024},
\texttt{exoUPRF} \citep{Radica2024c},
\texttt{ipython} \citep{PER-GRA:2007},
\texttt{jwst} \citep{bushouse_2023_8157276},
\texttt{matplotlib} \citep{Hunter:2007},
\texttt{MCMC} \citep{Foreman-Mackey2013EmceeHammer},
\texttt{numpy} \citep{harris2020array},
\texttt{scipy} \citep{2020SciPy-NMeth},
\texttt{Tiberius} \citep{Kirk2017RayleighHAT-P-18b,Kirk2021ACCESSWASP-103b,Ahrer2022LRG-BEASTS:NTT/EFOSC2}
}

\appendix
\setcounter{figure}{0}
\renewcommand{\thefigure}{A.\arabic{figure}}

\section{Joint Light Curve Fitting: Constraints on Eccentricity}
\label{sec: eccentricity fits}

\begin{figure}
    \centering
    \includegraphics[width=0.5\linewidth]{ 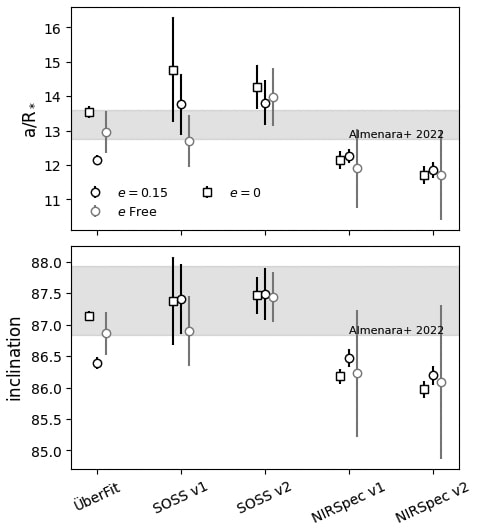}
    \caption{Comparison of derived orbital parameters using different datasets. The joint ``überfit'' uses all four observed transit white light curves, whereas the other four cases use the light curves from an individual visit. Three cases are shown for each fit: assuming a circular orbit, fixing the eccentricity to 0.15, and leaving the eccentricity free. The constraints from \citet{Almenara2022GJCharacterisation} are marked with grey shading. In general, NIRSpec prefers smaller values of $a/R_*$ and inclination (which are positively correlated). The joint fit values are consistent with the findings of \citet{Almenara2022GJCharacterisation}, except in the case where we fix $e=0.15$.}
    \label{fig:orbit compare}
\end{figure}

In their initial analysis of the GJ 3090\,b system, \citet{Almenara2022GJCharacterisation} are not able to give firm constraints on the orbital eccentricity of GJ 3090\,b, with Keplerian, and ``dynamical'' modelling suggesting a non-zero eccentricity between 0.16 and 0.18, which relaxes to $e<0.32$ at 3$\sigma$ confidence when considering the long term stability of the orbits of planets b and c. We, therefore, take an agnostic approach to the eccentricity in our fits, considering three separate cases: fixing $e=0$, fixing $e=0.15$ consistent with the dynamical and Keplerian modelling of \citet{Almenara2022GJCharacterisation}, and leaving the eccentricity free. Figure~\ref{fig:orbit compare} shows the impact of the eccentricity on the fitted orbital parameters (inclination and scaled semi-major axis). When fitting the transit white light curves of an individual visit, the eccentricity treatment does not have a significant impact on the other orbital parameters \citep{Seager2003}. However, there is a substantial impact when jointly fitting the four transits due to the eccentricity's impact on the time of transit. In the joint fitting case, the fixed zero and non-zero eccentricity results bracket the free eccentricity case, indicating that we do not have a significant constraint on the orbital eccentricity from this dataset. The best fitting eccentricity from the free-eccentricity joint fit is slightly-non-zero (Table~\ref{tab:fitted_lc_params}), though consistent with zero when considering the Lucy-Sweeney bias \citep{Lucy1971}.

\begin{figure}
    \centering
    \includegraphics[width=0.5\linewidth]{ 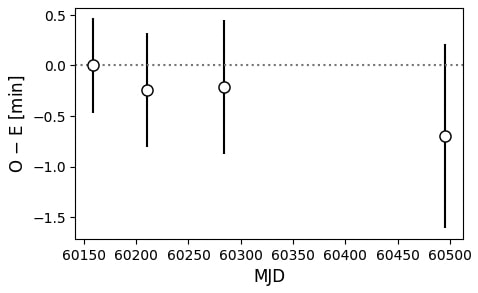}
    \caption{Mid-transit times for each visit compared to a linear ephemeris propagated from visit 1. There is a slight trend to earlier times, with the final visit occurs approximately 1 minute earlier than would be expected from the linear ephemeris (though still consistent with zero with the propagated error), potentially suggesting some evidence for a non-zero orbital eccentricity, or dynamical interactions with a potential planet c.}
    \label{fig:transit times}
\end{figure}

This is also demonstrated in Figure~\ref{fig:transit times}, where we show the times of mid-transit obtained from fitting each visit separately and assuming a circular orbit, compared to what would be expected given a circular orbit. Although all differences are consistent with zero within the propagated errors, there is a slight trend to earlier transit times, with the final visit occurring $\sim$1 minute earlier than predicted. This potentially indicates evidence of a non-zero eccentricity in \myplanet's orbit, or dynamical interactions with planet c. Moreover, we confirm that this tentative trend is not due to inaccuracy in the period reported by \citet{Almenara2022GJCharacterisation}. We repeat the fit described in Section~\ref{sec:joint-fit} but setting a Gaussian prior on the orbital period centered on the value from \citet{Almenara2022GJCharacterisation} and with a width of 10\,min, in order to check whether the period inferred from these four transits disagreed with that published by \citet{Almenara2022GJCharacterisation}. However, we find a period in perfect agreement with \citet{Almenara2022GJCharacterisation}. A more complete fit including the transit and RV data from \citet{Almenara2022GJCharacterisation} may be able to provide better constraints on the eccentricity. 

\section{Investigation into Stellar Activity Modelling}
\label{sec:appendix_stellar_flare_modelling}

To make sure our retrieved shape in the transmission spectrum is independent of the activity event captured in our transit light curves we explore fitting the flare using an asymmetric modified Lorentzian, an option within \texttt{Eureka!}'s Stage\,5 to account for stellar variability. The Lorentzian model $L(t)$ as a function of time $t$ is defined as 
\begin{equation}
    L(t) = 1 + \frac{A}{1 + x^p},
\end{equation}
such that $x = 2 (t - t_0)/\mathrm{HWHM}$, where $A$ is the amplitude, p the exponent (where $p=2$ for a standard Lorentzian), $\mathrm{HWHM}$ the half-width-at-half-maximum (HWHM). To create an asymmetric model representative of a flare, \texttt{Eureka!} allows for two independent Lorentzian models to fit the baseline before and after the flare mid-point $t_0$. Therefore we fit for two amplitudes, two HWHMs, one exponent and one flare mid-point at the white-light curve stage for both NRS1 and NRS2. The spectroscopic light curves we only fit for the two amplitudes before and after mid-flare, while we fix the other Lorentzian parameters as we assume that the shape of the flare is consistent across wavelengths. 

\begin{figure}
    \centering
    \includegraphics[width=.7\textwidth]{ 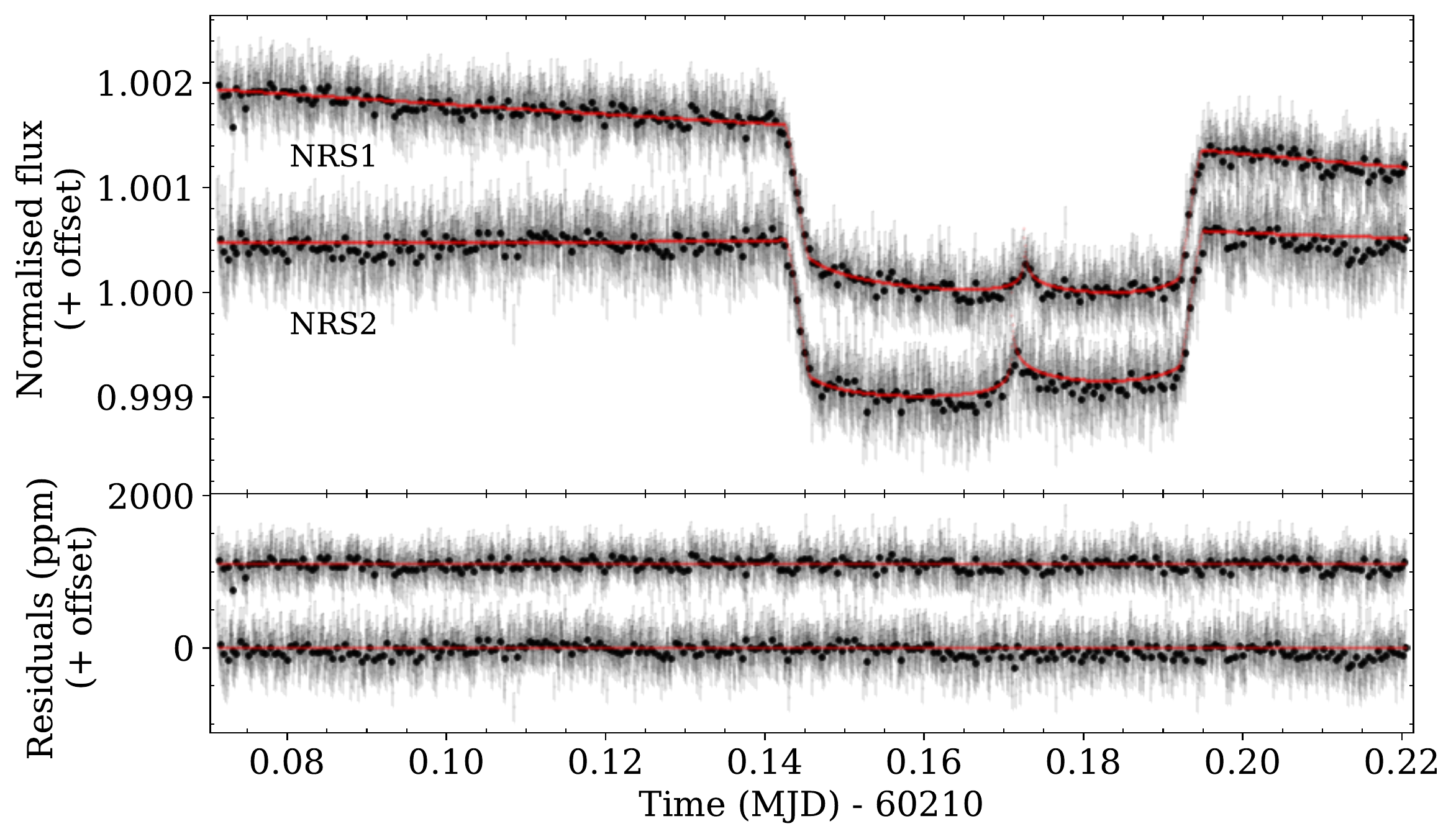}
    \caption{Visit 2: Broad-band (white) light curve of the transit of \myplanet with a stellar flare (grey), using NIRSpec's NRS1 and NRS2 detectors and the \texttt{Eureka!} pipeline. The black data points correspond to the flux binned (10 points) to aid visualisation. The fits to both light curves are shown in red and include a transit model, a linear model and a modified, asymmetric Lorentzian model to fit the flare. The residuals of the fits are shown in the bottom panel, displaying an offset in the data and the model after the flare. }
    \label{fig:flare}
\end{figure}

Figure~\ref{fig:flare} shows the white light curve of \myplanet's transit for NRS1 and NRS2 and the described stellar model. It does reasonably well in capturing the flare's shape, however, the residuals show that most data points lie below the model from just before the flare onwards. We investigated whether this step in the data could be caused by a mirror tilt event as has been previously seen in other JWST data sets \citep[e.g.][]{Alderson2023EarlyG395H}. We did not find any evidence for a tilt event occurring in our observations as we did not observe a change in the position or width of the trace. In addition, the guide star data also did not show any significant jumps during our observations \citep[as verified using the python tool \texttt{spelunker},][]{DealEspinoza2024spelunker}.

The retrieved transmission spectrum for Visit 2 does not show any significant differences compared to the spectrum where the flare is not fitted, however, the uncertainties are larger in the case where the flare is removed and a GP captures any leftover stellar variations. We conclude that the best fit to the light curves is achieved by the GP as it results in less residual noise and we argue that the larger uncertainties are more representative of the actual noise due to stellar variability in Visit 2. Therefore, moving forwards we use the transmission spectrum where the stellar variability in the light curves is modelled by the GP and the flare removed.

\section{Stellar Heterogeneity Modelling}

\subsection{Modelling the NIRISS Out-of-Transit Stellar Spectrum}
\label{sec:stellar_spectrum_modelling}

In order to verify that the stellar surface heterogeneities that we retrieve from our TLS analysis are robust, we also directly model the out-of-transit stellar spectrum from each NIRISS/SOSS visit \citep[e.g.,][]{wakeford_exoplanet_2019, Moran2023HighObservations}, closely following the methodology of \citet{radica_promise_2025}. We first flux calibrate the extracted wavelength-dependent stellar spectra from each visit following the methods of \citet{lim_atmospheric_2023}. We then median-combine the spectra along the temporal axis considering only the out-of-transit integrations; that is integrations 1--350 and 660--779 for both visits. Like \citet{Moran2023HighObservations}, we use the standard deviation of the flux in each wavelength bin as the flux uncertainty.

\begin{figure}
    \centering
    \includegraphics[width=0.95\textwidth]{ 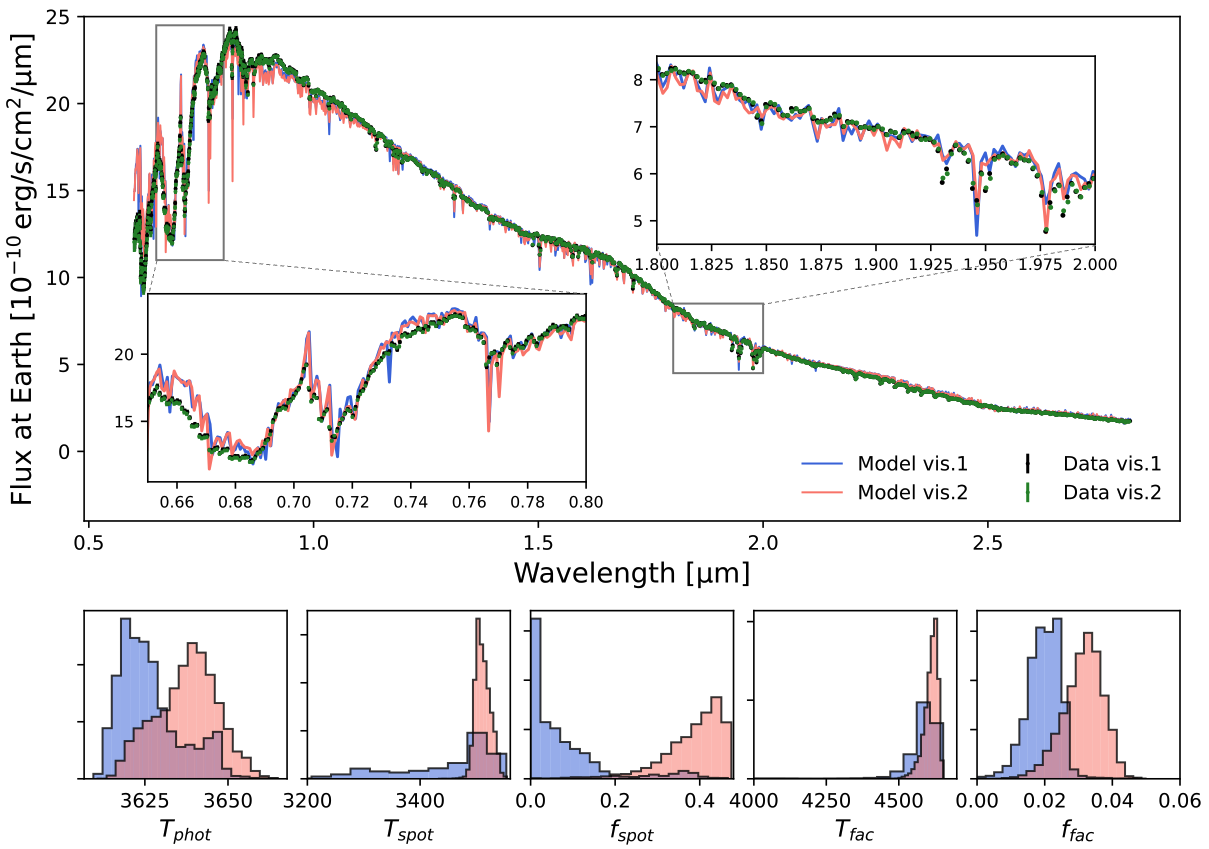}
    \caption{Constraints on stellar surface heterogeneities for both NIRISS/SOSS visits from the out-of-transit stellar spectra. 
    \emph{Top}: Flux-calibrated NIRISS/SOSS stellar spectra using only the out-of-transit integrations (visit 1; black, visit 2; green), and best fitting three-component PHOENIX model spectra (visit 1; blue, visit 2; red). Two $\sim$0.2\,µm wavelength ranges are shown zoomed-in in the insets to better visualize the higher-resolution structures. 
    \emph{Bottom}: Retrieved photosphere temperature, as well as spot and faculae temperatures and covering fractions. Posteriors for visit 1 are shown in blue and visit 2 in red. The out-of-transit results are broadly consistent with the in-transit analysis in Section~\ref{sec:tls_effect}.}
    \label{fig: stellar spectrum fit}
\end{figure}

We then fit one-, two-, and three-component stellar models to the data from each visit using the \texttt{StellarFit}\footnote{\url{https://github.com/radicamc/StellarFit}} package \citep{radica_promise_2025}. A one-component fit consists of a single PHOENIX stellar model \citep{husser_new_2013}, with a given effective temperature and surface gravity, to represent a homogeneous stellar photosphere. Two- or three-component fits also include spots or spots and faculae respectively to model surface heterogeneities. A heterogeneous model is constructed using a weighted linear combination of a photosphere as well as spots and/or faculae, where each component is weighted by their covering fraction, and the photosphere is required to compose $>$50\% of the stellar surface. All models are also scaled by $R_*^2/D^2$ using values from \citet{Almenara2022GJCharacterisation} for the stellar radius ($R_*$=0.516\,$R_\odot$) and the distance of the system to the Earth ($D$=22.45\,pc). 

Unlike previous works, we don't only consider the best-fitting stellar model, but fully sample the posterior space in order to obtain distributions on the acceptable ranges of heterogeneity parameters, analogous to our in-transit analysis. Moreover, as we use nested sampling, implemented via \texttt{dynesty} \citep{Speagle2020Dynesty:Evidences}, for the posterior exploration we can also robustly ascertain whether heterogeneous models are truly statistically preferred over a homogeneous photosphere model instead of relying on the $\chi^2$ statistic. 

One-component fits have four free parameters: the stellar effective temperature, $T_{phot}$, the surface gravity, as well as a spectrum scaling factor and multiplicative error inflation term. The photosphere temperature was allowed to vary from 2300 -- 5000\,K, the gravity from 3.5 -- 5, and the scale factor from 0.8 -- 1.2. Multi-component fits also include the spot/faculae temperature ($T_\mathrm{spot}$/$T_{fac}$), covering fraction ($f_\mathrm{spot}$/$f_\mathrm{fac}$), and gravity. Spot/faculae temperatures are required to be at least 100\,K and up to 1000\,K cooler/warmer than the photosphere, whereas their gravity must be within 1.0 of the photosphere value. The best-fitting three-component results are shown in Figure~\ref{fig: stellar spectrum fit}.

In general, our out-of-transit spectrum analysis is consistent with the in-transit heterogeneity parameters. Visit 2 prefers (at $>$3$\sigma$ confidence) the presence of both spots and faculae over a homogeneous photosphere, or spots/faculae alone. Like the in-transit analysis, we find the second visit to be dominated by spots, with minimal contributions from faculae. For visit 1 however, there is no strong statistical evidence for a heterogeneous photosphere -- homogeneous and heterogeneous models fit the data equally well. This finding is slightly discrepant from the in-transit analysis, which weakly prefers the inclusion of spots, however the retrieved spot parameters from the out-of-transit stellar spectrum are still broadly consistent with the in-transit findings. 

It is also important to note that none of the stellar models are particularly good fits to the data, with $\chi^2_\nu$-values generally $>$150. As shown in Figure~\ref{fig: stellar spectrum fit}, particularly at optical wavelengths the models are poor representations of the observed spectra. With this in mind, although both the out-of- and in-transit analyses rely on stellar models, since the in-transit analysis uses \textit{ratios} of models instead of \textit{absolute fits}, it may be more robust against absolute inaccuracies in stellar models. 

\subsection{\texttt{stctm} TLS-Only Modelling of the NIRISS Spectra}
\label{sec:appendix_TLS_modeling}

Stellar contamination can mimic atmospheric signatures in exoplanet transmission spectra, especially if they orbit M-type host stars \citep{Rackham2018ThePlanets}. Specifically, the presence of unocculted patches of the stellar surface that are colder than the photosphere (spots) can introduce spurious water absorption features and slopes of increasing transit depth towards the bluest wavelengths \citep{iyer_influence_2020,SAG23_effect_TLSE}. The complex superposition of multiple heterogeneity temperatures and covering fractions can partially erase or even reverse such slopes, while still altering the depth of any measured planetary absorption features and impacting the retrieved atmospheric properties. Our detection of a strong slope at short wavelengths in the spectra extracted from both NIRISS/SOSS visits motivates us to explore the impact of stellar contamination on our observations of \myplanet.

The \texttt{stctm}\footnote{\url{https://github.com/cpiaulet/stctm}} \citep{stctm_zenodo_temp} module performs forward modelling of stellar contamination spectra and retrievals of stellar heterogeneity properties from observations. The retrievals are run assuming the spectrum can be fully explained by unocculted faculae or spots on the stellar surface. We model the planetary atmosphere's contribution to the transmission spectrum as a wavelength-independent constant, $D$, and later perform joint retrievals including wavelength-dependent atmospheric signatures as well as stellar heterogeneities (Section \ref{sec:atm_joint_retrievals}). 

The short-wavelength differences between the two NIRISS/SOSS spectra (Figure~\ref{fig:niriss-transmission-spectrum-nameless-differences}) suggest important visit-to-visit variations in the TLS signature (Figure~\ref{fig:spots_niriss}). Meanwhile, the two NIRSpec visits preceded the first NIRISS visit by more than 2 months --- significantly longer than the rotation period of the star. Combined with the $\sim$100\,ppm offset between the NIRISS and NIRSpec spectra, this prompts us to believe that the background stellar spot and faculae distributions were different during the NIRSpec observations than that observed at the epochs of the NIRISS observations. Since the NIRSpec spectra lack the critical short-wavelength coverage necessary to constrain independent TLS parameters from the NIRISS spectra \citep[e.g.,][]{Moran2023HighObservations, MayMacdonald2023doubletrouble}, we choose to ignore the NIRSpec/G395H observations in our spot-only retrievals.

We perform a first set of retrievals assuming either a single heterogeneity population consisting of stellar spots cooler than the photosphere. We applied these retrievals to both NIRISS/SOSS visits individually, to determine whether they can be well-described by stellar contamination alone, and to search for potential visit-to-visit variations in the heterogeneity makeup of the stellar surface.

\begin{figure*}
    \centering
    \includegraphics[width=0.8\textwidth]{ 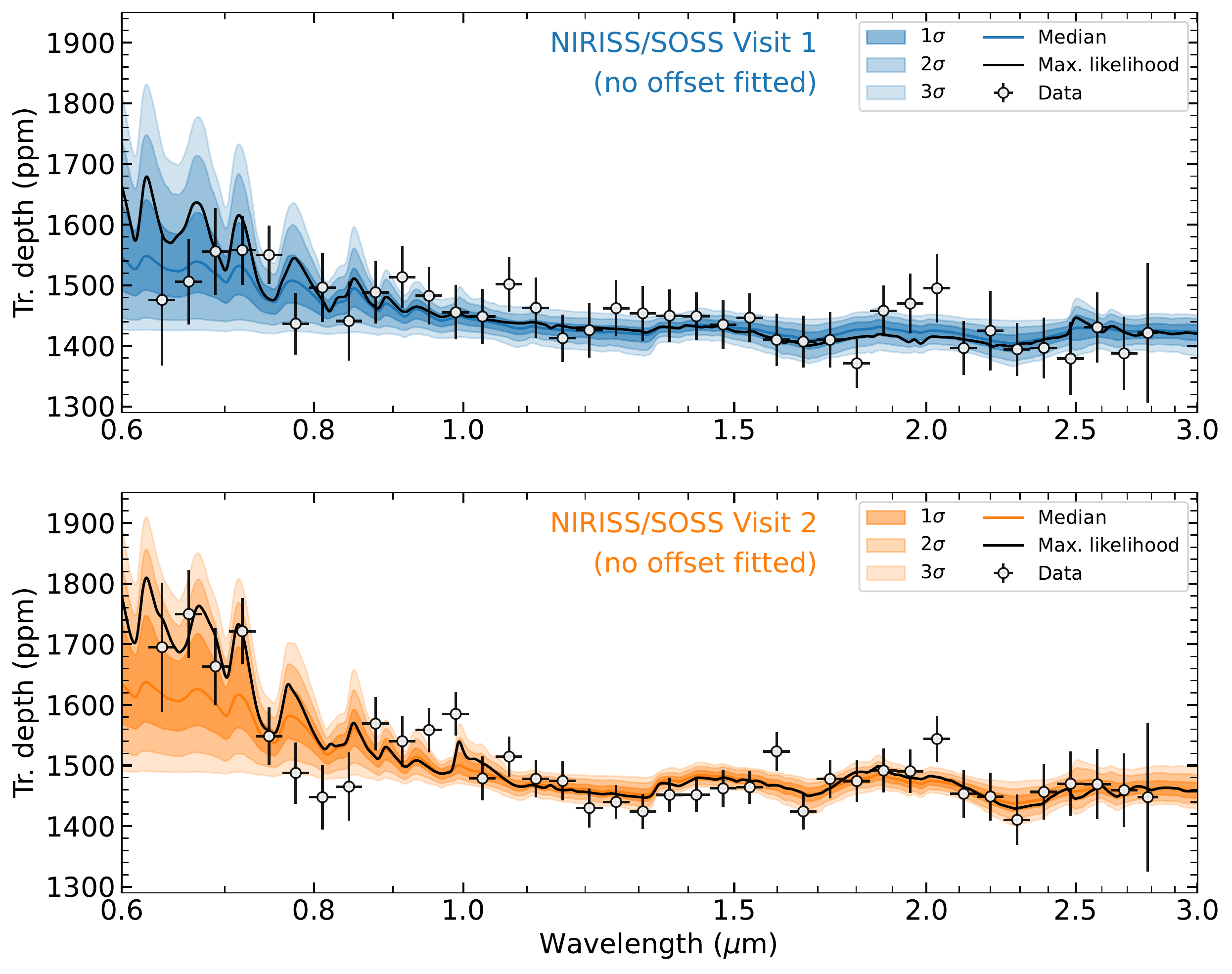}
    \caption{Results from the TLS-only fits to the transmission spectrum of each NIRISS/SOSS visit of GJ 3090 b from the \texttt{stctm} retrievals. We show the best-fitting unocculted heterogeneity model (black), as well as the median and 1, 2, and 3-$\sigma$ ranges obtained from sample posterior spectra (colour shading). The top panel showcases the results for the fit to the visit 1 spectrum, and the bottom panel is the same result for the visit 2 spectrum. We show the binned (four points together) points for each visit (black points), with no offset applied between order 1 and order 2. The NIRISS/SOSS spectrum of GJ 3090 b is well matched by stellar contamination alone, in line with the lack of information content on the atmosphere we obtain from joint TLS-atmosphere retrievals over this wavelength range.}
    \label{fig:TLS_only_fit}
\end{figure*}

\begin{figure*}
    \centering
    \includegraphics[width=0.8\textwidth]{ 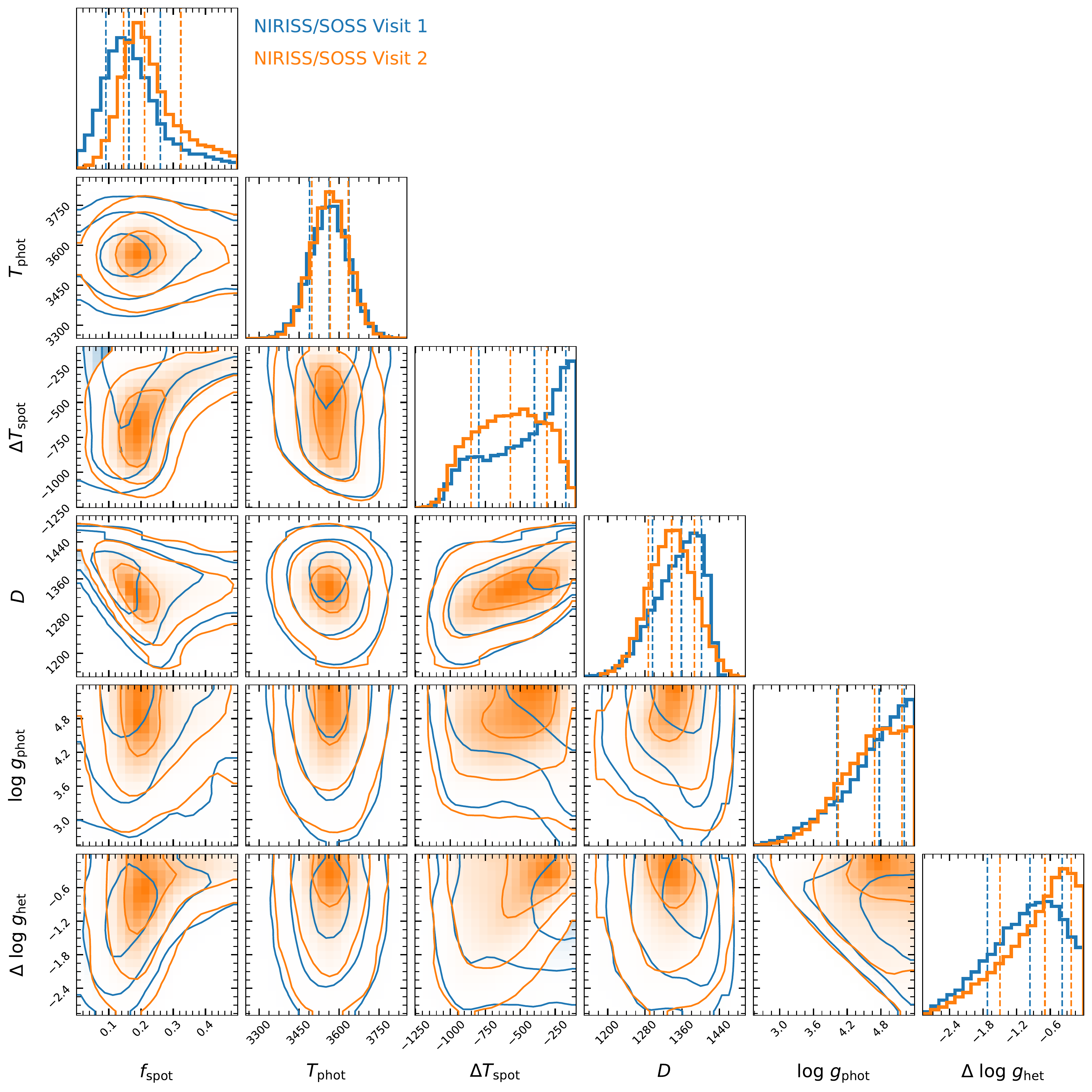}
    \caption{Joint and marginalized posterior distributions from the TLS-only fits to the transmission spectrum of each NIRISS/SOSS visit of GJ 3090 b performed using \texttt{stctm}. The contours correspond to 1, 2, and 3$\sigma$ limits, and the median and 1$\sigma$ interval is highlighted on each marginalized posterior distribution. The results for visit 1 (2) are shown in blue (orange).}
    \label{fig:TLS_only_corner}
\end{figure*}

We closely follow the steps outlined in \citet{Piaulet_Ghorayeb_2024} and \citet{radica_promise_2025} for the application of \texttt{stctm} to the TLS-only retrievals. The fitted parameters are the temperature differences between the heterogeneities and the photosphere: $\Delta T_\mathrm{spot}$, the photosphere temperature itself ($T_\mathrm{phot}$), and the covering fraction of spots ($f_\mathrm{spot}$). We calculate the spectra of each component by interpolating in the $T_\mathrm{eff}$ and $\log g$ dimensions over the PHOENIX stellar models grid \citep{husser_new_2013}, using the \texttt{MSG} module \citep{townsend_msg_2023}. Our priors are uniform between 0 and 50\% for the heterogeneity covering fraction, uniform from $-100$ to $-1000$\,K for the heterogeneity temperature contrast with the photosphere, and Gaussian for the photosphere temperature, with the mean and standard deviation informed by \citet{Almenara2022GJCharacterisation}. We also use the newly-introduced (Piaulet-Ghorayeb et al., in prep) fitting of the stellar photosphere $\log g$ ($\log g \in [2.5,5.5]$) and of a potential difference between the surface gravities representative of the spectrum of the photosphere and stellar spot component (enforcing a $\log g$ for the spots equal to or lower than that of the photosphere, see e.g., \citealp{fournier-tondreau_near-infrared_2024}). We sample the parameter space using the \texttt{emcee} implementation of Markov-Chain Monte Carlo \citep{foreman-mackey_emcee_2013}, with 20 times as many walkers as fitted parameters. Each chain is run for 5,000 steps, visually checked for convergence, and the first 60\% of each chain are discarded as burn-in to obtain the distribution of posterior samples (results in Figures \ref{fig:TLS_only_fit}, \ref{fig:TLS_only_corner}, and \ref{fig:spots_niriss}). When the photosphere $\log g$ is fitted, we obtain consistent results in terms of the stellar heterogeneity properties for each visit, and the preferred value for the surface gravity is larger than the literature value. 

\section{Atmosphere retrieval results}

\begin{figure*}
    \centering
    \includegraphics[width=0.8\textwidth]{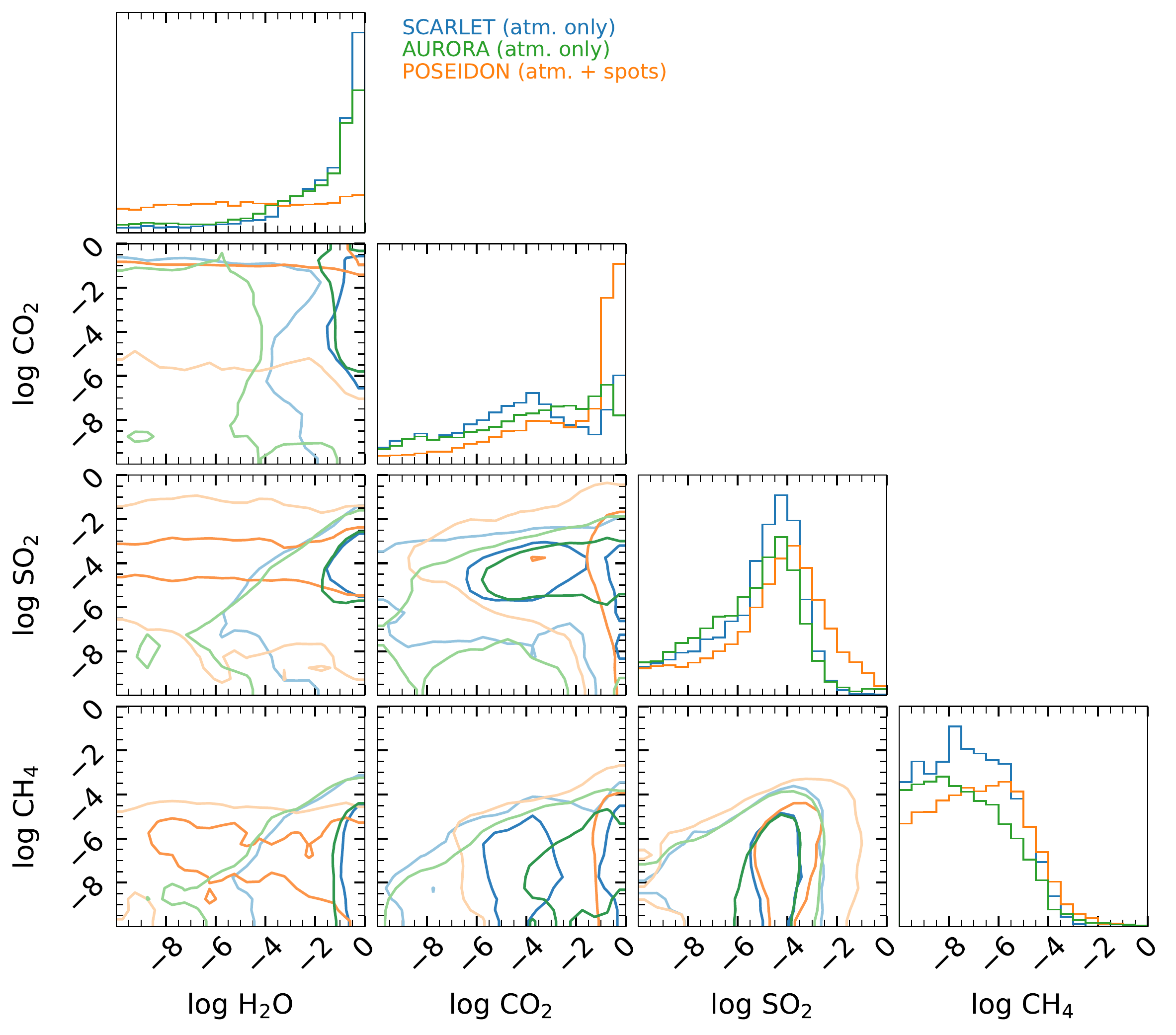}
    \caption{{  Joint and marginalized posterior distributions for the main atmospheric absorbers that NIRSpec/G395H is sensitive to, from the SCARLET (blue), POSEIDON (orange), and AURORA (green) retrievals performed on the visit 1+2 NIRSpec/G395H spectrum of GJ 3090 b. While POSEIDON marginalized over the potential contributions from unocculted spots, the SCARLET and AURORA results shown do not account for a TLS contribution, leading to overall higher water abundances. The contours correspond to 1 and 2$\sigma$ confidence (dark and light color shadings).}}
    \label{fig:corner_nirspec_molecules}
\end{figure*}

\begin{figure*}
    \centering
    \includegraphics[width=0.5\textwidth]{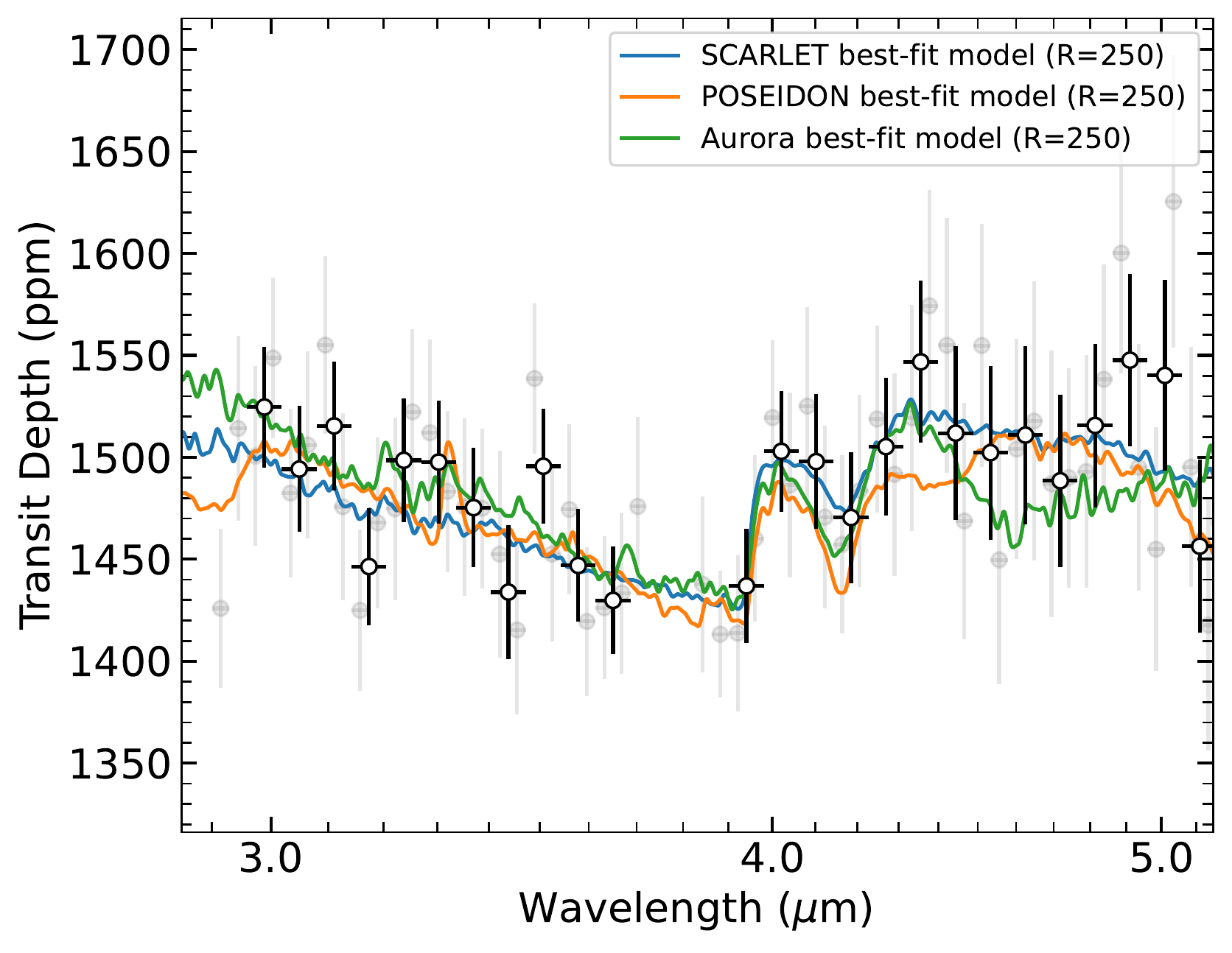}
    \caption{{ Best-fit models from the \texttt{SCARLET} (blue), \texttt{POSEIDON} (orange), and \texttt{Aurora} (green) retrievals performed on the NIRSpec/G395H transmission spectrum of \myplanet (gray points at R=100, binned in black), corresponding to the posterior plot shown in Fig.\,\ref{fig:corner_nirspec_molecules}. The three models were smoothed to a resolving power of 250. }}
    \label{fig:best_fit_models}
\end{figure*}

{ We report the constraints on atmospheric parameters from the atmosphere + TLS retrievals performed on the NIRISS/SOSS spectra in Table \ref{tab:niriss_allpara}, which are mostly unconstrained since the spectrum can be fully explained by stellar contamination alone. We also present a comparison between the posterior distributions on major absorbers obtained with the three retrieval frameworks from the retrievals performed on the NIRSpec/G395H transmission spectrum in Figure \ref{fig:corner_nirspec_molecules}. The corresponding best-fit models for each retrieval code are shown in Figure \ref{fig:best_fit_models}.}

\begin{table}
    \caption{\label{tab:niriss_allpara} {  Constraints on the atmospheric properties for the retrievals performed with \texttt{SCARLET} (model with uniform clouds and no hazes), \texttt{Aurora} (model including hazes and patchy clouds), and \texttt{POSEIDON} (model with hazes and uniform clouds) on each NIRISS/SOSS visit. Since the spectrum is best explained by stellar contamination alone, several atmospheric parameters are unconstrained with a posterior consistent with the prior: in these cases, we simply report `N/A' for the constraints, while `--' means that the parameter was not fitted in that retrieval configuration. Lower or upper limits are reported at 2$\sigma$.}}
    \centering
    \begin{tabular}{lcccccc}
    \hline
    \hline
    Parameter & \multicolumn{2}{c}{SCARLET} & \multicolumn{2}{c}{\texttt{Aurora}} & \multicolumn{2}{c}{\texttt{POSEIDON}} \\
    & Visit 1 & Visit 2 & Visit 1 & Visit 2 & Visit 1 & Visit 2\\
    \hline
    \multicolumn{1}{l}{\textbf{Atmosphere composition}}\\
    $\log_{10}$H$_2$O & $<-0.67$ & $<-1.05$ & N/A & $<-2.97$ & $<-1.26$ & $<-2.08$\\ 
    $\log_{10}$ CO$_2$ & N/A & N/A & N/A & $<-1.33$ & $<-0.93$ & N/A\\ 
    $\log_{10}$ CH$_4$ & $<-1.14$ & $<-1.14$ & $<-1.62$ & $<-4.25$ & $<-2.06$ & $<-2.76$\\ 
    $\log_{10}$ SO$_2$ & N/A & N/A & N/A & $<-1.07$ & $<-1.04$ & N/A\\ 
    $\log_{10}$ CO & N/A & N/A & -- & -- & $<-0.96$ & N/A\\ 
        $\log_{10}$ H$_2$S & -- & -- & N/A & N/A & $<-1.23$ & N/A\\ 
    $\log_{10}$ HCN & -- & -- & -- & -- & $<-1.02$ & $<-1.14$\\ 
    $\log_{10}$ NH$_3$ & -- & -- & -- & -- & $<-1.7$ & $<-2.64$\\ 

    \multicolumn{1}{l}{\textbf{Temperature}}\\

    T$_\mathrm{atm}$ [K] & $<895.45$ & $<892.93$ & $<332.7$ & $<351.53$ & $<489.58$ & $<542.93$\\ 
    \multicolumn{4}{l}{\textbf{Clouds and hazes}}\\
    $\log_{10} p_\mathrm{cloud}$ [bar] & $<-2.54$ & $<0.66$ & $-3.08^{+2.91}_{-3.14}$ & $-2.03^{+2.46}_{-3.62}$ & $-2.7^{+2.45}_{-2.17}$ & $-1.92^{+2.12}_{-2.35}$\\ 
    $f_\mathrm{cloud}$ [\%] & -- & -- & $66^{+20}_{-29}$ & $57^{+24}_{-27}$ & -- & --\\ 
    $\log_{10} a$ & -- & -- & $>-2.61$ & $>-2.21$ & $4.35^{+1.98}_{-3.8}$ & $1.67^{+3.16}_{-3.11}$\\ 
    $\gamma$ & -- & -- & $-7.46^{+4.15}_{-6.05}$ & $<-0.95$ & $-5.55^{+3.01}_{-6.24}$ & $-9.96^{+5.86}_{-5.78}$\\ 
    \multicolumn{1}{l}{\textbf{Reference pressure or radius}}\\
    $p_\mathrm{ref}$ [bar] & --$^{(a)}$ & --$^{(a)}$ & $-3.8^{+3.14}_{-2.84}$ & $-3.07^{+2.7}_{-3.09}$ & -- & --\\ 
    $R_\mathrm{ref}$ [R$_\oplus$] & -- & -- & $2.06^{+0.06}_{-0.05}$ & $2.01^{+0.07}_{-0.05}$ & $1.95^{+0.06}_{-0.1}$ & $2.0^{+0.03}_{-0.08}$\\ 

    \hline
    \hline
    \multicolumn{7}{l}{\footnotesize{$^{(a)}$  In the SCARLET retrievals, $p_\mathrm{ref}$ is fitted at each step to find the best match to the observed spectrum but is not recorded.}}\\
    \end{tabular}
    
\end{table}

\section{Additional Figures}
\label{sec:appendix_cut_integrations}

\begin{figure}
    \centering
    \includegraphics[width=0.8\linewidth]{ 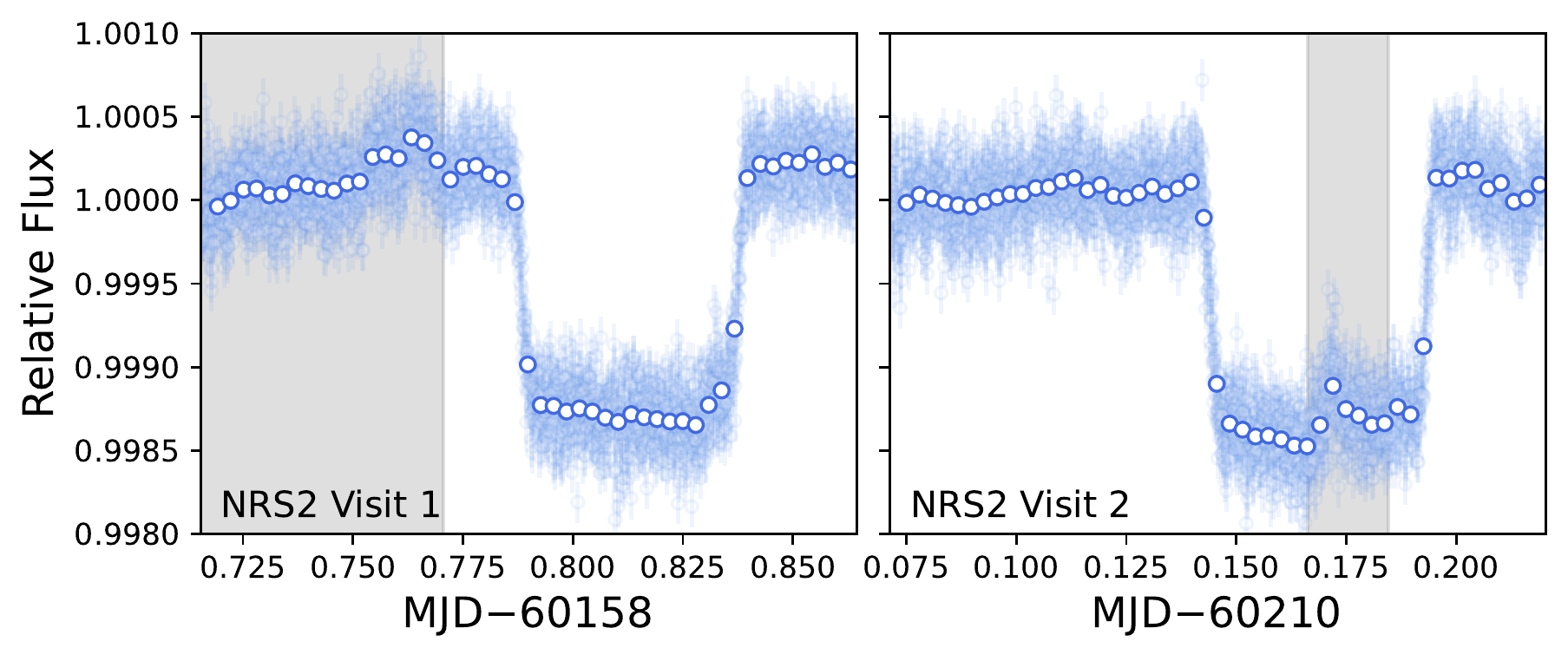}
    \caption{Raw NRS2 white light curves for each NIRSpec/G395H without any integrations cut as a result of systematics (visit 1) or flares (visit 2). The grey shaded regions denote the integrations that are cut in the light curve analysis presented in Section~\ref{sec:joint-fit}.}
    \label{fig:raw nirspec}
\end{figure}

Figure~\ref{fig:raw nirspec} shows the NIRSpec/G395H white light curves for NRS2 without any integrations cut as a result of systematics or flares.

\begin{figure}
    \centering
    \begin{minipage}{0.48\textwidth}
        \includegraphics[width=\textwidth]{ 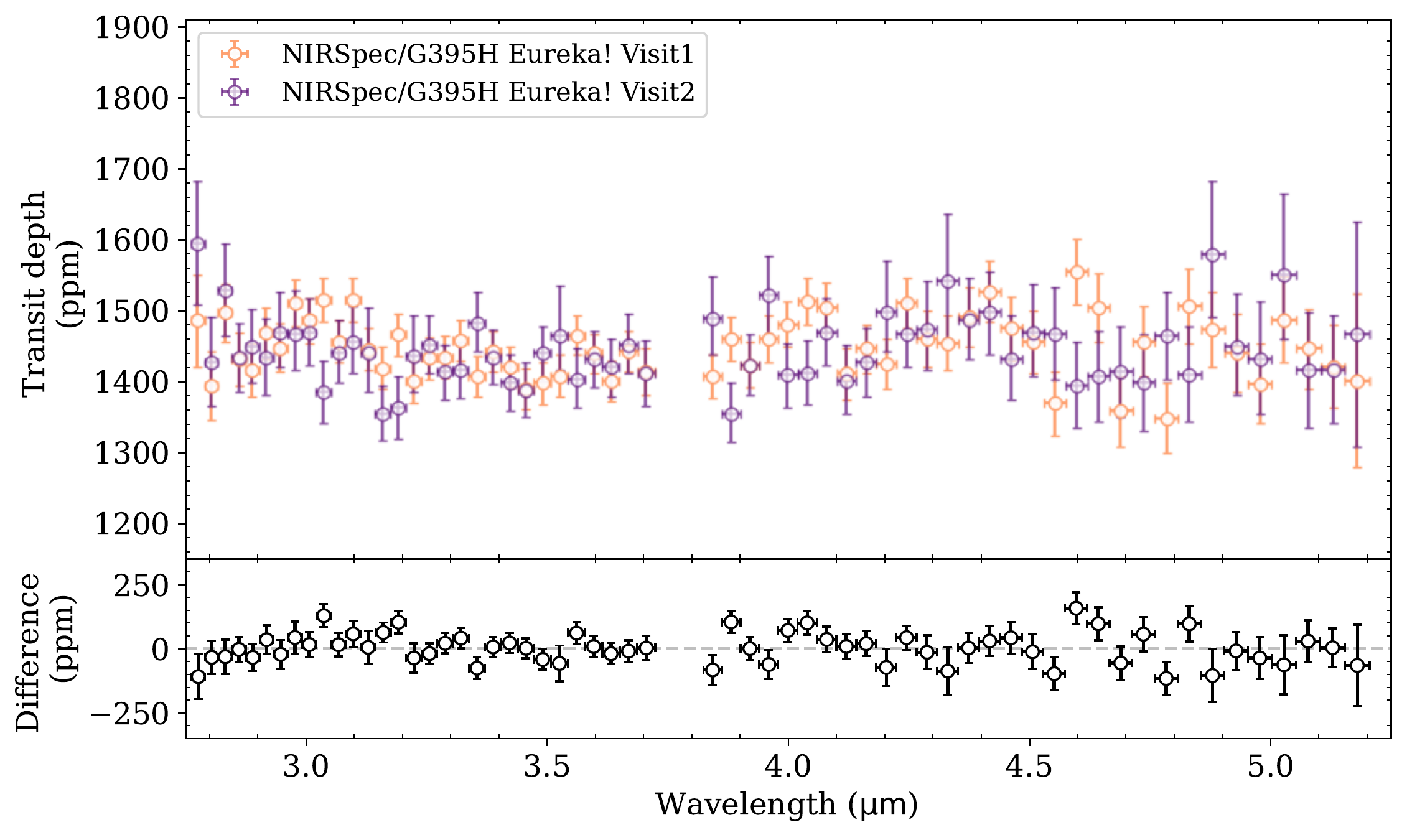}
    \end{minipage}
\hfill
    \begin{minipage}{0.48\textwidth}\includegraphics[width=\linewidth]{ 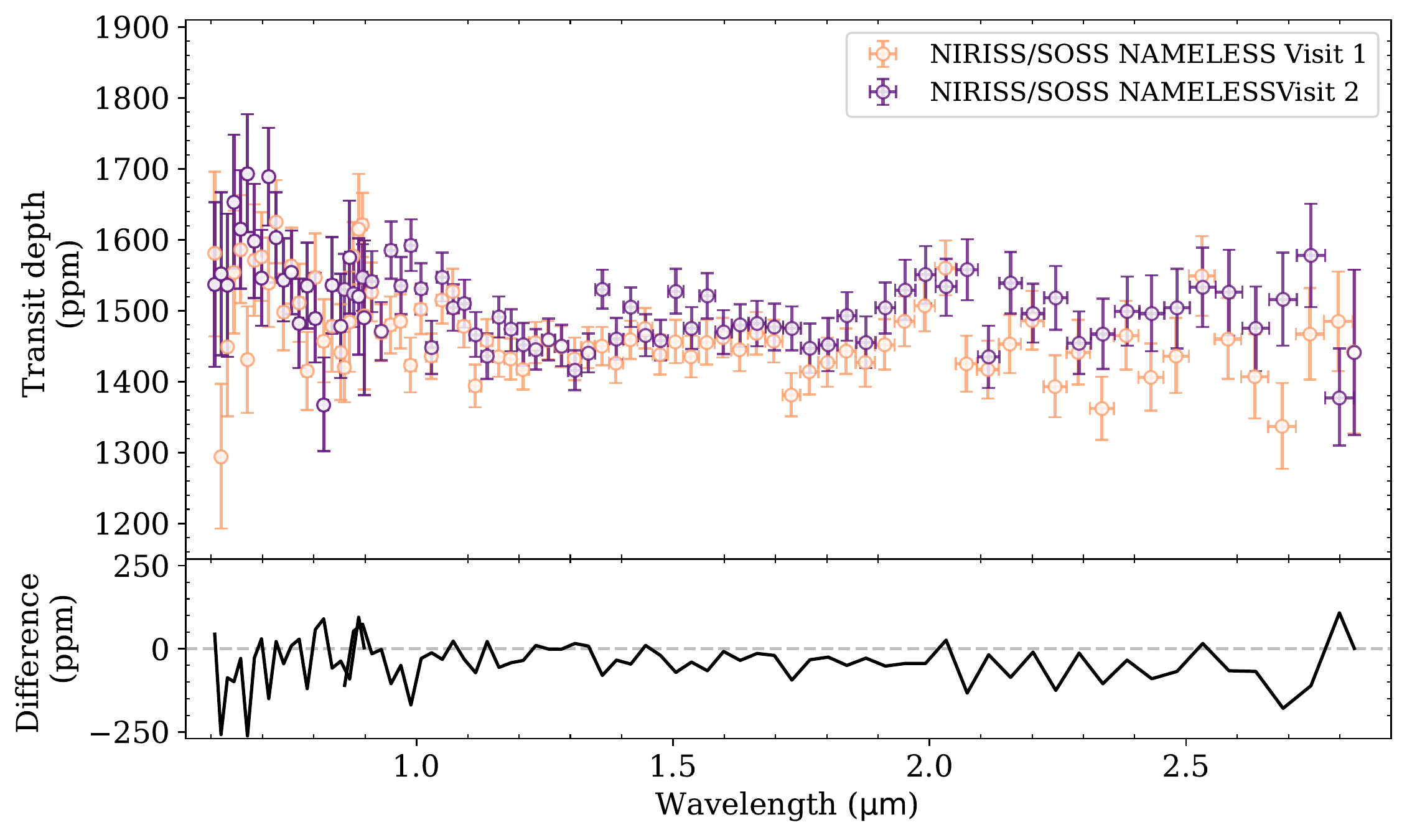}
    \end{minipage}
    
    \begin{minipage}{0.48\textwidth}
        \includegraphics[width=\textwidth]{ 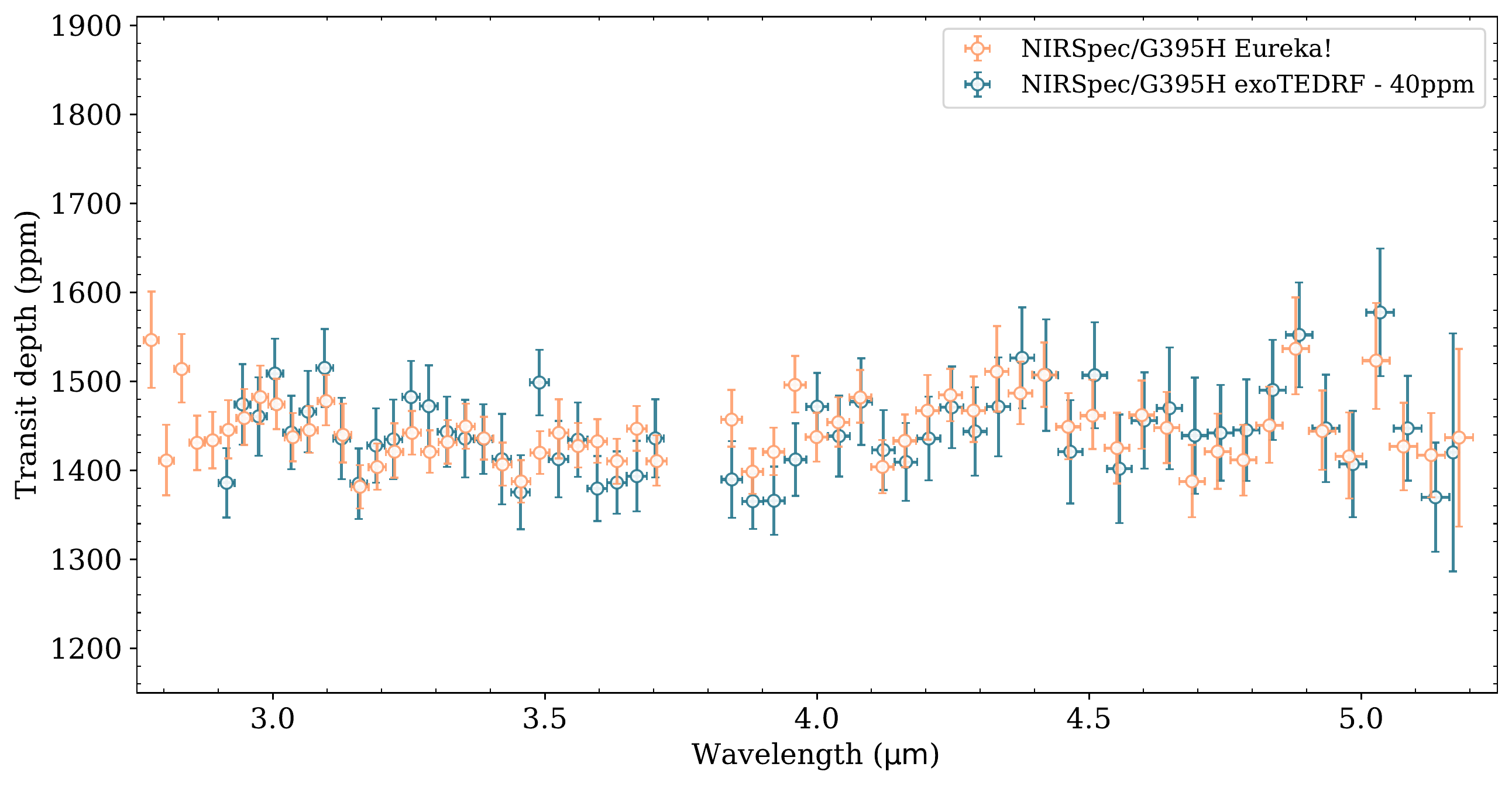}
    \end{minipage}
\hfill
    \begin{minipage}{0.48\textwidth}\includegraphics[width=\linewidth]{ 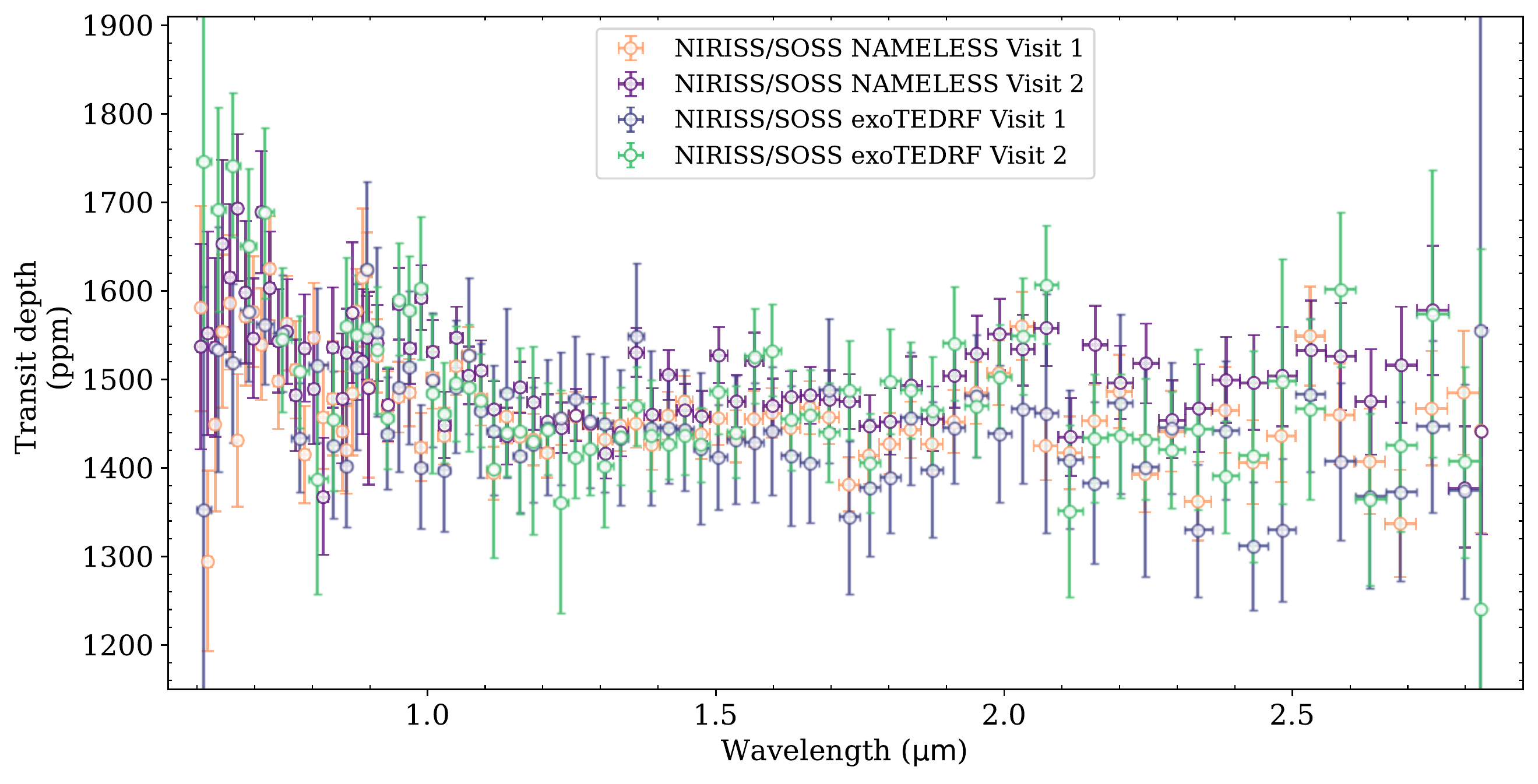}
    \end{minipage}
    \caption{Comparison of transmission spectra of \myplanet from different reduction pipelines. \textit{Top left:} \texttt{Eureka!} transmission spectra from both  NIRSpec/G395H visits. Visit 1 is displayed in the darker colour and Visit 2 in the lighter colour and the differences between them are shown in the bottom panel in black. 
    \textit{Bottom left:} Combined (visit 1 \& visit 2) NIRSpec transmission spectra from \texttt{Eureka!} (orange) and \texttt{exoTEDRF} (turquoise). The latter reduction is vertically offset by their average difference of 40\,ppm to allow for direct comparison. 
    \textit{Top right:} \texttt{NAMELESS} transmission spectra of \myplanet from both NIRISS/SOSS visits. Visit 1 is displayed in the darker purple colour and Visit 2 in the lighter orange. The differences between the two visits are shown in the bottom panel in black, demonstrating an $\sim$30\,ppm offset between the two visits. 
    \textit{Bottom right:} Comparison of the \texttt{NAMELESS} (orange, purple) and \texttt{exoTEDRF} (turquoise, dark blue) transmission spectra for both NIRISS/SOSS visits.}
    \label{fig:other-reductions-differences}
\end{figure}

In Figure~\ref{fig:other-reductions-differences} the second independent reductions for both NIRSpec and NIRISS are displayed. The left panels show the spectrum of the NIRSpec data retrieved using \texttt{Eureka!}, with the top panel comparing the two visits with each other and the bottom panel displaying the comparison between our nominal \texttt{exoTEDRF} reduction. 
Similarly, for NIRISS/SOSS we show the two visits of our second independent analysis using the \texttt{NAMELESS} pipeline for the data reduction and the \texttt{Tiberius} pipeline for light curve fitting in the right panels of Fig.\,\ref{fig:other-reductions-differences}. In the top panel we show both visits of the \texttt{NAMELESS} reduction, while the bottom panel compares both visits and both reductions directly, demonstrating that the GPs used in the light curve fitting in the \texttt{exoTEDRF} reduction significantly increase the uncertainties for the transmission spectrum. Nevertheless, both reductions show differences between the two visits that are consistent and showcase that the NIRISS/SOSS observations cannot be combined if the stellar activity level has changed in the time passed between the two visits.  

\begin{figure*}
    \centering
    \includegraphics[width=0.8\linewidth]{ 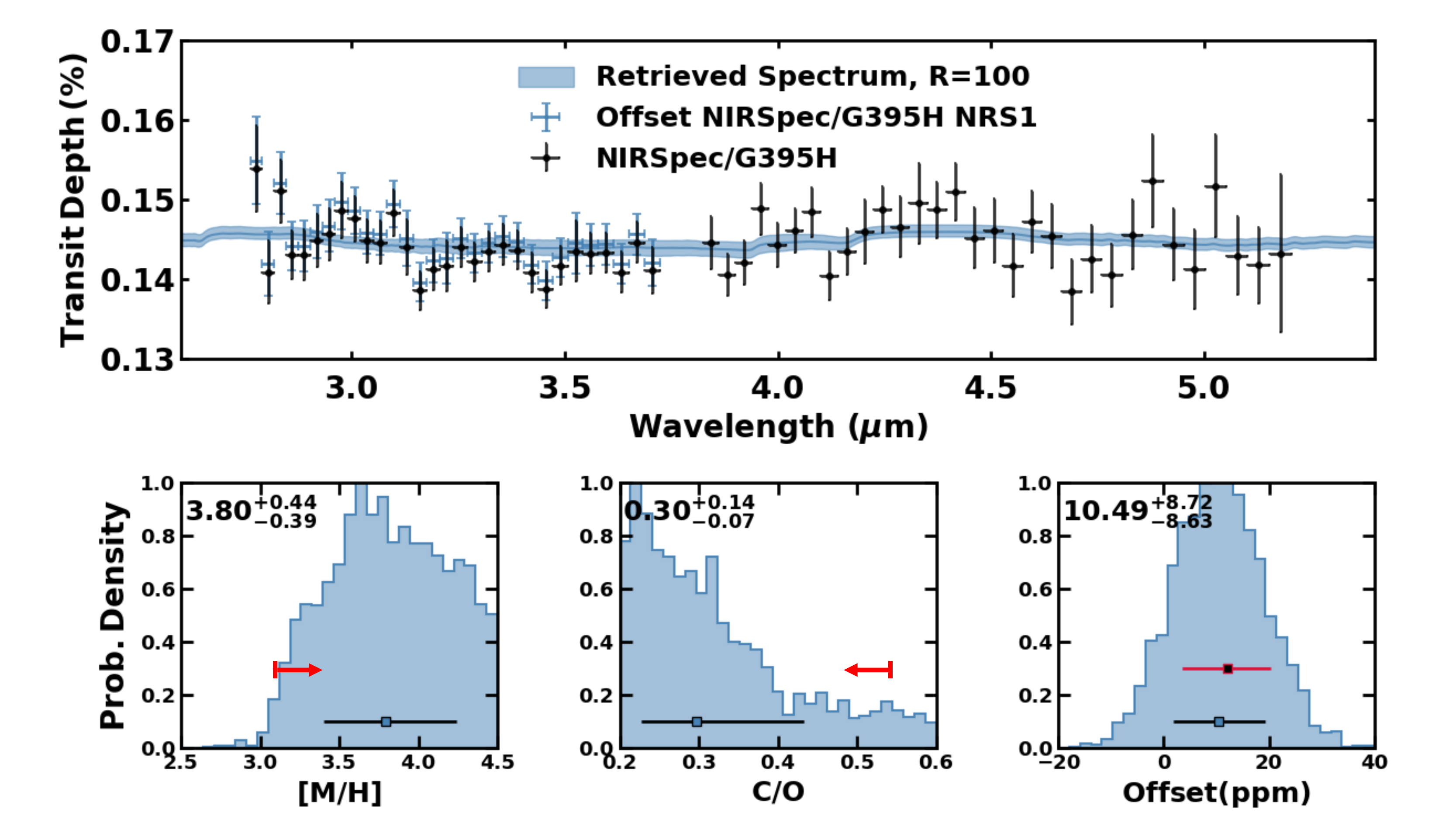} 
    \caption{\textit{Top:} The retrieved spectrum from the grid-based 1D radiative-convective-thermochemical equilibrium retrieval with \texttt{ScCHIMERA}. Shaded is the 1$\sigma$ confidence region about the median model, plotted as a solid line. The blue points are the NRS1 points with the median retrieved instrument offset of 10.49 ppm. \textit{Bottom panels:} The posterior probability density distributions for [M/H], where [] denotes log$_{10}$ relative to solar, C/O, and the instrument offset on NIRSpec/G395H NRS1. Points with error bars denote the median and 1$\sigma$ confidence region on each parameter. The red points show parameter constraints from the disequilibrium retrieval. Arrows mark the 2$\sigma$ lower limit for M/H, and the 1$\sigma$ preference for C/O.}
    \label{fig:gridretrieval}
\end{figure*}

Lastly, we show in Figure~\ref{fig:gridretrieval} the range of grid model spectra and parameters that match the NIRSpec/G395H spectrum of \myplanet.

\begin{thebibliography}{}
\expandafter\ifx\csname natexlab\endcsname\relax\def\natexlab#1{#1}\fi
\providecommand{\url}[1]{\href{#1}{#1}}
\providecommand{\dodoi}[1]{doi:~\href{http://doi.org/#1}{\nolinkurl{#1}}}
\providecommand{\doeprint}[1]{\href{http://ascl.net/#1}{\nolinkurl{http://ascl.net/#1}}}
\providecommand{\doarXiv}[1]{\href{https://arxiv.org/abs/#1}{\nolinkurl{https://arxiv.org/abs/#1}}}

\bibitem[{{Acu{\~n}a} {et~al.}(2021){Acu{\~n}a}, {Deleuil}, {Mousis}, {Marcq}, {Levesque}, \& {Aguichine}}]{acuna21}
{Acu{\~n}a}, L., {Deleuil}, M., {Mousis}, O., {et~al.} 2021, \aap, 647, A53, \dodoi{10.1051/0004-6361/202039885}

\bibitem[{{Acu{\~n}a} {et~al.}(2024){Acu{\~n}a}, {Kreidberg}, {Zhai}, \& {Molli{\`e}re}}]{acuna24}
{Acu{\~n}a}, L., {Kreidberg}, L., {Zhai}, M., \& {Molli{\`e}re}, P. 2024, \aap, 688, A60, \dodoi{10.1051/0004-6361/202450559}

\bibitem[{Acuna {et~al.}(2023)Acuna, Deleuil, \& Mousis}]{acuna_interior-atmosphere_2023}
Acuna, L., Deleuil, M., \& Mousis, O. 2023, Interior-atmosphere modelling to assess the observability of rocky planets with {JWST},  arXiv.
\newblock \url{http://arxiv.org/abs/2305.01250}

\bibitem[{Aguichine {et~al.}(2021)Aguichine, Mousis, Deleuil, \& Marcq}]{aguichine_mass-radius_2021}
Aguichine, A., Mousis, O., Deleuil, M., \& Marcq, E. 2021, The Astrophysical Journal, 914, 84, \dodoi{10.3847/1538-4357/abfa99}

\bibitem[{{Aguichine} {et~al.}(2020){Aguichine}, {Mousis}, {Devouard}, \& {Ronnet}}]{aguichine_craddles_20}
{Aguichine}, A., {Mousis}, O., {Devouard}, B., \& {Ronnet}, T. 2020, \apj, 901, 97, \dodoi{10.3847/1538-4357/abaf47}

\bibitem[{Ahrer {et~al.}(2022)Ahrer, Wheatley, Kirk, Gandhi, King, \& Louden}]{Ahrer2022LRG-BEASTS:NTT/EFOSC2}
Ahrer, E., Wheatley, P.~J., Kirk, J., {et~al.} 2022, Monthly Notices of the Royal Astronomical Society, 510, 4857, \dodoi{10.1093/mnras/stab3805}

\bibitem[{Ahrer {et~al.}(2023)Ahrer, Stevenson, Mansfield, Moran, Brande, Morello, Murray, Nikolov, Petit dit de~la Roche, Schlawin, Wheatley, Zieba, Batalha, Damiano, Goyal, Lendl, Lothringer, Mukherjee, Ohno, Batalha, Battley, Bean, Beatty, Benneke, Berta-Thompson, Carter, Cubillos, Daylan, Espinoza, Gao, Gibson, Gill, Harrington, Hu, Kreidberg, Lewis, Line, L{\'{o}}pez-Morales, Parmentier, Powell, Sing, Tsai, Wakeford, Welbanks, Alam, Alderson, Allen, Anderson, Barstow, Bayliss, Bell, Blecic, Bryant, Burleigh, Carone, Casewell, Changeat, Chubb, Crossfield, Crouzet, Decin, D{\'{e}}sert, Feinstein, Flagg, Fortney, Gizis, Heng, Iro, Kempton, Kendrew, Kirk, Knutson, Komacek, Lagage, Leconte, Lustig-Yaeger, MacDonald, Mancini, May, Mayne, Miguel, Mikal-Evans, Molaverdikhani, Palle, Piaulet, Rackham, Redfield, Rogers, Roy, Rustamkulov, Shkolnik, Sotzen, Taylor, Tremblin, Tucker, Turner, de~Val-Borro, Venot, \& Zhang}]{Ahrer2023EarlyNIRCam}
Ahrer, E.~M., Stevenson, K.~B., Mansfield, M., {et~al.} 2023, Nature, 614, 653, \dodoi{10.1038/s41586-022-05590-4}

\bibitem[{Albert {et~al.}(2023)Albert, Lafrenière, René, Artigau, Volk, Goudfrooij, Martel, Radica, Rowe, Espinoza, Roy, Filippazzo, Darveau-Bernier, Talens, Sivaramakrishnan, Willott, Fullerton, LaMassa, Hutchings, Rowlands, Vila, Zhou, Aldridge, Maszkiewicz, Beaulieu, Cook, Piaulet, Roy, Lamontagne, Morel, Frost, Salhi, Coulombe, Benneke, MacDonald, Johnstone, Turner, Fournier-Tondreau, Allart, \& Kaltenegger}]{albert_near_2023}
Albert, L., Lafrenière, D., René, D., {et~al.} 2023, Publications of the Astronomical Society of the Pacific, 135, 075001, \dodoi{10.1088/1538-3873/acd7a3}

\bibitem[{Alderson {et~al.}(2023)Alderson, Wakeford, Alam, Batalha, Lothringer, Redai, Barat, Brande, Damiano, Daylan, Espinoza, Flagg, Goyal, Grant, Hu, Inglis, Lee, Mikal-Evans, Ramos-Rosado, Roy, Wallack, Batalha, Bean, Benneke, Berta-Thompson, Carter, Changeat, Col{\'{o}}n, Crossfield, D{\'{e}}sert, Foreman-Mackey, Gibson, Kreidberg, Line, L{\'{o}}pez-Morales, Molaverdikhani, Moran, Morello, Moses, Mukherjee, Schlawin, Sing, Stevenson, Taylor, Aggarwal, Ahrer, Allen, Barstow, Bell, Blecic, Casewell, Chubb, Crouzet, Cubillos, Decin, Feinstein, Fortney, Harrington, Heng, Iro, Kempton, Kirk, Knutson, Krick, Leconte, Lendl, MacDonald, Mancini, Mansfield, May, Mayne, Miguel, Nikolov, Ohno, Palle, Parmentier, Petit dit de~la Roche, Piaulet, Powell, Rackham, Redfield, Rogers, Rustamkulov, Tan, Tremblin, Tsai, Turner, de~Val-Borro, Venot, Welbanks, Wheatley, \& Zhang}]{Alderson2023EarlyG395H}
Alderson, L., Wakeford, H.~R., Alam, M.~K., {et~al.} 2023, Nature, 614, 664, \dodoi{10.1038/S41586-022-05591-3}

\bibitem[{{Alderson} {et~al.}(2024){Alderson}, {Batalha}, {Wakeford}, {Wallack}, {Aguichine}, {Teske}, {Adams Redai}, {Alam}, {Batalha}, {Gao}, {Kirk}, {L{\'o}pez-Morales}, {Moran}, {Scarsdale}, {Wogan}, \& {Wolfgang}}]{Alderson2024}
{Alderson}, L., {Batalha}, N.~E., {Wakeford}, H.~R., {et~al.} 2024, \aj, 167, 216, \dodoi{10.3847/1538-3881/ad32c9}

\bibitem[{Alibert \& Benz(2017)}]{alibert_formation_2017}
Alibert, Y., \& Benz, W. 2017, Astronomy \& Astrophysics, 598, L5, \dodoi{10.1051/0004-6361/201629671}

\bibitem[{Allart {et~al.}(2019)Allart, Bourrier, Lovis, Ehrenreich, Aceituno, Guijarro, Pepe, Sing, Spake, \& Wyttenbach}]{allart_high-resolution_2019}
Allart, R., Bourrier, V., Lovis, C., {et~al.} 2019, arXiv:1901.08073 [astro-ph].
\newblock \url{da}

\bibitem[{Almenara {et~al.}(2022)Almenara, Bonfils, Otegi, Attia, Turbet, Astudillo-Defru, Collins, Polanski, Bourrier, Hellier, Ziegler, Bouchy, Briceno, Charbonneau, Cointepas, Collins, Crossfield, Delfosse, Diaz, Dorn, Doty, Forveille, Gaisn{\'{e}}, Gan, Helled, Hesse, Jenkins, Jensen, Latham, Law, Mann, Mao, McLean, Murgas, Myers, Seager, Shporer, Tan, Twicken, \& Winn}]{Almenara2022GJCharacterisation}
Almenara, J.~M., Bonfils, X., Otegi, J.~F., {et~al.} 2022, Astronomy and Astrophysics, 665, A91, \dodoi{10.1051/0004-6361/202243975}

\bibitem[{{Asplund} {et~al.}(2009){Asplund}, {Grevesse}, {Sauval}, \& {Scott}}]{asplund_2009}
{Asplund}, M., {Grevesse}, N., {Sauval}, A.~J., \& {Scott}, P. 2009, \araa, 47, 481, \dodoi{10.1146/annurev.astro.46.060407.145222}

\bibitem[{{Astropy Collaboration} {et~al.}(2013){Astropy Collaboration}, {Robitaille}, {Tollerud}, {Greenfield}, {Droettboom}, {Bray}, {Aldcroft}, {Davis}, {Ginsburg}, {Price-Whelan}, {Kerzendorf}, {Conley}, {Crighton}, {Barbary}, {Muna}, {Ferguson}, {Grollier}, {Parikh}, {Nair}, {Unther}, {Deil}, {Woillez}, {Conseil}, {Kramer}, {Turner}, {Singer}, {Fox}, {Weaver}, {Zabalza}, {Edwards}, {Azalee Bostroem}, {Burke}, {Casey}, {Crawford}, {Dencheva}, {Ely}, {Jenness}, {Labrie}, {Lim}, {Pierfederici}, {Pontzen}, {Ptak}, {Refsdal}, {Servillat}, \& {Streicher}}]{astropy:2013}
{Astropy Collaboration}, {Robitaille}, T.~P., {Tollerud}, E.~J., {et~al.} 2013, \aap, 558, A33, \dodoi{10.1051/0004-6361/201322068}

\bibitem[{{Astropy Collaboration} {et~al.}(2018){Astropy Collaboration}, {Price-Whelan}, {Sip{\H{o}}cz}, {G{\"u}nther}, {Lim}, {Crawford}, {Conseil}, {Shupe}, {Craig}, {Dencheva}, {Ginsburg}, {Vand erPlas}, {Bradley}, {P{\'e}rez-Su{\'a}rez}, {de Val-Borro}, {Aldcroft}, {Cruz}, {Robitaille}, {Tollerud}, {Ardelean}, {Babej}, {Bach}, {Bachetti}, {Bakanov}, {Bamford}, {Barentsen}, {Barmby}, {Baumbach}, {Berry}, {Biscani}, {Boquien}, {Bostroem}, {Bouma}, {Brammer}, {Bray}, {Breytenbach}, {Buddelmeijer}, {Burke}, {Calderone}, {Cano Rodr{\'\i}guez}, {Cara}, {Cardoso}, {Cheedella}, {Copin}, {Corrales}, {Crichton}, {D'Avella}, {Deil}, {Depagne}, {Dietrich}, {Donath}, {Droettboom}, {Earl}, {Erben}, {Fabbro}, {Ferreira}, {Finethy}, {Fox}, {Garrison}, {Gibbons}, {Goldstein}, {Gommers}, {Greco}, {Greenfield}, {Groener}, {Grollier}, {Hagen}, {Hirst}, {Homeier}, {Horton}, {Hosseinzadeh}, {Hu}, {Hunkeler}, {Ivezi{\'c}}, {Jain}, {Jenness}, {Kanarek}, {Kendrew}, {Kern}, {Kerzendorf}, {Khvalko}, {King}, {Kirkby}, {Kulkarni},
  {Kumar}, {Lee}, {Lenz}, {Littlefair}, {Ma}, {Macleod}, {Mastropietro}, {McCully}, {Montagnac}, {Morris}, {Mueller}, {Mumford}, {Muna}, {Murphy}, {Nelson}, {Nguyen}, {Ninan}, {N{\"o}the}, {Ogaz}, {Oh}, {Parejko}, {Parley}, {Pascual}, {Patil}, {Patil}, {Plunkett}, {Prochaska}, {Rastogi}, {Reddy Janga}, {Sabater}, {Sakurikar}, {Seifert}, {Sherbert}, {Sherwood-Taylor}, {Shih}, {Sick}, {Silbiger}, {Singanamalla}, {Singer}, {Sladen}, {Sooley}, {Sornarajah}, {Streicher}, {Teuben}, {Thomas}, {Tremblay}, {Turner}, {Terr{\'o}n}, {van Kerkwijk}, {de la Vega}, {Watkins}, {Weaver}, {Whitmore}, {Woillez}, {Zabalza}, \& {Astropy Contributors}}]{astropy:2018}
{Astropy Collaboration}, {Price-Whelan}, A.~M., {Sip{\H{o}}cz}, B.~M., {et~al.} 2018, \aj, 156, 123, \dodoi{10.3847/1538-3881/aabc4f}

\bibitem[{Azzam {et~al.}(2016)Azzam, Yurchenko, Tennyson, \& Naumenko}]{ExoMol_H2S}
Azzam, A. A.~A., Yurchenko, S.~N., Tennyson, J., \& Naumenko, O.~V. 2016, Mon. Not. R. Astron. Soc., 460, 4063, \dodoi{10.1093/mnras/stw1133}

\bibitem[{{Banerjee} {et~al.}(2024){Banerjee}, {Barstow}, {Gressier}, {Espinoza}, {Sing}, {Allen}, {Birkmann}, {Challener}, {Crouzet}, {Haswell}, {Lewis}, {Lewis}, \& {Yang}}]{Banerjee2024L98-59d}
{Banerjee}, A., {Barstow}, J.~K., {Gressier}, A., {et~al.} 2024, \apjl, 975, L11, \dodoi{10.3847/2041-8213/ad73d0}

\bibitem[{Barclay {et~al.}(2021)Barclay, Kostov, Colón, Quintana, Schlieder, Louie, Gilbert, \& Mullally}]{barclay_stellar_2021}
Barclay, T., Kostov, V.~B., Colón, K.~D., {et~al.} 2021, The Astronomical Journal, 162, 300, \dodoi{10.3847/1538-3881/ac2824}

\bibitem[{Bean {et~al.}(2013)Bean, Désert, Seifahrt, Madhusudhan, Chilingarian, Homeier, \& Szentgyorgyi}]{bean_ground-based_2013}
Bean, J.~L., Désert, J.-M., Seifahrt, A., {et~al.} 2013, The Astrophysical Journal, 771, 108, \dodoi{10.1088/0004-637X/771/2/108}

\bibitem[{Beatty {et~al.}(2024)Beatty, Welbanks, Schlawin, Bell, Line, Murphy, Edelman, Greene, Fortney, Henry, Mukherjee, Ohno, Parmentier, Rauscher, Wiser, \& Arnold}]{beatty_sulfur_2024}
Beatty, T.~G., Welbanks, L., Schlawin, E., {et~al.} 2024, Sulfur {Dioxide} and {Other} {Molecular} {Species} in the {Atmosphere} of the {Sub}-{Neptune} {GJ} 3470 b,  arXiv.
\newblock \url{http://arxiv.org/abs/2406.04450}

\bibitem[{{Beatty} {et~al.}(2024){Beatty}, {Welbanks}, {Schlawin}, {Bell}, {Line}, {Murphy}, {Edelman}, {Greene}, {Fortney}, {Henry}, {Mukherjee}, {Ohno}, {Parmentier}, {Rauscher}, {Wiser}, \& {Arnold}}]{Beatty2024GJ3470}
{Beatty}, T.~G., {Welbanks}, L., {Schlawin}, E., {et~al.} 2024, \apjl, 970, L10, \dodoi{10.3847/2041-8213/ad55e9}

\bibitem[{Bell {et~al.}(2022)Bell, Ahrer, Brande, Carter, Feinstein, Guzman~Caloca, Mansfield, Zieba, Piaulet, Benneke, Filippazzo, May, Roy, Kreidberg, \& Stevenson}]{Bell2022Eureka:Observations}
Bell, T.~J., Ahrer, E.-M., Brande, J., {et~al.} 2022, JOSS, 7, 4503

\bibitem[{Benneke(2015)}]{benneke_strict_2015}
Benneke, B. 2015, arXiv e-prints, 1504, arXiv:1504.07655.
\newblock \url{http://adsabs.harvard.edu/abs/2015arXiv150407655B}

\bibitem[{Benneke \& Seager(2012)}]{benneke_atmospheric_2012}
Benneke, B., \& Seager, S. 2012, The Astrophysical Journal, 753, 100, \dodoi{10.1088/0004-637X/753/2/100}

\bibitem[{Benneke \& Seager(2013)}]{benneke_how_2013}
---. 2013, The Astrophysical Journal, 778, 153, \dodoi{10.1088/0004-637X/778/2/153}

\bibitem[{Benneke {et~al.}(2019{\natexlab{a}})Benneke, Knutson, Lothringer, Crossfield, Moses, Morley, Kreidberg, Fulton, Dragomir, Howard, Wong, Désert, McCullough, Kempton, Fortney, Gilliland, Deming, \& Kammer}]{benneke_sub-neptune_2019}
Benneke, B., Knutson, H.~A., Lothringer, J., {et~al.} 2019{\natexlab{a}}, Nature Astronomy, 3, 813, \dodoi{10.1038/s41550-019-0800-5}

\bibitem[{Benneke {et~al.}(2019{\natexlab{b}})Benneke, Wong, Piaulet, Knutson, Lothringer, Morley, Crossfield, Gao, Greene, Dressing, Dragomir, Howard, McCullough, Kempton, Fortney, \& Fraine}]{benneke_water_2019}
Benneke, B., Wong, I., Piaulet, C., {et~al.} 2019{\natexlab{b}}, The Astrophysical Journal Letters, 887, L14, \dodoi{10.3847/2041-8213/ab59dc}

\bibitem[{Benneke {et~al.}(2024)Benneke, Roy, Coulombe, Radica, Piaulet, Ahrer, Pierrehumbert, Krissansen-Totton, Schlichting, Hu, Yang, Christie, Thorngren, Young, Pelletier, Knutson, Miguel, Evans-Soma, Dorn, Gagnebin, Fortney, Komacek, MacDonald, Raul, Cloutier, Acuna, Lafrenière, Cadieux, Doyon, Welbanks, \& Allart}]{benneke_jwst_2024}
Benneke, B., Roy, P.-A., Coulombe, L.-P., {et~al.} 2024, {JWST} {Reveals} {CH}\$\_4\$, {CO}\$\_2\$, and {H}\$\_2\${O} in a {Metal}-rich {Miscible} {Atmosphere} on a {Two}-{Earth}-{Radius} {Exoplanet},  arXiv.
\newblock \url{http://arxiv.org/abs/2403.03325}

\bibitem[{Berta {et~al.}(2012)Berta, Charbonneau, Désert, Kempton, McCullough, Burke, Fortney, Irwin, Nutzman, \& Homeier}]{berta_flat_2012}
Berta, Z.~K., Charbonneau, D., Désert, J.-M., {et~al.} 2012, The Astrophysical Journal, 747, 35, \dodoi{10.1088/0004-637X/747/1/35}

\bibitem[{{B{\'e}zard} {et~al.}(2022){B{\'e}zard}, {Charnay}, \& {Blain}}]{Bezard2022}
{B{\'e}zard}, B., {Charnay}, B., \& {Blain}, D. 2022, Nature Astronomy, 6, 537, \dodoi{10.1038/s41550-022-01678-z}

\bibitem[{{Biassoni} {et~al.}(2024){Biassoni}, {Caldiroli}, \& et~al.}]{BiassoniCaldiroli2024}
{Biassoni}, F., {Caldiroli}, A., \& et~al. 2024, Astronomy and Astrophysics, 682, A115, \dodoi{10.1051/0004-6361/202347517}

\bibitem[{Birkmann {et~al.}(2022)Birkmann, Ferruit, Giardino, Nielsen, García~Muñoz, Kendrew, Rauscher, Beck, Keyes, Valenti, Jakobsen, Dorner, Alves~de Oliveira, Arribas, Böker, Bunker, Charlot, de~Marchi, Kumari, López-Caniego, Lützgendorf, Maiolino, Manjavacas, Marston, Moseley, Prizkal, Proffitt, Rawle, Rix, te~Plate, Sabbi, Sirianni, Willott, \& Zeidler}]{birkmann_near-infrared_2022}
Birkmann, S.~M., Ferruit, P., Giardino, G., {et~al.} 2022, Astronomy and Astrophysics, 661, A83, \dodoi{10.1051/0004-6361/202142592}

\bibitem[{Bitsch {et~al.}(2021)Bitsch, Raymond, Buchhave, Bello-Arufe, Rathcke, \& Schneider}]{bitsch_dry_2021}
Bitsch, B., Raymond, S.~N., Buchhave, L.~A., {et~al.} 2021, arXiv e-prints, 2104, arXiv:2104.11631.
\newblock \url{http://adsabs.harvard.edu/abs/2021arXiv210411631B}

\bibitem[{Borucki {et~al.}(2010)Borucki, Koch, Basri, Batalha, Brown, Caldwell, Caldwell, Christensen-Dalsgaard, Cochran, DeVore, Dunham, Dupree, Gautier, Geary, Gilliland, Gould, Howell, Jenkins, Kondo, Latham, Marcy, Meibom, Kjeldsen, Lissauer, Monet, Morrison, Sasselov, Tarter, Boss, Brownlee, Owen, Buzasi, Charbonneau, Doyle, Fortney, Ford, Holman, Seager, Steffen, Welsh, Rowe, Anderson, Buchhave, Ciardi, Walkowicz, Sherry, Horch, Isaacson, Everett, Fischer, Torres, Johnson, Endl, MacQueen, Bryson, Dotson, Haas, Kolodziejczak, Van~Cleve, Chandrasekaran, Twicken, Quintana, Clarke, Allen, Li, Wu, Tenenbaum, Verner, Bruhweiler, Barnes, \& Prsa}]{borucki_kepler_2010}
Borucki, W.~J., Koch, D., Basri, G., {et~al.} 2010, Science, 327, 977, \dodoi{10.1126/science.1185402}

\bibitem[{{Bouchy} {et~al.}(2017){Bouchy}, {Doyon}, {Artigau}, {Melo}, {Hernandez}, {Wildi}, {Delfosse}, {Lovis}, {Figueira}, {Canto Martins}, {Gonz{\'a}lez Hern{\'a}ndez}, {Thibault}, {Reshetov}, {Pepe}, {Santos}, {de Medeiros}, {Rebolo}, {Abreu}, {Adibekyan}, {Bandy}, {Benz}, {Blind}, {Bohlender}, {Boisse}, {Bovay}, {Broeg}, {Brousseau}, {Cabral}, {Chazelas}, {Cloutier}, {Coelho}, {Conod}, {Cumming}, {Delabre}, {Genolet}, {Hagelberg}, {Jayawardhana}, {K{\"a}ufl}, {Lafreni{\`e}re}, {de Castro Le{\~a}o}, {Malo}, {de Medeiros Martins}, {Matthews}, {Metchev}, {Oshagh}, {Ouellet}, {Parro}, {Rasilla Pi{\~n}eiro}, {Santos}, {Sarajlic}, {Segovia}, {Sordet}, {Udry}, {Valencia}, {Vall{\'e}e}, {Venn}, {Wade}, \& {Saddlemyer}}]{Bouchy_2017NIRPS}
{Bouchy}, F., {Doyon}, R., {Artigau}, {\'E}., {et~al.} 2017, The Messenger, 169, 21, \dodoi{10.18727/0722-6691/5034}

\bibitem[{{Buchner} {et~al.}(2014){Buchner}, {Georgakakis}, {Nandra}, {Hsu}, {Rangel}, {Brightman}, {Merloni}, {Salvato}, {Donley}, \& {Kocevski}}]{Buchner2014}
{Buchner}, J., {Georgakakis}, A., {Nandra}, K., {et~al.} 2014, Astronomy \& Astrophysics, 564, A125, \dodoi{10.1051/0004-6361/201322971}

\bibitem[{Buchner {et~al.}(2014)Buchner, Georgakakis, Nandra, Hsu, Rangel, Brightman, Merloni, Salvato, Donley, \& Kocevski}]{Buchner2014X-rayCatalogue}
Buchner, J., Georgakakis, A., Nandra, K., {et~al.} 2014, Astronomy and Astrophysics, 564, \dodoi{10.1051/0004-6361/201322971}

\bibitem[{Burn {et~al.}(2024)Burn, Mordasini, Mishra, Haldemann, Venturini, Emsenhuber, \& Henning}]{burn_radius_2024}
Burn, R., Mordasini, C., Mishra, L., {et~al.} 2024, A radius valley between migrated steam worlds and evaporated rocky cores,  arXiv.
\newblock \url{http://arxiv.org/abs/2401.04380}

\bibitem[{Bushouse {et~al.}(2023)Bushouse, Eisenhamer, Dencheva, Davies, Greenfield, Morrison, Hodge, Simon, Grumm, Droettboom, Slavich, Sosey, Pauly, Miller, Jedrzejewski, Hack, Davis, Crawford, Law, Gordon, Regan, Cara, MacDonald, Bradley, Shanahan, Jamieson, Teodoro, \& Williams}]{bushouse_2023_8157276}
Bushouse, H., Eisenhamer, J., Dencheva, N., {et~al.} 2023, JWST Calibration Pipeline, 1.11.3,  Zenodo, \dodoi{10.5281/zenodo.8157276}

\bibitem[{{Cadieux} {et~al.}(2024){Cadieux}, {Doyon}, {MacDonald}, {Turbet}, {Artigau}, {Lim}, {Radica}, {Fauchez}, {Salhi}, {Dang}, {Albert}, {Coulombe}, {Cowan}, {Lafreni{\`e}re}, {L'Heureux}, {Piaulet-Ghorayeb}, {Benneke}, {Cloutier}, {Charnay}, {Cook}, {Fournier-Tondreau}, {Plotnykov}, \& {Valencia}}]{Cadieux2024}
{Cadieux}, C., {Doyon}, R., {MacDonald}, R.~J., {et~al.} 2024, \apjl, 970, L2, \dodoi{10.3847/2041-8213/ad5afa}

\bibitem[{{Caldiroli} {et~al.}(2021){Caldiroli}, {Haardt}, \& et~al.}]{CaldiroliHaardt2021}
{Caldiroli}, A., {Haardt}, F., \& et~al. 2021, Astronomy and Astrophysics, 655, A30, \dodoi{10.1051/0004-6361/202141497}

\bibitem[{{Chachan} \& {Stevenson}(2018)}]{Chachan2018magma}
{Chachan}, Y., \& {Stevenson}, D.~J. 2018, \apj, 854, 21, \dodoi{10.3847/1538-4357/aaa459}

\bibitem[{Charbonneau {et~al.}(2000)Charbonneau, Brown, Latham, \& Mayor}]{Charbonneau2000DetectionStar}
Charbonneau, D., Brown, T.~M., Latham, D.~W., \& Mayor, M. 2000, The Astrophysical Journal, 529, L45, \dodoi{10.1086/312457}

\bibitem[{{Cherubim} {et~al.}(2025){Cherubim}, {Wordsworth}, {Bower}, {Sossi}, {Adams}, \& {Hu}}]{cherubim_oxidation_2025}
{Cherubim}, C., {Wordsworth}, R., {Bower}, D., {et~al.} 2025, arXiv e-prints, arXiv:2503.05055, \dodoi{10.48550/arXiv.2503.05055}

\bibitem[{{Cherubim} {et~al.}(2024){Cherubim}, {Wordsworth}, {Hu}, \& {Shkolnik}}]{Cherubim2024}
{Cherubim}, C., {Wordsworth}, R., {Hu}, R., \& {Shkolnik}, E. 2024, \apj, 967, 139, \dodoi{10.3847/1538-4357/ad3e77}

\bibitem[{Cloutier \& Menou(2020)}]{cloutier_evolution_2020}
Cloutier, R., \& Menou, K. 2020, The Astronomical Journal, 159, 211, \dodoi{10.3847/1538-3881/ab8237}

\bibitem[{Coles {et~al.}(2019)Coles, , Yurchenko, \& Tennyson}]{ExoMol_NH3}
Coles, P.~A., , Yurchenko, S.~N., \& Tennyson, J. 2019, Mon. Not. R. Astron. Soc., 490, 4638, \dodoi{10.1093/mnras/stz2778}

\bibitem[{{Cooke} \& {Madhusudhan}(2024)}]{Cooke2024}
{Cooke}, G.~J., \& {Madhusudhan}, N. 2024, \apj, 977, 209, \dodoi{10.3847/1538-4357/ad8cda}

\bibitem[{{Coulombe} {et~al.}(2024){Coulombe}, {Roy}, \& {Benneke}}]{Coulombe2024LDbiases}
{Coulombe}, L.-P., {Roy}, P.-A., \& {Benneke}, B. 2024, \aj, 168, 227, \dodoi{10.3847/1538-3881/ad7aef}

\bibitem[{Coulombe {et~al.}(2023)Coulombe, Benneke, Challener, Piette, Wiser, Mansfield, MacDonald, Beltz, Feinstein, Radica, Savel, Dos~Santos, Bean, Parmentier, Wong, Rauscher, Komacek, Kempton, Tan, Hammond, Lewis, Line, Lee, Shivkumar, Crossfield, Nixon, Rackham, Wakeford, Welbanks, Zhang, Batalha, Berta-Thompson, Changeat, Désert, Espinoza, Goyal, Harrington, Knutson, Kreidberg, López-Morales, Shporer, Sing, Stevenson, Aggarwal, Ahrer, Alam, Bell, Blecic, Caceres, Carter, Casewell, Crouzet, Cubillos, Decin, Fortney, Gibson, Heng, Henning, Iro, Kendrew, Lagage, Leconte, Lendl, Lothringer, Mancini, Mikal-Evans, Molaverdikhani, Nikolov, Ohno, Palle, Piaulet, Redfield, Roy, Tsai, Venot, \& Wheatley}]{coulombe_broadband_2023}
Coulombe, L.-P., Benneke, B., Challener, R., {et~al.} 2023, Nature, 620, 292, \dodoi{10.1038/s41586-023-06230-1}

\bibitem[{{Coulombe} {et~al.}(2025){Coulombe}, {Radica}, {Benneke}, {D'Aoust}, {Dang}, {Cowan}, {Parmentier}, {Albert}, {Lafreni{\`e}re}, {Taylor}, {Roy}, {Pelletier}, {Allart}, {Artigau}, {Doyon}, {Jayawardhana}, {Johnstone}, {Kaltenegger}, {Langeveld}, {MacDonald}, {Rowe}, \& {Turner}}]{Coulombe2025highlyreflectivewhiteclouds}
{Coulombe}, L.-P., {Radica}, M., {Benneke}, B., {et~al.} 2025, arXiv e-prints, arXiv:2501.14016.
\newblock \doarXiv{2501.14016}

\bibitem[{{Darveau-Bernier} {et~al.}(2022){Darveau-Bernier}, {Albert}, {Talens}, {Lafreni{\`e}re}, {Radica}, {Doyon}, {Cook}, {Rowe}, {Allart}, {Artigau}, {Benneke}, {Cowan}, {Dang}, {Espinoza}, {Johnstone}, {Kaltenegger}, {Lim}, {Pauly}, {Pelletier}, {Piaulet}, {Roy}, {Roy}, {Splinter}, {Taylor}, \& {Turner}}]{Darveau-Bernier2022}
{Darveau-Bernier}, A., {Albert}, L., {Talens}, G.~J., {et~al.} 2022, \pasp, 134, 094502, \dodoi{10.1088/1538-3873/ac8a77}

\bibitem[{{Deal} \& {Espinoza}(2024)}]{DealEspinoza2024spelunker}
{Deal}, D., \& {Espinoza}, N. 2024, The Journal of Open Source Software, 9, 6202, \dodoi{10.21105/joss.06202}

\bibitem[{{Dos Santos} {et~al.}(2023){Dos Santos}, {Alam}, {Espinoza}, \& {Vissapragada}}]{DosSantos2023JWSTAtmosphericEscape}
{Dos Santos}, L.~A., {Alam}, M.~K., {Espinoza}, N., \& {Vissapragada}, S. 2023, \aj, 165, 244, \dodoi{10.3847/1538-3881/accf10}

\bibitem[{{Dos Santos} {et~al.}(2022){Dos Santos}, {Vidotto}, \& et~al.}]{DosSantosVidotto2022}
{Dos Santos}, L.~A., {Vidotto}, A.~A., \& et~al. 2022, Astronomy and Astrophysics, 659, A62, \dodoi{10.1051/0004-6361/202142038}

\bibitem[{{Doyon} {et~al.}(2023){Doyon}, {Willott}, {Hutchings}, {Sivaramakrishnan}, {Albert}, {Lafreni{\`e}re}, {Rowlands}, {Bego{\~n}a Vila}, {Martel}, {LaMassa}, {Aldridge}, {Artigau}, {Cameron}, {Chayer}, {Cook}, {Cooper}, {Darveau-Bernier}, {Dupuis}, {Earnshaw}, {Espinoza}, {Filippazzo}, {Fullerton}, {Gaudreau}, {Gawlik}, {Goudfrooij}, {Haley}, {Kammerer}, {Kendall}, {Lambros}, {Ignat}, {Maszkiewicz}, {McColgan}, {Morishita}, {Ouellette}, {Pacifici}, {Philippi}, {Radica}, {Ravindranath}, {Rowe}, {Roy}, {Roy}, {Saad}, {Sohn}, {Talens}, {Touahri}, {Thatte}, {Taylor}, {Vandal}, {Volk}, {Wander}, {Warner}, {Zheng}, {Zhou}, {Abraham}, {Beaulieu}, {Benneke}, {Ferrarese}, {Jayawardhana}, {Johnstone}, {Kaltenegger}, {Meyer}, {Pipher}, {Rameau}, {Rieke}, {Salhi}, \& {Sawicki}}]{Doyon2023}
{Doyon}, R., {Willott}, C.~J., {Hutchings}, J.~B., {et~al.} 2023, \pasp, 135, 098001, \dodoi{10.1088/1538-3873/acd41b}

\bibitem[{{Edwards} {et~al.}(2021){Edwards}, {Changeat}, {Mori}, {Anisman}, {Morvan}, {Yip}, {Tsiaras}, {Al-Refaie}, {Waldmann}, \& {Tinetti}}]{edwards_2021_hubble}
{Edwards}, B., {Changeat}, Q., {Mori}, M., {et~al.} 2021, \aj, 161, 44, \dodoi{10.3847/1538-3881/abc6a5}

\bibitem[{Ehrenreich {et~al.}(2015)Ehrenreich, Bourrier, Wheatley, Etangs, Hébrard, Udry, Bonfils, Delfosse, Désert, Sing, \& Vidal-Madjar}]{Ehrenreich2015}
Ehrenreich, D., Bourrier, V., Wheatley, P., {et~al.} 2015, Nature, 522, 459, \dodoi{10.1038/nature14501}

\bibitem[{{Erkaev} {et~al.}(2007){Erkaev}, {Kulikov}, {Lammer}, {Selsis}, {Langmayr}, {Jaritz}, \& {Biernat}}]{Erkaev2007}
{Erkaev}, N.~V., {Kulikov}, Y.~N., {Lammer}, H., {et~al.} 2007, \aap, 472, 329, \dodoi{10.1051/0004-6361:20066929}

\bibitem[{Feinstein {et~al.}(2023)Feinstein, Radica, Welbanks, Murray, Ohno, Coulombe, Espinoza, Bean, Teske, Benneke, Line, Rustamkulov, Saba, Tsiaras, Barstow, Fortney, Gao, Knutson, MacDonald, Mikal-Evans, Rackham, Taylor, Parmentier, Batalha, Berta-Thompson, Carter, Changeat, dos Santos, Gibson, Goyal, Kreidberg, López-Morales, Lothringer, Miguel, Molaverdikhani, Moran, Morello, Mukherjee, Sing, Stevenson, Wakeford, Ahrer, Alam, Alderson, Allen, Batalha, Bell, Blecic, Brande, Caceres, Casewell, Chubb, Crossfield, Crouzet, Cubillos, Decin, Désert, Harrington, Heng, Henning, Iro, Kempton, Kendrew, Kirk, Krick, Lagage, Lendl, Mancini, Mansfield, May, Mayne, Nikolov, Palle, Petit dit de~la Roche, Piaulet, Powell, Redfield, Rogers, Roman, Roy, Nixon, Schlawin, Tan, Tremblin, Turner, Venot, Waalkes, Wheatley, \& Zhang}]{feinstein_early_2023}
Feinstein, A.~D., Radica, M., Welbanks, L., {et~al.} 2023, Nature, 614, 670, \dodoi{10.1038/s41586-022-05674-1}

\bibitem[{Feroz {et~al.}(2009)Feroz, Hobson, \& Bridges}]{Feroz2009MULTINEST:Physics}
Feroz, F., Hobson, M.~P., \& Bridges, M. 2009, Monthly Notices of the Royal Astronomical Society, 398, 1601, \dodoi{10.1111/j.1365-2966.2009.14548.x}

\bibitem[{Foreman-Mackey {et~al.}(2017)Foreman-Mackey, Agol, Ambikasaran, \& Angus}]{foreman-mackey_fast_2017}
Foreman-Mackey, D., Agol, E., Ambikasaran, S., \& Angus, R. 2017, The Astronomical Journal, 154, 220, \dodoi{10.3847/1538-3881/aa9332}

\bibitem[{Foreman-Mackey {et~al.}(2013)Foreman-Mackey, Hogg, Lang, \& Goodman}]{foreman-mackey_emcee_2013}
Foreman-Mackey, D., Hogg, D.~W., Lang, D., \& Goodman, J. 2013, Publications of the Astronomical Society of the Pacific, 125, 306, \dodoi{10.1086/670067}

\bibitem[{{Foreman-Mackey} {et~al.}(2013){Foreman-Mackey}, {Hogg}, {Lang}, \& {Goodman}}]{Foremak-Mackey2013}
{Foreman-Mackey}, D., {Hogg}, D.~W., {Lang}, D., \& {Goodman}, J. 2013, \pasp, 125, 306, \dodoi{10.1086/670067}

\bibitem[{Foreman-Mackey {et~al.}(2013)Foreman-Mackey, Hogg, Lang, \& Goodman}]{Foreman-Mackey2013EmceeHammer}
Foreman-Mackey, D., Hogg, D.~W., Lang, D., \& Goodman, J. 2013, Publications of the Astronomical Society of the Pacific, 125, 306, \dodoi{10.1086/670067}

\bibitem[{Fortney {et~al.}(2007)Fortney, Marley, \& Barnes}]{fortney_planetary_2007}
Fortney, J.~J., Marley, M.~S., \& Barnes, J.~W. 2007, The Astrophysical Journal, 659, 1661, \dodoi{10.1086/512120}

\bibitem[{{Fortney} {et~al.}(2013){Fortney}, {Mordasini}, {Nettelmann}, {Kempton}, {Greene}, \& {Zahnle}}]{Fortney2013}
{Fortney}, J.~J., {Mordasini}, C., {Nettelmann}, N., {et~al.} 2013, \apj, 775, 80, \dodoi{10.1088/0004-637X/775/1/80}

\bibitem[{Fortney {et~al.}(2013)Fortney, Mordasini, Nettelmann, Kempton, Greene, \& Zahnle}]{fortney_framework_2013}
Fortney, J.~J., Mordasini, C., Nettelmann, N., {et~al.} 2013, The Astrophysical Journal, 775, 80, \dodoi{10.1088/0004-637X/775/1/80}

\bibitem[{Fortney {et~al.}(2020)Fortney, Visscher, Marley, Hood, Line, Thorngren, Freedman, \& Lupu}]{fortney_beyond_2020}
Fortney, J.~J., Visscher, C., Marley, M.~S., {et~al.} 2020, The Astronomical Journal, 160, 288, \dodoi{10.3847/1538-3881/abc5bd}

\bibitem[{{Fossati} {et~al.}(2023){Fossati}, {Pillitteri}, {Shaikhislamov}, {Bonfanti}, {Borsa}, {Carleo}, {Guilluy}, \& {Rumenskikh}}]{fossati_possible_2023}
{Fossati}, L., {Pillitteri}, I., {Shaikhislamov}, I.~F., {et~al.} 2023, \aap, 673, A37, \dodoi{10.1051/0004-6361/202245667}

\bibitem[{Fournier-Tondreau {et~al.}(2024)Fournier-Tondreau, MacDonald, Radica, Lafrenière, Welbanks, Piaulet, Coulombe, Allart, Morel, Artigau, Albert, Lim, Doyon, Benneke, Rowe, Darveau-Bernier, Cowan, Lewis, Cook, Flagg, Genest, Pelletier, Johnstone, Dang, Kaltenegger, Taylor, \& Turner}]{fournier-tondreau_near-infrared_2024}
Fournier-Tondreau, M., MacDonald, R.~J., Radica, M., {et~al.} 2024, Monthly Notices of the Royal Astronomical Society, 528, 3354, \dodoi{10.1093/mnras/stad3813}

\bibitem[{Fu {et~al.}(2023)Fu, Espinoza, Sing, Lothringer, Dos~Santos, Rustamkulov, Deming, Kempton, Komacek, Knutson, Albert, Pontoppidan, Volk, \& Filippazzo}]{fu_water_2023}
Fu, G., Espinoza, N., Sing, D., {et~al.} 2023, 241, 159.04.
\newblock \url{https://ui.adsabs.harvard.edu/abs/2023AAS...24115904F}

\bibitem[{Fulton \& Petigura(2018)}]{fulton_california-kepler_2018}
Fulton, B.~J., \& Petigura, E.~A. 2018, The Astronomical Journal, 156, 264, \dodoi{10.3847/1538-3881/aae828}

\bibitem[{Fulton {et~al.}(2017)Fulton, Petigura, Howard, Isaacson, Marcy, Cargile, Hebb, Weiss, Johnson, Morton, Sinukoff, Crossfield, \& Hirsch}]{fulton_california-kepler_2017}
Fulton, B.~J., Petigura, E.~A., Howard, A.~W., {et~al.} 2017, The Astronomical Journal, 154, 109, \dodoi{10.3847/1538-3881/aa80eb}

\bibitem[{{Gao} {et~al.}(2023){Gao}, {Piette}, {Steinrueck}, {Nixon}, {Zhang}, {Kempton}, {Bean}, {Rauscher}, {Parmentier}, {Batalha}, {Savel}, {Arnold}, {Roman}, {Malsky}, \& {Taylor}}]{Gao2023GJ1214}
{Gao}, P., {Piette}, A. A.~A., {Steinrueck}, M.~E., {et~al.} 2023, \apj, 951, 96, \dodoi{10.3847/1538-4357/acd16f}

\bibitem[{Ginzburg {et~al.}(2016)Ginzburg, Schlichting, \& Sari}]{ginzburg_super-earth_2016}
Ginzburg, S., Schlichting, H.~E., \& Sari, R. 2016, The Astrophysical Journal, 825, 29, \dodoi{10.3847/0004-637X/825/1/29}

\bibitem[{Ginzburg {et~al.}(2018)Ginzburg, Schlichting, \& Sari}]{ginzburg_core-powered_2018}
---. 2018, Monthly Notices of the Royal Astronomical Society, 476, 759, \dodoi{10.1093/mnras/sty290}

\bibitem[{{Glein}(2024)}]{glein2024}
{Glein}, C.~R. 2024, \apjl, 964, L19, \dodoi{10.3847/2041-8213/ad3079}

\bibitem[{Gordon \& Mcbride(1994)}]{GordonMcbride1994}
Gordon, S., \& Mcbride, B.~J. 1994, Computer program for calculation of complex chemical equilibrium compositions and applications. Part 1: Analysis, Tech. Rep. 19950013764, NASA Lewis Research Center

\bibitem[{{Grant} \& {Wakeford}(2024)}]{Grant2024}
{Grant}, D., \& {Wakeford}, H. 2024, The Journal of Open Source Software, 9, 6816, \dodoi{10.21105/joss.06816}

\bibitem[{{Gressier} {et~al.}(2024){Gressier}, {Espinoza}, {Allen}, {Sing}, {Banerjee}, {Barstow}, {Valenti}, {Lewis}, {Birkmann}, {Challener}, {Manjavacas}, {Alves de Oliveira}, {Crouzet}, \& {Beck}}]{Gressier2024L98-59d}
{Gressier}, A., {Espinoza}, N., {Allen}, N.~H., {et~al.} 2024, \apjl, 975, L10, \dodoi{10.3847/2041-8213/ad73d1}

\bibitem[{Grimm \& Heng(2015)}]{grimm_helios-k_2015}
Grimm, S.~L., \& Heng, K. 2015, The Astrophysical Journal, 808, 182, \dodoi{10.1088/0004-637X/808/2/182}

\bibitem[{{Guilluy} {et~al.}(2024){Guilluy}, {D'Arpa}, {Bonomo}, {Spinelli}, {Biassoni}, {Fossati}, {Maggio}, {Giacobbe}, {Lanza}, {Sozzetti}, {Borsa}, {Rainer}, {Micela}, {Affer}, {Andreuzzi}, {Bignamini}, {Boschin}, {Carleo}, {Cecconi}, {Desidera}, {Fardella}, {Ghedina}, {Mantovan}, {Mancini}, {Nascimbeni}, {Knapic}, {Pedani}, {Petralia}, {Pino}, {Scandariato}, {Sicilia}, {Stangret}, \& {Zingales}}]{Guilley2024Heliumsurvey}
{Guilluy}, G., {D'Arpa}, M.~C., {Bonomo}, A.~S., {et~al.} 2024, \aap, 686, A83, \dodoi{10.1051/0004-6361/202348997}

\bibitem[{Gupta \& Schlichting(2019)}]{gupta_sculpting_2019}
Gupta, A., \& Schlichting, H.~E. 2019, Monthly Notices of the Royal Astronomical Society, 487, 24, \dodoi{10.1093/mnras/stz1230}

\bibitem[{{Gupta} \& {Schlichting}(2020)}]{Gupta2020corepowered_signatures}
{Gupta}, A., \& {Schlichting}, H.~E. 2020, \mnras, 493, 792, \dodoi{10.1093/mnras/staa315}

\bibitem[{Haldemann {et~al.}(2020)Haldemann, Alibert, Mordasini, \& Benz}]{haldemann_aqua_2020}
Haldemann, J., Alibert, Y., Mordasini, C., \& Benz, W. 2020, Astronomy \& Astrophysics, 643, A105, \dodoi{10.1051/0004-6361/202038367}

\bibitem[{Handley {et~al.}(2015{\natexlab{a}})Handley, Hobson, \& Lasenby}]{Handley2015PolyChord:Sampling}
Handley, W.~J., Hobson, M.~P., \& Lasenby, A.~N. 2015{\natexlab{a}}, Monthly Notices of the Royal Astronomical Society, 453, 4384, \dodoi{10.1093/mnras/stv1911}

\bibitem[{Handley {et~al.}(2015{\natexlab{b}})Handley, Hobson, \& Lasenby}]{Handley2015PolyChord:Cosmology}
---. 2015{\natexlab{b}}, Monthly Notices of the Royal Astronomical Society: Letters, 450, L61, \dodoi{10.1093/mnrasl/slv047}

\bibitem[{{Hargreaves} {et~al.}(2020){Hargreaves}, {Gordon}, {Rey}, {Nikitin}, {Tyuterev}, {Kochanov}, \& {Rothman}}]{HargreavesEtal2020apjsHitempCH4}
{Hargreaves}, R.~J., {Gordon}, I.~E., {Rey}, M., {et~al.} 2020, \apjs, 247, 55, \dodoi{10.3847/1538-4365/ab7a1a}

\bibitem[{{Hargreaves} {et~al.}(2019){Hargreaves}, {Gordon}, {Rothman}, {Tashkun}, {Perevalov}, {Lukashevskaya}, {Yurchenko}, {Tennyson}, \& {M{\"u}ller}}]{Hargreaves2019}
{Hargreaves}, R.~J., {Gordon}, I.~E., {Rothman}, L.~S., {et~al.} 2019, \jqsrt, 232, 35, \dodoi{10.1016/j.jqsrt.2019.04.040}

\bibitem[{Harris {et~al.}(2020)Harris, Millman, van~der Walt, Gommers, Virtanen, Cournapeau, Wieser, Taylor, Berg, Smith, Kern, Picus, Hoyer, van Kerkwijk, Brett, Haldane, del R{\'{i}}o, Wiebe, Peterson, G{\'{e}}rard-Marchant, Sheppard, Reddy, Weckesser, Abbasi, Gohlke, \& Oliphant}]{harris2020array}
Harris, C.~R., Millman, K.~J., van~der Walt, S.~J., {et~al.} 2020, Nature, 585, 357, \dodoi{10.1038/s41586-020-2649-2}

\bibitem[{{Harris} {et~al.}(2006){Harris}, {Tennyson}, {Kaminsky}, {Pavlenko}, \& {Jones}}]{Harris_HCN_2006}
{Harris}, G.~J., {Tennyson}, J., {Kaminsky}, B.~M., {Pavlenko}, Y.~V., \& {Jones}, H.~R.~A. 2006, \mnras, 367, 400, \dodoi{10.1111/j.1365-2966.2005.09960.x}

\bibitem[{{Helled} \& {Howard}(2024)}]{SS_interiors_Helled}
{Helled}, R., \& {Howard}, S. 2024, arXiv e-prints, arXiv:2407.05853, \dodoi{10.48550/arXiv.2407.05853}

\bibitem[{Henry {et~al.}(2000)Henry, Marcy, Butler, \& Vogt}]{Henry2000APlanet}
Henry, G.~W., Marcy, G.~W., Butler, R.~P., \& Vogt, S.~S. 2000, The Astrophysical Journal, 529, \dodoi{10.1086/312458}

\bibitem[{Ho {et~al.}(2024)Ho, Rogers, Van~Eylen, Owen, \& Schlichting}]{ho_shallower_2024}
Ho, C. S.~K., Rogers, J.~G., Van~Eylen, V., Owen, J.~E., \& Schlichting, H.~E. 2024, Shallower radius valley around low-mass hosts provides evidence for icy planets or collisions,  arXiv.
\newblock \url{http://arxiv.org/abs/2401.12378}

\bibitem[{Holmberg \& Madhusudhan(2024)}]{holmberg_possible_2024}
Holmberg, M., \& Madhusudhan, N. 2024, Astronomy and Astrophysics, 683, L2, \dodoi{10.1051/0004-6361/202348238}

\bibitem[{{Howard} {et~al.}(2023){Howard}, {Kowalski}, {Flagg}, {MacGregor}, {Lim}, {Radica}, {Piaulet}, {Roy}, {Lafreni{\`e}re}, {Benneke}, {Brown}, {Espinoza}, {Doyon}, {Coulombe}, {Johnstone}, {Cowan}, {Jayawardhana}, {Turner}, \& {Dang}}]{Howard2023}
{Howard}, W.~S., {Kowalski}, A.~F., {Flagg}, L., {et~al.} 2023, \apj, 959, 64, \dodoi{10.3847/1538-4357/acfe75}

\bibitem[{Hunter(2007)}]{Hunter:2007}
Hunter, J.~D. 2007, Computing in Science \& Engineering, 9, 90, \dodoi{10.1109/MCSE.2007.55}

\bibitem[{Husser {et~al.}(2013)Husser, Wende-von Berg, Dreizler, Homeier, Reiners, Barman, \& Hauschildt}]{husser_new_2013}
Husser, T.-O., Wende-von Berg, S., Dreizler, S., {et~al.} 2013, Astronomy \&amp; Astrophysics, Volume 553, id.A6, {\textless}NUMPAGES{\textgreater}9{\textless}/NUMPAGES{\textgreater} pp., 553, A6, \dodoi{10.1051/0004-6361/201219058}

\bibitem[{Innes {et~al.}(2023)Innes, Tsai, \& Pierrehumbert}]{innes_runaway_2023}
Innes, H., Tsai, S.-M., \& Pierrehumbert, R.~T. 2023, The Astrophysical Journal, 953, 168, \dodoi{10.3847/1538-4357/ace346}

\bibitem[{Iyer \& Line(2020)}]{iyer_influence_2020}
Iyer, A.~R., \& Line, M.~R. 2020, The Astrophysical Journal, 889, 78, \dodoi{10.3847/1538-4357/ab612e}

\bibitem[{{Iyer} {et~al.}(2023){Iyer}, {Line}, {Muirhead}, {Fortney}, \& {Gharib-Nezhad}}]{Iyer2023}
{Iyer}, A.~R., {Line}, M.~R., {Muirhead}, P.~S., {Fortney}, J.~J., \& {Gharib-Nezhad}, E. 2023, \apj, 944, 41, \dodoi{10.3847/1538-4357/acabc2}

\bibitem[{Kempton {et~al.}(2023)Kempton, Zhang, Bean, Steinrueck, Piette, Parmentier, Malsky, Roman, Rauscher, Gao, Bell, Xue, Taylor, Savel, Arnold, Nixon, Stevenson, Mansfield, Kendrew, Zieba, Ducrot, Dyrek, Lagage, Stassun, Henry, Barman, Lupu, Malik, Kataria, Ih, Fu, Welbanks, \& McGill}]{kempton_reflective_2023}
Kempton, E. M.~R., Zhang, M., Bean, J.~L., {et~al.} 2023, Nature, 620, 67, \dodoi{10.1038/s41586-023-06159-5}

\bibitem[{Kirk {et~al.}(2020)Kirk, Alam, Lopez-Morales, \& Zeng}]{Kirk2020ConfirmationII/NIRSPEC}
Kirk, J., Alam, M.~K., Lopez-Morales, M., \& Zeng, L. 2020, The Astronomical Journal, 159, 115.
\newblock \url{http://arxiv.org/abs/2001.07667}

\bibitem[{Kirk {et~al.}(2017)Kirk, Wheatley, Louden, Doyle, Skillen, McCormac, Irwin, \& Karjalainen}]{Kirk2017RayleighHAT-P-18b}
Kirk, J., Wheatley, P.~J., Louden, T., {et~al.} 2017, Monthly Notices of the Royal Astronomical Society, 468, 3907, \dodoi{10.1093/mnras/stx752}

\bibitem[{Kirk {et~al.}(2021)Kirk, Rackham, MacDonald, L{\'{o}}pez-Morales, Espinoza, Lendl, Wilson, Osip, Wheatley, Skillen, Apai, Bixel, Gibson, Jord{\'{a}}n, Lewis, Louden, McGruder, Nikolov, Rodler, \& Weaver}]{Kirk2021ACCESSWASP-103b}
Kirk, J., Rackham, B.~V., MacDonald, R.~J., {et~al.} 2021, The Astronomical Journal, 162, 34, \dodoi{10.3847/1538-3881/abfcd2}

\bibitem[{Kite {et~al.}(2019)Kite, {Bruce Fegley Jr.}, Schaefer, \& Ford}]{kite_superabundance_2019}
Kite, E.~S., {Bruce Fegley Jr.}, Schaefer, L., \& Ford, E.~B. 2019, The Astrophysical Journal Letters, 887, L33, \dodoi{10.3847/2041-8213/ab59d9}

\bibitem[{Kite \& Ford(2018)}]{kite_habitability_2018}
Kite, E.~S., \& Ford, E.~B. 2018, The Astrophysical Journal, 864, 75, \dodoi{10.3847/1538-4357/aad6e0}

\bibitem[{Kite \& Schaefer(2021)}]{kite_water_2021}
Kite, E.~S., \& Schaefer, L. 2021, The Astrophysical Journal, 909, L22, \dodoi{10.3847/2041-8213/abe7dc}

\bibitem[{Kostogryz {et~al.}(2023)Kostogryz, Shapiro, Witzke, Grant, Wakeford, Stevenson, Solanki, \& Gizon}]{kostogryz2023mps}
Kostogryz, N., Shapiro, A., Witzke, V., {et~al.} 2023, Research Notes of the AAS, 7, 39

\bibitem[{Kreidberg(2015)}]{Kreidberg2015BatmanPython}
Kreidberg, L. 2015, Publications of the Astronomical Society of the Pacific, 127, 1161, \dodoi{10.1086/683602}

\bibitem[{Kreidberg {et~al.}(2014)Kreidberg, Bean, Désert, Benneke, Deming, Stevenson, Seager, Berta-Thompson, Seifahrt, \& Homeier}]{kreidberg_clouds_2014}
Kreidberg, L., Bean, J.~L., Désert, J.-M., {et~al.} 2014, Nature, 505, 69, \dodoi{10.1038/nature12888}

\bibitem[{{Lammer} {et~al.}(2003){Lammer}, {Selsis}, {Ribas}, {Guinan}, {Bauer}, \& {Weiss}}]{Lammer2003}
{Lammer}, H., {Selsis}, F., {Ribas}, I., {et~al.} 2003, \apjl, 598, L121, \dodoi{10.1086/380815}

\bibitem[{{Leconte} {et~al.}(2024){Leconte}, {Spiga}, {Cl{\'e}ment}, {Guerlet}, {Selsis}, {Milcareck}, {Cavali{\'e}}, {Moreno}, {Lellouch}, {Carri{\'o}n-Gonz{\'a}lez}, {Charnay}, \& {Lef{\`e}vre}}]{Leconte_moist_convection_2024}
{Leconte}, J., {Spiga}, A., {Cl{\'e}ment}, N., {et~al.} 2024, \aap, 686, A131, \dodoi{10.1051/0004-6361/202348928}

\bibitem[{Lee \& Chiang(2016)}]{lee_breeding_2016}
Lee, E.~J., \& Chiang, E. 2016, The Astrophysical Journal, 817, 90, \dodoi{10.3847/0004-637X/817/2/90}

\bibitem[{Lichtenberg {et~al.}(2021)Lichtenberg, Bower, Hammond, Boukrouche, Sanan, Tsai, \& Pierrehumbert}]{lichtenberg_vertically_2021}
Lichtenberg, T., Bower, D.~J., Hammond, M., {et~al.} 2021, Journal of Geophysical Research: Planets, \dodoi{10.1029/2020JE006711}

\bibitem[{Lichtenberg \& Miguel(2024)}]{lichtenberg_super-earths_2024}
Lichtenberg, T., \& Miguel, Y. 2024, Super-{Earths} and {Earth}-like {Exoplanets},  arXiv, \dodoi{10.48550/arXiv.2405.04057}

\bibitem[{Lim {et~al.}(2023)Lim, Benneke, Doyon, MacDonald, Piaulet, Artigau, Coulombe, Radica, L'Heureux, Albert, Rackham, de~Wit, Salhi, Roy, Flagg, Fournier-Tondreau, Taylor, Cook, Lafrenière, Cowan, Kaltenegger, Rowe, Espinoza, Dang, \& Darveau-Bernier}]{lim_atmospheric_2023}
Lim, O., Benneke, B., Doyon, R., {et~al.} 2023, The Astrophysical Journal, 955, L22, \dodoi{10.3847/2041-8213/acf7c4}

\bibitem[{Line \& Parmentier(2016)}]{Line2016THESPECTRA}
Line, M.~R., \& Parmentier, V. 2016, The Astrophysical Journal, 820, 78, \dodoi{10.3847/0004-637x/820/1/78}

\bibitem[{Lopez \& Fortney(2014)}]{lopez_understanding_2014}
Lopez, E.~D., \& Fortney, J.~J. 2014, The Astrophysical Journal, 792, 1, \dodoi{10.1088/0004-637X/792/1/1}

\bibitem[{Lopez {et~al.}(2012)Lopez, Fortney, \& Miller}]{lopez_how_2012}
Lopez, E.~D., Fortney, J.~J., \& Miller, N. 2012, The Astrophysical Journal, 761, 59, \dodoi{10.1088/0004-637X/761/1/59}

\bibitem[{{Lucy} \& {Sweeney}(1971)}]{Lucy1971}
{Lucy}, L.~B., \& {Sweeney}, M.~A. 1971, \aj, 76, 544, \dodoi{10.1086/111159}

\bibitem[{Luger {et~al.}(2015)Luger, Barnes, Lopez, Fortney, Jackson, \& Meadows}]{luger_habitable_2015}
Luger, R., Barnes, R., Lopez, E., {et~al.} 2015, Astrobiology, 15, 57, \dodoi{10.1089/ast.2014.1215}

\bibitem[{Luo {et~al.}(2024)Luo, Dorn, \& Deng}]{luo_majority_2024}
Luo, H., Dorn, C., \& Deng, J. 2024, Majority of water hides deep in the interiors of exoplanets,  arXiv.
\newblock \url{http://arxiv.org/abs/2401.16394}

\bibitem[{Luque \& Pallé(2022)}]{luque_density_2022}
Luque, R., \& Pallé, E. 2022, Science, 377, 1211, \dodoi{10.1126/science.abl7164}

\bibitem[{Léger {et~al.}(2004)Léger, Selsis, Sotin, Guillot, Despois, Mawet, Ollivier, Labèque, Valette, Brachet, Chazelas, \& Lammer}]{leger_new_2004}
Léger, A., Selsis, F., Sotin, C., {et~al.} 2004, Icarus, 169, 499, \dodoi{10.1016/j.icarus.2004.01.001}

\bibitem[{{MacDonald}(2023)}]{macdonald_poseidon_2023}
{MacDonald}, R.~J. 2023, The Journal of Open Source Software, 8, 4873, \dodoi{10.21105/joss.04873}

\bibitem[{{MacDonald} \& {Lewis}(2022)}]{macdonald_trident_2022}
{MacDonald}, R.~J., \& {Lewis}, N.~K. 2022, \apj, 929, 20, \dodoi{10.3847/1538-4357/ac47fe}

\bibitem[{{MacDonald} \& {Madhusudhan}(2017)}]{macdonald_hd_209458b_2017}
{MacDonald}, R.~J., \& {Madhusudhan}, N. 2017, \mnras, 469, 1979, \dodoi{10.1093/mnras/stx804}

\bibitem[{Madhusudhan {et~al.}(2021)Madhusudhan, Piette, \& Constantinou}]{madhusudhan_habitability_2021}
Madhusudhan, N., Piette, A. A.~A., \& Constantinou, S. 2021, The Astrophysical Journal, 918, 1, \dodoi{10.3847/1538-4357/abfd9c}

\bibitem[{Madhusudhan {et~al.}(2023)Madhusudhan, Sarkar, Constantinou, Holmberg, Piette, \& Moses}]{madhusudhan_carbon-bearing_2023}
Madhusudhan, N., Sarkar, S., Constantinou, S., {et~al.} 2023, The Astrophysical Journal Letters, 956, L13, \dodoi{10.3847/2041-8213/acf577}

\bibitem[{{Madhusudhan} \& {Seager}(2009)}]{MadhusudhanSeager2009apjRetrieval}
{Madhusudhan}, N., \& {Seager}, S. 2009, \apj, 707, 24, \dodoi{10.1088/0004-637X/707/1/24}

\bibitem[{{Mah} {et~al.}(2024){Mah}, {Savvidou}, \& {Bitsch}}]{Mah21}
{Mah}, J., {Savvidou}, S., \& {Bitsch}, B. 2024, \aap, 686, L17, \dodoi{10.1051/0004-6361/202450322}

\bibitem[{Malsky {et~al.}(2022)Malsky, Rogers, Kempton, \& Marounina}]{malsky_helium_2022}
Malsky, I., Rogers, L., Kempton, E. M.~R., \& Marounina, N. 2022, Nature Astronomy, 7, 57, \dodoi{10.1038/s41550-022-01823-8}

\bibitem[{{Mansfield} {et~al.}(2018){Mansfield}, {Bean}, {Oklop{\v{c}}i{\'c}}, {Kreidberg}, {D{\'e}sert}, {Kempton}, {Line}, {Fortney}, {Henry}, {Mallonn}, {Stevenson}, {Dragomir}, {Allart}, \& {Bourrier}}]{Mansfield2018HeliumHATP11}
{Mansfield}, M., {Bean}, J.~L., {Oklop{\v{c}}i{\'c}}, A., {et~al.} 2018, \apjl, 868, L34, \dodoi{10.3847/2041-8213/aaf166}

\bibitem[{{Mansfield} {et~al.}(2021){Mansfield}, {Line}, {Bean}, {Fortney}, {Parmentier}, {Wiser}, {Kempton}, {Gharib-Nezhad}, {Sing}, {L{\'o}pez-Morales}, {Baxter}, {D{\'e}sert}, {Swain}, \& {Roudier}}]{Mansfield2021}
{Mansfield}, M., {Line}, M.~R., {Bean}, J.~L., {et~al.} 2021, Nature Astronomy, 5, 1224, \dodoi{10.1038/s41550-021-01455-4}

\bibitem[{{May} {et~al.}(2023){May}, {MacDonald}, {Bennett}, {Moran}, {Wakeford}, {Peacock}, {Lustig-Yaeger}, {Highland}, {Stevenson}, {Sing}, {Mayorga}, {Batalha}, {Kirk}, {L{\'o}pez-Morales}, {Valenti}, {Alam}, {Alderson}, {Fu}, {Gonzalez-Quiles}, {Lothringer}, {Rustamkulov}, \& {Sotzen}}]{MayMacdonald2023doubletrouble}
{May}, E.~M., {MacDonald}, R.~J., {Bennett}, K.~A., {et~al.} 2023, \apjl, 959, L9, \dodoi{10.3847/2041-8213/ad054f}

\bibitem[{Mayor \& Queloz(1995)}]{Mayor1995AStar}
Mayor, M., \& Queloz, D. 1995, Nature, 378, 355, \dodoi{10.1038/378355a0}

\bibitem[{Mayor {et~al.}(2003)Mayor, Pepe, Queloz, Bouchy, Rupprecht, Lo~Curto, Avila, Benz, Bertaux, Bonfils, Dall, Dekker, Delabre, Eckert, Fleury, Gilliotte, Gojak, Guzman, Kohler, Lizon, Longinotti, Lovis, Megevand, Pasquini, Reyes, Sivan, Sosnowska, Soto, Udry, van Kesteren, Weber, \& Weilenmann}]{Mayor2003SettingHARPS}
Mayor, M., Pepe, F., Queloz, D., {et~al.} 2003, The Messenger, 114, 20

\bibitem[{{Mikal-Evans} {et~al.}(2023){Mikal-Evans}, {Madhusudhan}, {Dittmann}, {G{\"u}nther}, {Welbanks}, {Van Eylen}, {Crossfield}, {Daylan}, \& {Kreidberg}}]{mikal-evans_hubble_2023}
{Mikal-Evans}, T., {Madhusudhan}, N., {Dittmann}, J., {et~al.} 2023, \aj, 165, 84, \dodoi{10.3847/1538-3881/aca90b}

\bibitem[{Mollière {et~al.}(2015)Mollière, van Boekel, Dullemond, Henning, \& Mordasini}]{molliere_model_2015}
Mollière, P., van Boekel, R., Dullemond, C., Henning, T., \& Mordasini, C. 2015, The Astrophysical Journal, 813, 47, \dodoi{10.1088/0004-637X/813/1/47}

\bibitem[{Moran {et~al.}(2023)Moran, Stevenson, Sing, MacDonald, Kirk, Lustig-Yaeger, Peacock, Mayorga, Bennett, L{\'{o}}pez-Morales, May, Rustamkulov, Valenti, Adams~Redai, Alam, Batalha, Fu, Gonzalez-Quiles, Highland, Kruse, Lothringer, Ortiz~Ceballos, Sotzen, Wakeford, Moran, Stevenson, Sing, MacDonald, Kirk, Lustig-Yaeger, Peacock, Mayorga, Bennett, L{\'{o}}pez-Morales, May, Rustamkulov, Valenti, Adams~Redai, Alam, Batalha, Fu, Gonzalez-Quiles, Highland, Kruse, Lothringer, Ortiz~Ceballos, Sotzen, \& Wakeford}]{Moran2023HighObservations}
Moran, S.~E., Stevenson, K.~B., Sing, D.~K., {et~al.} 2023, The Astrophysical Journal Letters, 948, L11, \dodoi{10.48550/ARXIV.2305.00868}

\bibitem[{Mousis {et~al.}(2020)Mousis, Deleuil, Aguichine, Marcq, Naar, Aguirre, Brugger, \& Gonçalves}]{mousis_irradiated_2020}
Mousis, O., Deleuil, M., Aguichine, A., {et~al.} 2020, The Astrophysical Journal, 896, L22, \dodoi{10.3847/2041-8213/ab9530}

\bibitem[{{Mousis} {et~al.}(2019){Mousis}, {Ronnet}, \& {Lunine}}]{Mousis19_icelines}
{Mousis}, O., {Ronnet}, T., \& {Lunine}, J.~I. 2019, \apj, 875, 9, \dodoi{10.3847/1538-4357/ab0a72}

\bibitem[{{Nortmann} {et~al.}(2018){Nortmann}, {Pall{\'e}}, {Salz}, {Sanz-Forcada}, {Nagel}, {Alonso-Floriano}, {Czesla}, {Yan}, {Chen}, {Snellen}, {Zechmeister}, {Schmitt}, {L{\'o}pez-Puertas}, {Casasayas-Barris}, {Bauer}, {Amado}, {Caballero}, {Dreizler}, {Henning}, {Lamp{\'o}n}, {Montes}, {Molaverdikhani}, {Quirrenbach}, {Reiners}, {Ribas}, {S{\'a}nchez-L{\'o}pez}, {Schneider}, \& {Zapatero Osorio}}]{Nortmann2018Helium}
{Nortmann}, L., {Pall{\'e}}, E., {Salz}, M., {et~al.} 2018, Science, 362, 1388, \dodoi{10.1126/science.aat5348}

\bibitem[{{Ohno} {et~al.}(2024){Ohno}, {Schlawin}, {Bell}, {Murphy}, {Beatty}, {Welbanks}, {Greene}, {Fortney}, {Parmentier}, {Edelman}, {Mehta}, \& {Rieke}}]{Ohno2024GJ1214b}
{Ohno}, K., {Schlawin}, E., {Bell}, T.~J., {et~al.} 2024, arXiv e-prints, arXiv:2410.10186, \dodoi{10.48550/arXiv.2410.10186}

\bibitem[{{Oklop{\v{c}}i{\'c}}(2019)}]{oklopcic_helium_2019}
{Oklop{\v{c}}i{\'c}}, A. 2019, \apj, 881, 133, \dodoi{10.3847/1538-4357/ab2f7f}

\bibitem[{{Oklop{\v{c}}i{\'c}} \& {Hirata}(2018)}]{oklopcic_2018_new}
{Oklop{\v{c}}i{\'c}}, A., \& {Hirata}, C.~M. 2018, \apjl, 855, L11, \dodoi{10.3847/2041-8213/aaada9}

\bibitem[{{Orell-Miquel} {et~al.}(2024){Orell-Miquel}, {Murgas}, {Pall{\'e}}, {Mallorqu{\'\i}n}, {L{\'o}pez-Puertas}, {Lamp{\'o}n}, {Sanz-Forcada}, {Nortmann}, {Czesla}, {Nagel}, {Ribas}, {Stangret}, {Livingston}, {Knudstrup}, {Albrecht}, {Carleo}, {Caballero}, {Dai}, {Esparza-Borges}, {Fukui}, {Heng}, {Henning}, {Kagetani}, {Lesjak}, {de Leon}, {Montes}, {Morello}, {Narita}, {Quirrenbach}, {Amado}, {Reiners}, {Schweitzer}, \& {Vico Linares}}]{Orell-Miquel2024Helium}
{Orell-Miquel}, J., {Murgas}, F., {Pall{\'e}}, E., {et~al.} 2024, \aap, 689, A179, \dodoi{10.1051/0004-6361/202449411}

\bibitem[{Otegi {et~al.}(2020)Otegi, Dorn, Helled, Bouchy, Haldemann, \& Alibert}]{otegi_impact_2020}
Otegi, J.~F., Dorn, C., Helled, R., {et~al.} 2020, Astronomy \& Astrophysics, 640, A135, \dodoi{10.1051/0004-6361/202038006}

\bibitem[{{Owen} \& {Jackson}(2012)}]{owen_planetary_2012}
{Owen}, J.~E., \& {Jackson}, A.~P. 2012, \mnras, 425, 2931, \dodoi{10.1111/j.1365-2966.2012.21481.x}

\bibitem[{{Owen} \& {Murray-Clay}(2018)}]{owen_metallicity_2018}
{Owen}, J.~E., \& {Murray-Clay}, R. 2018, \mnras, 480, 2206, \dodoi{10.1093/mnras/sty1943}

\bibitem[{Owen \& Wu(2017)}]{owen_evaporation_2017}
Owen, J.~E., \& Wu, Y. 2017, The Astrophysical Journal, 847, 29, \dodoi{10.3847/1538-4357/aa890a}

\bibitem[{Parker {et~al.}(2025)Parker, Mendonça, Diamond-Lowe, Birkby, Meech, Vaughan, Brogi, Fisher, Buchhave, Bello-Arufe, Kreidberg, \& Dittmann}]{Parker2025GJ3090}
Parker, L.~T., Mendonça, J.~M., Diamond-Lowe, H., {et~al.} 2025, Monthly Notices of the Royal Astronomical Society, staf469, \dodoi{10.1093/mnras/staf469}

\bibitem[{Pelletier {et~al.}(2021)Pelletier, Benneke, Darveau-Bernier, Boucher, Cook, Piaulet, Coulombe, Artigau, Lafrenière, Delisle, Allart, Doyon, Donati, Fouqué, Moutou, Cadieux, Delfosse, Hébrard, Martins, Martioli, \& Vandal}]{pelletier_where_2021}
Pelletier, S., Benneke, B., Darveau-Bernier, A., {et~al.} 2021, The Astronomical Journal, 162, 73, \dodoi{10.3847/1538-3881/ac0428}

\bibitem[{P\'erez \& Granger(2007)}]{PER-GRA:2007}
P\'erez, F., \& Granger, B.~E. 2007, Computing in Science and Engineering, 9, 21, \dodoi{10.1109/MCSE.2007.53}

\bibitem[{Petigura {et~al.}(2022)Petigura, Rogers, Isaacson, Owen, Kraus, Winn, MacDougall, Howard, Fulton, Kosiarek, Weiss, Behmard, \& Blunt}]{petigura_california-kepler_2022}
Petigura, E.~A., Rogers, J.~G., Isaacson, H., {et~al.} 2022, arXiv:2201.10020 [astro-ph].
\newblock \url{http://arxiv.org/abs/2201.10020}

\bibitem[{Piaulet {et~al.}(2023)Piaulet, Benneke, Almenara, Dragomir, Knutson, Thorngren, Peterson, Crossfield, Kempton, Kubyshkina, Howard, Angus, Isaacson, Weiss, Beichman, Fortney, Fossati, Lammer, McCullough, Morley, \& Wong}]{piaulet_evidence_2023}
Piaulet, C., Benneke, B., Almenara, J.~M., {et~al.} 2023, Nature Astronomy, 7, 206, \dodoi{10.1038/s41550-022-01835-4}

\bibitem[{{Piaulet-Ghorayeb}(2024)}]{stctm_zenodo_temp}
{Piaulet-Ghorayeb}, C. 2024, {First release of stellar contamination modeling and retrieval code}, v1.0.0,  Zenodo, \dodoi{10.5281/zenodo.13153252}

\bibitem[{Piaulet-Ghorayeb {et~al.}(2024)Piaulet-Ghorayeb, Benneke, Radica, Raul, Coulombe, Ahrer, Kubyshkina, Howard, Krissansen-Totton, MacDonald, Roy, Louca, Christie, Fournier-Tondreau, Allart, Miguel, Schlichting, Welbanks, Cadieux, Dorn, Evans-Soma, Fortney, Pierrehumbert, Lafrenière, Acuña, Komacek, Innes, Beatty, Cloutier, Doyon, Gagnebin, Gapp, \& Knutson}]{Piaulet_Ghorayeb_2024}
Piaulet-Ghorayeb, C., Benneke, B., Radica, M., {et~al.} 2024, The Astrophysical Journal Letters, 974, L10, \dodoi{10.3847/2041-8213/ad6f00}

\bibitem[{Pinhas {et~al.}(2018)Pinhas, Rackham, Madhusudhan, \& Apai}]{Pinhas2018RetrievalAURA}
Pinhas, A., Rackham, B.~V., Madhusudhan, N., \& Apai, D. 2018, Monthly Notices of the Royal Astronomical Society, 480, \dodoi{10.1093/MNRAS/STY2209}

\bibitem[{Polyansky {et~al.}(2018)Polyansky, Kyuberis, Zobov, Tennyson, Yurchenko, \& Lodi}]{ExoMol_H2O}
Polyansky, O.~L., Kyuberis, A.~A., Zobov, N.~F., {et~al.} 2018, Mon. Not. R. Astron. Soc., 480, 2597, \dodoi{10.1093/mnras/sty1877}

\bibitem[{Rackham {et~al.}(2018)Rackham, Apai, \& Giampapa}]{Rackham2018ThePlanets}
Rackham, B.~V., Apai, D., \& Giampapa, M.~S. 2018, The Astrophysical Journal, 853, 122, \dodoi{10.3847/1538-4357/aaa08c}

\bibitem[{Rackham {et~al.}(2019)Rackham, Apai, \& Giampapa}]{Rackham2019TheStars}
---. 2019, The Astronomical Journal, 157, 96, \dodoi{10.3847/1538-3881/aaf892}

\bibitem[{{Rackham} {et~al.}(2023){Rackham}, {Espinoza}, {Berdyugina}, {Korhonen}, {MacDonald}, {Montet}, {Morris}, {Oshagh}, {Shapiro}, {Unruh}, {Quintana}, {Zellem}, {Apai}, {Barclay}, {Barstow}, {Bruno}, {Carone}, {Casewell}, {Cegla}, {Criscuoli}, {Fischer}, {Fournier}, {Giampapa}, {Giles}, {Iyer}, {Kopp}, {Kostogryz}, {Krivova}, {Mallonn}, {McGruder}, {Molaverdikhani}, {Newton}, {Panja}, {Peacock}, {Reardon}, {Roettenbacher}, {Scandariato}, {Solanki}, {Stassun}, {Steiner}, {Stevenson}, {Tregloan-Reed}, {Valio}, {Wedemeyer}, {Welbanks}, {Yu}, {Alam}, {Davenport}, {Deming}, {Dong}, {Ducrot}, {Fisher}, {Gilbert}, {Kostov}, {L{\'o}pez-Morales}, {Line}, {Mo{\v{c}}nik}, {Mullally}, {Paudel}, {Ribas}, \& {Valenti}}]{SAG23_effect_TLSE}
{Rackham}, B.~V., {Espinoza}, N., {Berdyugina}, S.~V., {et~al.} 2023, RAS Techniques and Instruments, 2, 148, \dodoi{10.1093/rasti/rzad009}

\bibitem[{Radica(2024)}]{Radica2024b}
Radica, M. 2024, Journal of Open Source Software, 9, 6898, \dodoi{10.21105/joss.06898}

\bibitem[{{Radica}(2024)}]{Radica2024c}
{Radica}, M. 2024, {radicamc/exoUPRF: v1.0.1}, v1.0.1,  Zenodo, \dodoi{10.5281/zenodo.12628066}

\bibitem[{{Radica} {et~al.}(2022){Radica}, {Albert}, {Taylor}, {Lafreni{\`e}re}, {Coulombe}, {Darveau-Bernier}, {Doyon}, {Cook}, {Cowan}, {Espinoza}, {Johnstone}, {Kaltenegger}, {Piaulet}, {Roy}, \& {Talens}}]{Radica2022}
{Radica}, M., {Albert}, L., {Taylor}, J., {et~al.} 2022, \pasp, 134, 104502, \dodoi{10.1088/1538-3873/ac9430}

\bibitem[{{Radica} {et~al.}(2023){Radica}, {Welbanks}, {Espinoza}, {Taylor}, {Coulombe}, {Feinstein}, {Goyal}, {Scarsdale}, {Albert}, {Baghel}, {Bean}, {Blecic}, {Lafreni{\`e}re}, {MacDonald}, {Zamyatina}, {Allart1}, {Artigau}, {Batalha}, {Cook}, {Cowan}, {Dang}, {Doyon}, {Fournier-Tondreau}, {Johnstone}, {Line}, {Moran}, {Mukherjee}, {Pelletier}, {Roy}, {Talens}, {Filippazzo}, {Pontoppidan}, \& {Volk}}]{Radica2023}
{Radica}, M., {Welbanks}, L., {Espinoza}, N., {et~al.} 2023, \mnras, 524, 835, \dodoi{10.1093/mnras/stad1762}

\bibitem[{Radica {et~al.}(2024)Radica, Coulombe, Taylor, Albert, Allart, Benneke, Cowan, Dang, Lafrenière, Thorngren, Artigau, Doyon, Flagg, Johnstone, Pelletier, \& Roy}]{radica_muted_2024}
Radica, M., Coulombe, L.-P., Taylor, J., {et~al.} 2024, The Astrophysical Journal, 962, L20, \dodoi{10.3847/2041-8213/ad20e4}

\bibitem[{{Radica} {et~al.}(2025){Radica}, {Piaulet-Ghorayeb}, {Taylor}, {Coulombe}, {Benneke}, {Albert}, {Artigau}, {Cowan}, {Doyon}, {Lafreni{\`e}re}, {L'Heureux}, \& {Lim}}]{radica_promise_2025}
{Radica}, M., {Piaulet-Ghorayeb}, C., {Taylor}, J., {et~al.} 2025, \apjl, 979, L5, \dodoi{10.3847/2041-8213/ada381}

\bibitem[{Ricker {et~al.}(2015)Ricker, Winn, Vanderspek, Latham, Bakos, Bean, Berta-Thompson, Brown, Buchhave, Butler, Butler, Chaplin, Charbonneau, Christensen-Dalsgaard, Clampin, Deming, Doty, De~Lee, Dressing, Dunham, Endl, Fressin, Ge, Henning, Holman, Howard, Ida, Jenkins, Jernigan, Johnson, Kaltenegger, Kawai, Kjeldsen, Laughlin, Levine, Lin, Lissauer, MacQueen, Marcy, McCullough, Morton, Narita, Paegert, Palle, Pepe, Pepper, Quirrenbach, Rinehart, Sasselov, Sato, Seager, Sozzetti, Stassun, Sullivan, Szentgyorgyi, Torres, Udry, \& Villasenor}]{Ricker2015TransitingTESS}
Ricker, G.~R., Winn, J.~N., Vanderspek, R., {et~al.} 2015, Journal of Astronomical Telescopes, Instruments, and Systems, Volume 1, id. 014003 (2015)., 1, 014003, \dodoi{10.1117/1.JATIS.1.1.014003}

\bibitem[{{Rigby} {et~al.}(2024){Rigby}, {Pica-Ciamarra}, {Holmberg}, {Madhusudhan}, {Constantinou}, {Schaefer}, {Deng}, {Lee}, \& {Moses}}]{Rigby2024}
{Rigby}, F.~E., {Pica-Ciamarra}, L., {Holmberg}, M., {et~al.} 2024, \apj, 975, 101, \dodoi{10.3847/1538-4357/ad6c38}

\bibitem[{Rogers {et~al.}(2021)Rogers, Gupta, Owen, \& Schlichting}]{rogers_photoevaporation_2021}
Rogers, J.~G., Gupta, A., Owen, J.~E., \& Schlichting, H.~E. 2021, Monthly Notices of the Royal Astronomical Society, 508, 5886, \dodoi{10.1093/mnras/stab2897}

\bibitem[{{Rogers} {et~al.}(2023){Rogers}, {Schlichting}, \& {Owen}}]{Rogers2023}
{Rogers}, J.~G., {Schlichting}, H.~E., \& {Owen}, J.~E. 2023, \apjl, 947, L19, \dodoi{10.3847/2041-8213/acc86f}

\bibitem[{{Rogers} {et~al.}(2024){Rogers}, {Schlichting}, \& {Young}}]{Rogers2024}
{Rogers}, J.~G., {Schlichting}, H.~E., \& {Young}, E.~D. 2024, \apj, 970, 47, \dodoi{10.3847/1538-4357/ad5287}

\bibitem[{Rogers \& Seager(2010)}]{rogers_framework_2010}
Rogers, L.~A., \& Seager, S. 2010, The Astrophysical Journal, 712, 974, \dodoi{10.1088/0004-637X/712/2/974}

\bibitem[{Roy {et~al.}(2023)Roy, Benneke, Piaulet, Gully-Santiago, Crossfield, Morley, Kreidberg, Mikal-Evans, Brande, Delisle, Greene, Hardegree-Ullman, Barman, Christiansen, Dragomir, Fortney, Howard, Kosiarek, \& Lothringer}]{roy_water_2023}
Roy, P.-A., Benneke, B., Piaulet, C., {et~al.} 2023, The Astrophysical Journal Letters, 954, L52, \dodoi{10.3847/2041-8213/acebf0}

\bibitem[{{Salz} {et~al.}(2016){Salz}, {Czesla}, {Schneider}, \& {Schmitt}}]{Salz2016}
{Salz}, M., {Czesla}, S., {Schneider}, P.~C., \& {Schmitt}, J.~H.~M.~M. 2016, \aap, 586, A75, \dodoi{10.1051/0004-6361/201526109}

\bibitem[{{Schlawin} {et~al.}(2024){Schlawin}, {Ohno}, {Bell}, {Murphy}, {Welbanks}, {Beatty}, {Greene}, {Fortney}, {Parmentier}, {Edelman}, {Gill}, {Anderson}, {Wheatley}, {Henry}, {Mehta}, {Kreidberg}, \& {Rieke}}]{Schlawin2024GJ1214b}
{Schlawin}, E., {Ohno}, K., {Bell}, T.~J., {et~al.} 2024, \apjl, 974, L33, \dodoi{10.3847/2041-8213/ad7fef}

\bibitem[{Schlichting \& Young(2022)}]{schlichting_chemical_2022}
Schlichting, H.~E., \& Young, E.~D. 2022, The Planetary Science Journal, 3, 127, \dodoi{10.3847/PSJ/ac68e6}

\bibitem[{{Schmidt} {et~al.}(2025){Schmidt}, {MacDonald}, {Tsai}, {Radica}, {Wang}, {Ahrer}, {Bell}, {Fisher}, {Thorngren}, {Wogan}, {May}, {Ferrari}, {Bennett}, {Rustamkulov}, {L{\'o}pez-Morales}, \& {Sing}}]{Schmidt2025}
{Schmidt}, S.~P., {MacDonald}, R.~J., {Tsai}, S.-M., {et~al.} 2025, arXiv e-prints, arXiv:2501.18477, \dodoi{10.48550/arXiv.2501.18477}

\bibitem[{{Seager} \& {Mall{\'e}n-Ornelas}(2003)}]{Seager2003}
{Seager}, S., \& {Mall{\'e}n-Ornelas}, G. 2003, \apj, 585, 1038, \dodoi{10.1086/346105}

\bibitem[{{Shorttle} {et~al.}(2024){Shorttle}, {Jordan}, {Nicholls}, {Lichtenberg}, \& {Bower}}]{Shorttle2024K2-18b}
{Shorttle}, O., {Jordan}, S., {Nicholls}, H., {Lichtenberg}, T., \& {Bower}, D.~J. 2024, \apjl, 962, L8, \dodoi{10.3847/2041-8213/ad206e}

\bibitem[{Skilling(2004)}]{skilling_nested_2004}
Skilling, J. 2004, 735, 395, \dodoi{10.1063/1.1835238}

\bibitem[{Skilling(2006)}]{skilling_nested_2006}
---. 2006, Bayesian Analysis, 1, 833, \dodoi{10.1214/06-BA127}

\bibitem[{Skrutskie {et~al.}(2006)Skrutskie, Cutri, Stiening, Weinberg, Schneider, Carpenter, Beichman, Capps, Chester, Elias, Huchra, Liebert, Lonsdale, Monet, Price, Seitzer, Jarrett, Kirkpatrick, Gizis, Howard, Evans, Fowler, Fullmer, Hurt, Light, Kopan, Marsh, McCallon, Tam, Van~Dyk, \& Wheelock}]{Skrutskie2006The2MASS}
Skrutskie, M.~F., Cutri, R.~M., Stiening, R., {et~al.} 2006, The Astronomical Journal, 131, 1163, \dodoi{10.1086/498708}

\bibitem[{Spake {et~al.}(2021)Spake, Oklopčić, \& Hillenbrand}]{spake_posttransit_2021}
Spake, J.~J., Oklopčić, A., \& Hillenbrand, L.~A. 2021, The Astronomical Journal, 162, 284, \dodoi{10.3847/1538-3881/ac178a}

\bibitem[{Spake {et~al.}(2018)Spake, Sing, Evans, {Oklopčić}, ~, Bourrier, Kreidberg, Rackham, Irwin, Ehrenreich, Wyttenbach, Wakeford, Zhou, Chubb, Nikolov, Goyal, Henry, Williamson, Blumenthal, Anderson, Hellier, Charbonneau, Udry, \& Madhusudhan}]{spake_helium_2018}
Spake, J.~J., Sing, D.~K., Evans, T.~M., {et~al.} 2018, Nature, 557, 68, \dodoi{10.1038/s41586-018-0067-5}

\bibitem[{Speagle(2020)}]{Speagle2020Dynesty:Evidences}
Speagle, J.~S. 2020, Monthly Notices of the Royal Astronomical Society, 493, 3132, \dodoi{10.1093/mnras/staa278}

\bibitem[{{Teske} {et~al.}(2025){Teske}, {Batalha}, {Wallack}, {Kirk}, {Wogan}, {Gordon}, {Alam}, {Aguichine}, {Wolfgang}, {Wakeford}, {Scarsdale}, {Adams Redai}, {Moran}, {L{\'o}pez-Morales}, {Meech}, {Gao}, {Batalha}, {Alderson}, \& {Gagnebin}}]{teske_toi776c_2025}
{Teske}, J., {Batalha}, N.~E., {Wallack}, N.~L., {et~al.} 2025, arXiv e-prints, arXiv:2502.20501, \dodoi{10.48550/arXiv.2502.20501}

\bibitem[{Townsend \& Lopez(2023)}]{townsend_msg_2023}
Townsend, R., \& Lopez, A. 2023, Journal of Open Source Software, 8, 4602, \dodoi{10.21105/joss.04602}

\bibitem[{Trotta(2008)}]{trotta_bayes_2008}
Trotta, R. 2008, Contemporary Physics, 49, 71, \dodoi{10.1080/00107510802066753}

\bibitem[{{Tsai} {et~al.}(2022){Tsai}, {Lee}, \& {Pierrehumbert}}]{Tsai2022}
{Tsai}, S.-M., {Lee}, E. K.~H., \& {Pierrehumbert}, R. 2022, Astronomy \& Astrophysics, 664, A82, \dodoi{10.1051/0004-6361/202142816}

\bibitem[{{Tsai} {et~al.}(2017){Tsai}, {Lyons}, {Grosheintz}, {Rimmer}, {Kitzmann}, \& {Heng}}]{Tsai2017}
{Tsai}, S.-M., {Lyons}, J.~R., {Grosheintz}, L., {et~al.} 2017, {VULCAN: Chemical Kinetics For Exoplanetary Atmospheres}, Astrophysics Source Code Library, record ascl:1704.011

\bibitem[{Tsai {et~al.}(2023)Tsai, Lee, Powell, Gao, Zhang, Moses, H{\'{e}}brard, Venot, Parmentier, Jordan, Hu, Alam, Alderson, Batalha, Bean, Benneke, Bierson, Brady, Carone, Carter, Chubb, Inglis, Leconte, Line, L{\'{o}}pez-Morales, Miguel, Molaverdikhani, Rustamkulov, Sing, Stevenson, Wakeford, Yang, Aggarwal, Baeyens, Barat, de~Val-Borro, Daylan, Fortney, France, Goyal, Grant, Kirk, Kreidberg, Louca, Moran, Mukherjee, Nasedkin, Ohno, Rackham, Redfield, Taylor, Tremblin, Visscher, Wallack, Welbanks, Youngblood, Ahrer, Batalha, Behr, Berta-Thompson, Blecic, Casewell, Crossfield, Crouzet, Cubillos, Decin, D{\'{e}}sert, Feinstein, Gibson, Harrington, Heng, Henning, Kempton, Krick, Lagage, Lendl, Lothringer, Mansfield, Mayne, Mikal-Evans, Palle, Schlawin, Shorttle, Wheatley, \& Yurchenko}]{Tsai2023PhotochemicallyWASP-39b}
Tsai, S.~M., Lee, E.~K., Powell, D., {et~al.} 2023, Nature, 617, 483, \dodoi{10.1038/S41586-023-05902-2}

\bibitem[{Tsiaras {et~al.}(2019)Tsiaras, Waldmann, Tinetti, Tennyson, \& Yurchenko}]{tsiaras_water_2019}
Tsiaras, A., Waldmann, I.~P., Tinetti, G., Tennyson, J., \& Yurchenko, S.~N. 2019, Nature Astronomy, 3, 1086, \dodoi{10.1038/s41550-019-0878-9}

\bibitem[{Underwood {et~al.}(2016)Underwood, Tennyson, Yurchenko, Huang, Schwenke, Lee, Clausen, \& Fateev}]{ExoMol_SO2}
Underwood, D.~S., Tennyson, J., Yurchenko, S.~N., {et~al.} 2016, Mon. Not. R. Astron. Soc., 459, 3890, \dodoi{10.1093/mnras/stw849}

\bibitem[{Valencia(2010)}]{valencia_composition_2010}
Valencia, D. 2010, Proceedings of the International Astronomical Union, 6, 181, \dodoi{10.1017/S1743921311020151}

\bibitem[{{Vazan} {et~al.}(2022){Vazan}, {Sari}, \& {Kessel}}]{vazan22}
{Vazan}, A., {Sari}, R., \& {Kessel}, R. 2022, \apj, 926, 150, \dodoi{10.3847/1538-4357/ac458c}

\bibitem[{Venturini {et~al.}(2020)Venturini, Guilera, Haldemann, Ronco, \& Mordasini}]{venturini_nature_2020}
Venturini, J., Guilera, O.~M., Haldemann, J., Ronco, M.~P., \& Mordasini, C. 2020, Astronomy \& Astrophysics, 643, L1, \dodoi{10.1051/0004-6361/202039141}

\bibitem[{Venturini {et~al.}(2024)Venturini, Ronco, Guilera, Haldemann, Mordasini, \& Bertolami}]{venturini_fading_2024}
Venturini, J., Ronco, M.~P., Guilera, O.~M., {et~al.} 2024, A fading radius valley towards {M}-dwarfs, a persistent density valley across stellar types,  arXiv.
\newblock \url{http://arxiv.org/abs/2404.01967}

\bibitem[{{Vidal-Madjar} {et~al.}(2003){Vidal-Madjar}, {Lecavelier des Etangs}, {D{\'e}sert}, {Ballester}, {Ferlet}, {H{\'e}brard}, \& {Mayor}}]{VidalMadjar2003}
{Vidal-Madjar}, A., {Lecavelier des Etangs}, A., {D{\'e}sert}, J.~M., {et~al.} 2003, \nat, 422, 143, \dodoi{10.1038/nature01448}

\bibitem[{Virtanen {et~al.}(2020)Virtanen, Gommers, Oliphant, Haberland, Reddy, Cournapeau, Burovski, Peterson, Weckesser, Bright, {van der Walt}, Brett, Wilson, Millman, Mayorov, Nelson, Jones, Kern, Larson, Carey, Polat, Feng, Moore, {VanderPlas}, Laxalde, Perktold, Cimrman, Henriksen, Quintero, Harris, Archibald, Ribeiro, Pedregosa, {van Mulbregt}, \& {SciPy 1.0 Contributors}}]{2020SciPy-NMeth}
Virtanen, P., Gommers, R., Oliphant, T.~E., {et~al.} 2020, Nature Methods, 17, 261, \dodoi{10.1038/s41592-019-0686-2}

\bibitem[{Vissapragada {et~al.}(2024)Vissapragada, Greklek-McKeon, Linssen, MacLeod, Thorngren, Gao, Knutson, Latham, López-Morales, Oklopčić, González, Saidel, Tumborang, \& Yoshida}]{vissapragada_helium_2024}
Vissapragada, S., Greklek-McKeon, M., Linssen, D., {et~al.} 2024, Helium in the {Extended} {Atmosphere} of the {Warm} {Super}-{Puff} {TOI}-1420b,  arXiv.
\newblock \url{http://arxiv.org/abs/2403.05614}

\bibitem[{{Vissapragada} {et~al.}(2024){Vissapragada}, {McCreery}, {Dos Santos}, {Espinoza}, {McWilliam}, {Matsunaga}, {Redai}, {Behr}, {France}, {Hamano}, {Hull}, {Ikeda}, {Katoh}, {Kawakita}, {L{\'o}pez-Morales}, {Ortiz Ceballos}, {Otsubo}, {Sarugaku}, \& {Takeuchi}}]{Vissapragada2024}
{Vissapragada}, S., {McCreery}, P., {Dos Santos}, L.~A., {et~al.} 2024, \apjl, 962, L19, \dodoi{10.3847/2041-8213/ad23cf}

\bibitem[{{von Zahn} \& {Hunten}(1996)}]{vonZahn1996}
{von Zahn}, U., \& {Hunten}, D.~M. 1996, Science, 272, 849, \dodoi{10.1126/science.272.5263.849}

\bibitem[{{von Zahn} {et~al.}(1998){von Zahn}, {Hunten}, \& {Lehmacher}}]{vonzahn_helium_1998}
{von Zahn}, U., {Hunten}, D.~M., \& {Lehmacher}, G. 1998, \jgr, 103, 22815, \dodoi{10.1029/98JE00695}

\bibitem[{Wakeford {et~al.}(2019)Wakeford, Wilson, Stevenson, \& Lewis}]{wakeford_exoplanet_2019}
Wakeford, H.~R., Wilson, T.~J., Stevenson, K.~B., \& Lewis, N.~K. 2019, Research Notes of the American Astronomical Society, 3, 7, \dodoi{10.3847/2515-5172/aafc63}

\bibitem[{Wallack {et~al.}(2024)Wallack, Batalha, Alderson, Scarsdale, Redai, Aguichine, Alam, Gao, Wolfgang, Batalha, Kirk, López-Morales, Moran, Teske, Wakeford, \& Wogan}]{wallack_jwst_2024}
Wallack, N.~L., Batalha, N.~E., Alderson, L., {et~al.} 2024, {JWST} {COMPASS}: {A} {NIRSpec}/{G395H} {Transmission} {Spectrum} of the {Sub}-{Neptune} {TOI}-836c,  arXiv.
\newblock \url{http://arxiv.org/abs/2404.01264}

\bibitem[{{Wallack} {et~al.}(2024){Wallack}, {Batalha}, {Alderson}, {Scarsdale}, {Adams Redai}, {Aguichine}, {Alam}, {Gao}, {Wolfgang}, {Batalha}, {Kirk}, {L{\'o}pez-Morales}, {Moran}, {Teske}, {Wakeford}, \& {Wogan}}]{wallack_toi836_2024}
{Wallack}, N.~L., {Batalha}, N.~E., {Alderson}, L., {et~al.} 2024, \aj, 168, 77, \dodoi{10.3847/1538-3881/ad3917}

\bibitem[{{Watson} {et~al.}(1981){Watson}, {Donahue}, \& {Walker}}]{Watson1981}
{Watson}, A.~J., {Donahue}, T.~M., \& {Walker}, J.~C.~G. 1981, \icarus, 48, 150, \dodoi{10.1016/0019-1035(81)90101-9}

\bibitem[{Welbanks \& Madhusudhan(2019)}]{welbanks_degeneracies_2019}
Welbanks, L., \& Madhusudhan, N. 2019, The Astronomical Journal, 157, 206, \dodoi{10.3847/1538-3881/ab14de}

\bibitem[{Welbanks \& Madhusudhan(2021)}]{Welbanks2021Aurora:Spectra}
---. 2021, The Astrophysical Journal, 913, \dodoi{10.3847/1538-4357/abee94}

\bibitem[{{Welbanks} {et~al.}(2024){Welbanks}, {Bell}, {Beatty}, {Line}, {Ohno}, {Fortney}, {Schlawin}, {Greene}, {Rauscher}, {McGill}, {Murphy}, {Parmentier}, {Tang}, {Edelman}, {Mukherjee}, {Wiser}, {Lagage}, {Dyrek}, \& {Arnold}}]{Welbanks2024}
{Welbanks}, L., {Bell}, T.~J., {Beatty}, T.~G., {et~al.} 2024, \nat, 630, 836, \dodoi{10.1038/s41586-024-07514-w}

\bibitem[{Wogan {et~al.}(2024)Wogan, Batalha, Zahnle, Krissansen-Totton, Tsai, \& Hu}]{wogan_jwst_2024}
Wogan, N.~F., Batalha, N.~E., Zahnle, K., {et~al.} 2024, {JWST} observations of {K2}-18b can be explained by a gas-rich mini-{Neptune} with no habitable surface,  arXiv, \dodoi{10.48550/arXiv.2401.11082}

\bibitem[{{Xue} {et~al.}(2024){Xue}, {Bean}, {Zhang}, {Welbanks}, {Lunine}, \& {August}}]{Xue2024HD209}
{Xue}, Q., {Bean}, J.~L., {Zhang}, M., {et~al.} 2024, \apjl, 963, L5, \dodoi{10.3847/2041-8213/ad2682}

\bibitem[{{Youngblood} {et~al.}(2016){Youngblood}, {France}, {Loyd}, {Linsky}, {Redfield}, {Schneider}, {Wood}, {Brown}, {Froning}, {Miguel}, {Rugheimer}, \& {Walkowicz}}]{Youngblood2016ApJ...824..101Y}
{Youngblood}, A., {France}, K., {Loyd}, R.~O.~P., {et~al.} 2016, \apj, 824, 101, \dodoi{10.3847/0004-637X/824/2/101}

\bibitem[{Yurchenko {et~al.}(2020)Yurchenko, Mellor, Freedman, \& Tennyson}]{ExoMol_CO2}
Yurchenko, S.~N., Mellor, T.~M., Freedman, R.~S., \& Tennyson, J. 2020, Mon. Not. R. Astron. Soc., 496, 5282, \dodoi{10.1093/mnras/staa1874}

\bibitem[{{Zamyatina} {et~al.}(2023){Zamyatina}, {H{\'e}brard}, {Drummond}, {Mayne}, {Manners}, {Christie}, {Tremblin}, {Sing}, \& {Kohary}}]{zamyatina_2023}
{Zamyatina}, M., {H{\'e}brard}, E., {Drummond}, B., {et~al.} 2023, \mnras, 519, 3129, \dodoi{10.1093/mnras/stac3432}

\bibitem[{{Zamyatina} {et~al.}(2024){Zamyatina}, {Christie}, {H{\'e}brard}, {Mayne}, {Radica}, {Taylor}, {Baskett}, {Moore}, {Lils}, {Sergeev}, {Ahrer}, {Manners}, {Kohary}, \& {Feinstein}}]{Zamyatina2024WASP-96bCH4}
{Zamyatina}, M., {Christie}, D.~A., {H{\'e}brard}, E., {et~al.} 2024, \mnras, 529, 1776, \dodoi{10.1093/mnras/stae600}

\bibitem[{{Zhang} {et~al.}(2023){Zhang}, {Knutson}, {Dai}, {Wang}, {Ricker}, {Schwarz}, {Mann}, \& {Collins}}]{zhang_detection_2023}
{Zhang}, M., {Knutson}, H.~A., {Dai}, F., {et~al.} 2023, \aj, 165, 62, \dodoi{10.3847/1538-3881/aca75b}

\bibitem[{{Zhang} {et~al.}(2022{\natexlab{a}}){Zhang}, {Knutson}, {Wang}, {Dai}, \& {Barrag{\'a}n}}]{ZhangTOI560}
{Zhang}, M., {Knutson}, H.~A., {Wang}, L., {Dai}, F., \& {Barrag{\'a}n}, O. 2022{\natexlab{a}}, \aj, 163, 67, \dodoi{10.3847/1538-3881/ac3fa7}

\bibitem[{{Zhang} {et~al.}(2022{\natexlab{b}}){Zhang}, {Knutson}, {Wang}, {Dai}, {dos Santos}, {Fossati}, {Henry}, {Ehrenreich}, {Alibert}, {Hoyer}, {Wilson}, \& {Bonfanti}}]{Zhang2022}
{Zhang}, M., {Knutson}, H.~A., {Wang}, L., {et~al.} 2022{\natexlab{b}}, \aj, 163, 68, \dodoi{10.3847/1538-3881/ac3f3b}

\bibitem[{{Zhang} {et~al.}(2024){Zhang}, {Bean}, {Wilson}, {Duvvuri}, {Schneider}, {Knutson}, {Dai}, {Collins}, {Watkins}, {Schwarz}, {Barkaoui}, {Shporer}, {Horne}, {Sefako}, {Murgas}, \& {Palle}}]{Zhang2024}
{Zhang}, M., {Bean}, J.~L., {Wilson}, D., {et~al.} 2024, arXiv e-prints, arXiv:2409.08318, \dodoi{10.48550/arXiv.2409.08318}

\bibitem[{{Zieba} {et~al.}(2023){Zieba}, {Kreidberg}, {Ducrot}, {Gillon}, {Morley}, {Schaefer}, {Tamburo}, {Koll}, {Lyu}, {Acu{\~n}a}, {Agol}, {Iyer}, {Hu}, {Lincowski}, {Meadows}, {Selsis}, {Bolmont}, {Mandell}, \& {Suissa}}]{Zieba2023Trappist1c}
{Zieba}, S., {Kreidberg}, L., {Ducrot}, E., {et~al.} 2023, \nat, 620, 746, \dodoi{10.1038/s41586-023-06232-z}

\end{thebibliography}

\bibliographystyle{aasjournal}

\end{document}